\documentclass[a4paper,11pt,reqno]{article}
\usepackage{epsfig}
\usepackage{hyperref,amsmath,mathrsfs}
\usepackage{graphicx}
\usepackage{parskip}
\usepackage{pgf,tikz,tikz-cd,pgfplots,physics}
\usepackage{amsfonts}
\usepackage{pdflscape}
\usepackage{amssymb}
\usepackage[bbgreekl]{mathbbol}
\usepackage{bbm}

\usepackage{a4wide}

\usepackage{multirow} 
\usepackage{color} 
\definecolor{rougef}{rgb}{0.7,0,0}
\definecolor{vertf}{rgb}{0,0.6,0}
\definecolor{bleuf}{rgb}{0,0,0.9}

\newcommand{\be}{\begin{equation}}
\newcommand{\ee}{\end{equation}}
\newcommand{\bea}{\begin{eqnarray}}
\newcommand{\eea}{\end{eqnarray}}

\newcommand{\eq}[1]{(\ref{#1})}

\def\y{\eta}
\def\a{\alpha} \def\ad{\dot{\a}} \def\ua{{\underline \a}}

\def\b{\beta}  \def\bd{\dot{\b}} 
\def\c{\gamma} 
\def\C{\Gamma}
\def\d{\delta} 
\def\D{\Delta}
\def\e{\epsilon}

\def\k{\kappa}

\def\l{\lambda}
\def\L{\Lambda}

\def\s{\sigma}

\def\y{\eta}

\def\sb{{\bar\s}}

\def\cP{{\cal P}}

\def\yb{{\bar y}}
\def\zb{{\bar z}}

\def\ft#1#2{{\textstyle{{\scriptstyle #1}
\over {\scriptstyle #2}}}} 

\def\mso{\mathfrak{so}}
\def\msu{\mathfrak{su}}
\def\msl{\mathfrak{sl}}
\def\msp{\mathfrak{sp}}
\def\miso{\mathfrak{iso}}

\def\Real{{\mathbb R}}
\def\Comp{{\mathbb C}}

\def\ket#1{|#1\rangle}
\def\bra#1{\langle#1|}
\numberwithin{equation}{section}
\usepackage[utf8]{inputenc}
\begin{document}

\begin{center}
\thispagestyle{empty}

{\LARGE Fractional Spins, Unfolding, and Holography:}\\[10pt] {\LARGE I. Parent field equations for dual higher-spin gravity reductions}

\vskip .6cm

{Felipe Diaz${}^{a,}$\footnote{\href{mailto:f.diazmartinez@uandresbello.edu }{\texttt{f.diazmartinez@uandresbello.edu}}}, Carlo Iazeolla${}^{b,}$\footnote{\href{mailto:c.iazeolla@gmail.com  }{\texttt{c.iazeolla@gmail.com }}}, and Per Sundell${}^{c,d,}$\footnote{
\href{mailto:per.anders.sundell@gmail.com }{\texttt{per.anders.sundell@gmail.com}}}}
\vskip .2cm
$^{a}$ \textit{\small  Departamento de Ciencias F\'isicas, Universidad Andres Bello, Sazi\'e 2212, Santiago, Chile}\\
$^{b}$\textit{\small Dipartimento di Scienze Ingegneristiche, G. Marconi University -- Via Plinio 44, 00193, Roma, Italy \& Sezione INFN Roma “Tor Vergata” -- Via della Ricerca Scientifica 1, 00133, Roma, Italy}\\
${}^{c}$\textit{\small Instituto de Ciencias Exactas y Naturales, Universidad Arturo Prat, Playa Brava 3265, 1111346 Iquique, Chile}\\
${}^{d}$\textit{\small
Facultad de Ciencias, Universidad Arturo Prat, Avenida Arturo Prat Chacón 2120, 1110939 Iquique, Chile}

\end{center}
\vskip 1cm

\paragraph{Abstract:} 
In this work and in the companion paper arXiv:2403.02301, we initiate an approach to holography based on the AKSZ formalism.
As the first step, we refine Vasiliev's holography proposal in arXiv:1203.5554 by obtaining 4D higher-spin gravity (HSG) and 3D coloured conformal higher-spin gravity (CCHSG) --- i.e., coloured conformal matter fields coupled to conformal higher-spin gauge fields and colour gauge fields --- as two distinct and classically consistent reductions of a single parent theory. 
The latter consists, on-shell, of a flat superconnection valued in a fractional-spin extension of Vasiliev's higher-spin algebra. 
The HSG and CCHSG reductions are characterized by dual structure groups and two-form cohomology elements, and their embedding in a common parent model provides a rationale for deriving holographic relations from multi-dimensional AKSZ partition functions on cylinders with dual boundary conditions, to appear separately.
In this work we i) construct the underlying non-commutative geometry as a metaplectic operator algebra represented in a Hermitian module of a pair of conformal particles; ii) identify a discrete modular group, arising from twisted boundary conditions of the first-quantized system, and connecting different boundary conditions of the second-quantized system; and iii) identify the holonomies, structure groups and two-form cohomology elements that characterize the HSG and CCHSG reductions, and equate the dual second Chern classes.

\newpage

\clearpage
\setcounter{page}{1}

\tableofcontents

\section{Introduction}

\subsection{General motivations and approaches to higher-spin gravity}

Higher-spin gravities (HSG) are minimal extensions of general relativity with gauge fields of spin higher than $2$, referred to as higher-spin gauge fields, which are interesting for several reasons \cite{snowmass}: first of all, they may be stand-alone contenders for theories for quantum gravity, with Einstein's theory (possibly coupled to lower-spin fields) appearing in broken phases; they also appear naturally as subsectors of tensionless limits of string theories, associated to the first Regge trajectory; moreover, since the underlying HS gauge symmetries arise naturally as symmetries of non-commutative symplectic manifolds of underlying first-quantized systems, there is an interesting possible connection between HSG, non-commutative geometry, and non-linear extensions of quantum mechanics \cite{Misha,Arias:2015wha,Bonezzi:2015lfa,Sharapov:2023erv}; finally, they provide a particularly tractable window into holography, given the large amount of symmetry governing both bulk and boundary dynamics. 

The study of higher-spin fields has a long history, dating back to the birth of relativistic quantum mechanics; for a review and references, see \cite{Bengtsson1,Bengtsson2}. 
While the current quantum field theory (QFT) paradigm applies well to lower-spin fields, its application to unifying the standard models of particles and cosmology conflicts with observational data. In questioning the
QFT framework, HSG provides an interesting theoretical laboratory, as beyond the spin-two barrier non-abelian higher-spin gauge symmetries trigger spacetime non-localities, providing a purely theoretical motivation for reconsidering basic assumptions. 
To this end, the reformulation of classical, relativistic field theories (containing local degrees of freedom) in terms of free differential algebras (FDA) of forms, initiated by Cartan \cite{Bryant} and developed further in the contexts of supergravity \cite{DAuria:1982uck,vanNieuwenhuizen:1982zf,DAuria:1982mkx}, and Vasiliev's unfolded formulation of HSG \cite{Vasiliev:1988xc,MV,Vasiliev:1992gr,Vasiliev:2005zu,review99,Bekaert:2005vh} may serve as a guide towards a more fundamental principle for QFT --- applicable not only to HSG but also to supergravity, gravity and lower-spin fields as well. 

Currently, the development of classical and quantum HSG has entered an interesting phase in which several approaches are being pursued.
Among these, the ones more relevant to our investigation are (for an extensive list of approaches see \cite{snowmass}):

\begin{enumerate}

\item The \emph{Fronsdal approach}, which is a generalization of the Gupta-Feynman approach to general relativity, in which the interacting theory is built by deforming Fronsdal actions for free fields \cite{Fronsdal:1978rb,Fronsdal:1978vb} in a perturbative expansion around maximally symmetric spacetimes. This approach has several sub-branches employing different deformation techniques, mainly the Noether procedure, the Fradkin-Vasiliev frame-like version thereof using underlying non-abelian higher-spin algebras, and light-cone methods (see \cite{snowmass,Bengtsson1,Bengtsson2} and references therein\footnote{A note on terminology: what we call Fronsdal approach here does not coincide with the approach referred to as ``Fronsdal program'' in \cite{Bengtsson2}, but also includes part of the approach there called ``Vasiliev program'' (precisely, the study of the cubic action in frame-like approach).}). Despite its initial transparency, this perturbative approach to defining HSG leads to vertices arising in a double-perturbative expansions in a dimensionless gauge coupling and a mass parameter set by the cosmological constant, which blurs spacetime locality properties as well as the geometric origin of the non-abelian higher-spin gauge symmetries \cite{Bekaert:2015tva,Sleight:2017pcz}, together with its background-independent origin.

\item The \emph{Vasiliev approach}, based on Vasiliev's classical, fully non-linear, background-independent, unfolded field equations \cite{Vasiliev90,properties,more,Vasiliev03} (see \cite{review99,Bekaert:2005vh,Didenko:2014dwa} for reviews), whereby all nonlinear corrections are encoded as solutions to evolution equations along auxiliary, non-commutative directions.
The resulting on-shell formulation is based on a novel approach to relativistic field theory akin to topological string field theory, in which fundamental fields are operators obeying first-order equations of motion encoding deformations of non-commutative symplectic geometries. 
This approach in turn bifurcates into two different approaches, depending on how the physical observables are extracted:

\begin{enumerate}

\item[2.1] The \emph{spin-local Vasiliev approach} \cite{Vasiliev:2017cae,Gelfond:2018vmi,Didenko:2018fgx,Didenko:2019xzz,Gelfond:2019tac,Didenko:2020bxd,Vasiliev:2022med}, which proceeds via perturbatively fixing a field frame and a gauge for the connection along the auxiliary directions, such that elimination of the auxiliary coordinates yields an on-shell version of the Fronsdal approach, with vertices subject to a generalized locality criterion (spin-locality); in particular, in the context of HS/CFT correspondences, to be discussed in more detail below, spin-locality is meant to ensure that holographic quantities can be then computed using the GKPW prescription.

\item[2.2] The \emph{AKSZ approach}, which is a stand-alone, off-shell formulation of higher-spin gravity based on a natural generalization of deformation quantization of Poisson manifolds \cite{Kontsevich:1997vb,Cattaneo:1999fm} to spaces of classical HSG solutions, with the goal of second-quantizing the theory as an Alexandrov--Kontsevich--Schwarz--Zaboronsky (AKSZ) sigma model \cite{Alexandrov:1995kv,Grigoriev,Barnich:2006pc,BarnichGrigoriev,Sezgin:2011hq,Boulanger:2011dd,Boulanger:2015kfa,Bonezzi:2016ttk}. In this approach, quantum field theories are formulated in terms of differential form fields introduced via Vasiliev's unfolding procedure and physical observables arise using BRST methods.
In the HSG context, the physical information is thus extracted via gauge-invariant, closed differential forms of the full theory in the extended space; in particular, this approach naturally incorporates the computation of higher-spin amplitudes from generalized Chern classes evaluated on the internal space (which are zero-forms on the original spacetime manifold) \cite{Colombo:2010fu,Colombo:2012jx,Didenko:2012tv,Bonezzi:2017vha}. Importantly, the AKSZ approach facilitates the Vasiliev system's off-shell extension, by including all its auxiliary fields into a path-integral measure, and treating Vasiliev's on-shell master fields as a boundary state for an AKSZ sigma model on manifolds with boundaries. 
In particular, in the context of HS/CFT correspondences, the AKSZ approach to HSG does \emph{not} rely on extracting spacetime vertices for Fronsdal fields from the Vasiliev system, but rather on imposing asymptotically free boundary conditions on the master fields (describing unfolded Fronsdal fields on-shell) \cite{COMST} after which holographic quantities are meant to be computed from gauge-invariant functionals on the entire correspondence space (rather than using the GKPW prescription).

\end{enumerate}

\item More recently, the Fronsdal and Vasiliev approaches have cascaded further, resulting in anti-holographic formulations of deformed Fronsdal actions based on boundary correlation functions \cite{Sleight:2016dba}, derivations of Vasiliev-like systems from 
renormalization group equations \cite{Douglas:2010rc}, and novel formulations of HSG based on fundamental, bilocal dipole fields \cite{Jevicki:2011ss,bilocal2,bilocalrecent,Aharony:2020omh}; for a related, anti-holographic approach that highlights the delicate dependence of locality properties on signatures, see \cite{Neiman:2023orj}. Moreover, underlying non-commutative geometries have been unveiled by systematizing the unfolding procedure within the context of formal HSG in \cite{formal} and references therein. 
Furthermore, the light-cone approach \cite{Metsaev} has led to interesting, perturbatively complete chiral HSG models \cite{Ponomarev:2016lrm,Sharapov:2022faa,Sharapov:2022awp,Didenko:2022qga,Sharapov:2023erv}, that admit flat space limits. The possible embedding of such models within the Vasiliev system was studied in \cite{Didenko:2022qga}. 

\end{enumerate}

Our working hypothesis is that i) the Fronsdal and Vasiliev approaches coincide provided that correct boundary conditions are imposed and suitable observables, such as higher-spin amplitudes on anti-de Sitter backgrounds, are considered, possibly in appropriate limits; and ii) the Vasiliev and AKSZ approaches can be matched because Vasiliev's equations describe boundary configurations of saddle points of the AKSZ sigma model. 
Thus, we hypothesise that one may start from a classical solution space to the Vasiliev equations, which can then, on the one hand, be brought via Vasiliev's spin-locality approach to a set of Fronsdal field vertices, from which correlation functions can be extracted via the GKPW prescription; on the other hand, the same classical solution space can be quantized using the AKSZ approach leading to an operator algebra that may be used for the same purpose according to new principles.
Indeed, in this paper, we initiate a formulation of holographic duality involving HSG within the AKSZ approach.  

\subsection{Higher-spin gravity and holography}

Interconnections between HSG in four dimensions and free conformal field theories (CFT) in three dimensions were pointed out already in \cite{Bergshoeff:1988jm} by Bergshoeff, Salam, Sezgin and Tanii within the context of the supermembrane theory on $AdS_4\times S^7$.
The topic resurfaced in extrapolations of Maldacena's conjecture within the above context \cite{Sezgin1998} and that of tensionless superstrings on $AdS_5\times S^5$ for which Sundborg \cite{Sundborg:2000wp} proposed four-dimensional (4D), free super-Yang--Mills theory as a holographic dual in a large N limit. 
The role of large N in relating weakly coupled CFTs to Vasiliev-style HSG coupled to matter was later stressed in \cite{WittenJHS60,Sezgin:2002rt}, with Vasiliev's 4D models playing a prominent role due to their relative simplicity while propagating local degrees of freedom.
Refined proposals\footnote{In this work, as well as in Paper II, we focus on the holographic correspondence involving 4D bulk HSG theory, though $AdS_3/CFT_2$ theories involving bulk HSG have also been extensively studied, see e.g. \cite{Henneaux:2010xg,Campoleoni:2010zq,Gaberdiel:2010pz,Gaberdiel:2011wb,Gaberdiel:2011zw,Chang:2011mz,Kraus:2011ds,Ammon:2011ua,Perlmutter:2012ds,Campoleoni:2013lma,Iazeolla:2015tca}.} appeared subsequently in \cite{KP,LeighPetkou} and tests were performed at the level of scalar \cite{Sezgin:2003pt} and more general \cite{GiombiYin,GiombiYin2} three-point functions by applying the on-shell GKPW prescription to Vasiliev's equations using a specific field frame and gauge choice.
However, these choices lead to subtle divergencies \cite{Boulanger:2015ova}, which prompted the development of the spin-local approach. The resulting refined cubic vertices were holographically tested in \cite{Didenko:2017lsn,Sezgin:2017jgm}. In parallel, the complete three-point functions of vector models have been computed more recently using the chiral theory in \cite{Skvortsov:2018uru}. 

There are two main difficulties in applying the standard GKPW prescription to the Vasiliev system. First, the non-commutative nature of the auxiliary directions implies that the homotopy contraction of a relatively simple full system yields a highly non-linear theory of Fronsdal fields with non-local interactions (even at low perturbative orders) that may cause boundary correlation functions to diverge \cite{GiombiYin,Boulanger:2015ova}. 
Thus, to maintain predictability, it is crucial to specify which class of functions of the non-commutative coordinates the master fields belong to, i.e., which operator algebra they represent.
Naturally, however, this problem, which in effect amounts to choosing a non-commutative geometry, translates into imposing appropriate boundary conditions on the master fields, which is expected in describing a dynamical system starting from a formally background-independent set of equations. Thus, in extracting boundary CFT correlation functions, our main hypothesis, spelt out in \cite{COMST}, is that asymptotically anti-de Sitter boundary conditions lead to well-defined equivalence classes of operators whose gauge artefacts factor out at the level of classical observables arising naturally within the AKSZ approach. 

The second difficulty is that so far no action with Lagrangian spacetime density built from Fronsdal fields alone has been found for the Vasiliev equations, which further complicates implementing the usual AdS/CFT prescriptions. On the other hand, applying the AKSZ formalism \cite{Alexandrov:1995kv} to the full Vasiliev system yields a BV master action of covariant Hamiltonian type \cite{Boulanger:2011dd,Boulanger:2015kfa,Bonezzi:2016ttk} formulated on non-commutative manifolds. The additional bonus of this setup is the possibility of deforming the action but not the field equations, such that the resulting on-shell action produces holographic correlation functions, even though the action does not contain any Fronsdal kinetic terms; to be more precise, the AKSZ actions are formulated on homotopy classes of open manifolds, and the resulting boundary field equations are identified with the Vasiliev system.

Even from a pure holography perspective, it would be desirable to have access to an intrinsic bulk formulation of HSG, such as Vasiliev's equations and their AKSZ extension, inducing, via AdS/CFT correspondences, various 3D CFTs. 
In this respect, the Maldacena-Zhiboedov (MZh) theorem \cite{MZh} --- which states that a conformal, unitary, local theory possessing at least one conserved HS current is free --- seems to restrict the possible dual boundary theories. However, as already noted in \cite{Misha}, bypassing at least one of the theorem's assumptions widens the potential reach of the holographic dualities, extending it to more general conformal boundary theories. While the dual free vector models (and even more so the critical vector models \cite{KP,Giombi:2011ya,Mzh2}) already encode non-trivial physics, such an enlargement seems somehow natural given the vastness of the classical solution space of Vasiliev's equations. In particular, more general moduli spaces of the boundary theory seem to be required to naturally encode the degenerate conformal geometries arising in the asymptotic regions of 4D BTZ-like vacua of Vasiliev's bulk theory \cite{BTZ}. 

In this and a companion paper \cite{paperII}, working within the AKSZ approach, we shall argue that the holographic dual contains additional degrees of freedom beyond the conformal matter fields along the lines of Vasiliev's proposal of 2012 \cite{Misha}, which introduces topological, conformal HS gauge fields coupled to the dual CFT, thus unsettling the assumptions of the MZh theorem and enabling the boundary dual to encode degenerate conformal geometries. 

Indeed, returning to the original ideas of \cite{Bergshoeff:1988jm} and \cite{Sundborg:2000wp}, and thinking of both Vasiliev's theory and its boundary dual as arising in a tensionless limit of supermembrane theory, it is natural to view the boundary theory as a matter system coupled to a tensionless extension of the 3D Polyakov metric, for which topological, conformal HSG is a natural candidate.

While we will not attempt to justify this statement in this paper, we find it reasonable to expect that such an extended holographic dual reproduces the vector model dual proposal in the leading and first subleading order in the $1/N$ expansion and deviate beyond, as integrating out the topological HS fields may introduce double-trace deformations of the type proposed in \cite{Sezgin:2002rt} as part of a dynamical membrane with fluctuating induced geometry, and whose relevant piece is the $O(N)$-model deformation of \cite{KP}.

\subsection{Vasiliev's holography proposal}

The fully non-linear HSG equations written down by Vasiliev constitute a compact Cartan integrable system (CIS) of constraints on the exterior derivatives of a set of differential forms, referred to as \emph{master fields}, living on a fibered, non-commutative extension of the spacetime manifold. 
Thus, Vasiliev's formulation of HSG is an example of a general approach to dynamical systems described by partial differential equations, in which the equations of motion take the form of zero-curvature and covariant constancy conditions, referred to as \emph{unfolded dynamics}, which can be thought of as a field-theoretic version of Hamiltonian dynamics without the need for introducing any time-slicing.

Unfolded field equations implement a duality between spacetime and fibre twistor coordinates that the master fields depend on, much akin to a Penrose transform \cite{Misha,Didenko:2021vui,corfu19}). Thus, the local spacetime features of an unfolded field configuration are stored on-shell in the fibre dependence, i.e., spacetime is locally an accessory. This means that, in principle, a given dynamical content may be equivalently described on spacetimes of different dimensions, which was used in \cite{Misha} to obtain holographically dual interpretations of given twistor-space configurations, thus described via different yet equivalent spacetime equations related by a non-trivial mapping. 

In particular, in \cite{Misha}, the HS/CFT duality was approached this way\footnote{The exploration of holographic properties of gravity often exploits the traits of Euclidean signature with its simpler boundary structures, Green-function singularity structures, stable saddle points and rich thermodynamic interpretations. Lorentzian signature is kept in most works concerning HS/CFT, including \cite{Misha} and the present paper. The resulting real-time holography leads to more involved boundary structures including initial and final states \cite{bdhm,svr,svr2,silva,corfu19}, null infinities \cite{celestial}, and singularity structures drawn from quantum field theory. The HS extension introduces: i) a formulation of quantum field theories akin to topological field theories, more lenient towards topology change, facilitating Lorentzian formulations on manifolds with simpler boundary structures that corresponding gravitational backgrounds \cite{BTZ}; ii) resolutions of Green-function singularities in the boundary \cite{Colombo:2012jx,Neiman:2022enh} as well as the bulk \cite{BTZ,corfu19} essentially as the result of first summing over spins and then taking limits at the level of master fields on non-commutative spaces.},
by reducing linearized 4D HSG field equations on 3D Minkowski leaves embedded in the Poincar\'e patch of $AdS_4$. 
In this setup, the holographic duality occurs on any 3D surface embedded in the bulk, not just at the boundary.
The reduction rearranges the bulk Weyl tensors into boundary currents, in such a way that the resulting unfolded system describes the coupling of 3D conformal higher-spin currents to corresponding 3D gauge fields plus current conservation laws, as expected in a Fefferman-Graham scheme\footnote{For the derivation of an analogous relation between bulk and boundary fields approached using the ambient space formulation, and its connection to unfolding and Fefferman-Graham expansion, see \cite{Bekaert:2012vt,Bekaert:2013zya,Bekaert:2017bpy}.}. The structure of the resulting equations prompted Vasiliev to propose that the holographic dual of 4D HSG is given by a 3D theory of \emph{conformal higher-spin gravity}  (CHSG) \emph{coupled to conformal matter fields} via conserved currents\footnote{Holographic duals of 4D HSG containing 3D CHSG arising as modifications of the already existing duality to free/critical $O(N)$ vector models due to altering the boundary conditions on the bulk gauge fields were proposed at the linearized level in \cite{LP2} and studied more fully in \cite{Giombi:2013yva}.}. 
The Chern--Simons formulation of the pure gauge sector was studied in \cite{Pope:1989vj,F&L}\footnote{The identification 3D CHSG as the parity anomaly of 3D fermions coupled to background CHSG fields was proposed in \cite{Grigoriev:2019xmp}, together with new classes of CHSGs. There are also the approaches of \cite{Segal,Tseytlin:2002gz} to CHSG which are natural HS extensions of conformal gravity. However, as these include local degrees of freedom in four and higher dimensions it is unclear whether they are relevant for deforming holographic duals of HSG in five and higher dimensions. Indeed, as the CHSG sector introduced in our approach is akin to an HS extension of the Polyakov metric on the worldvolume of gravity membrane, we expect it to admit a natural extension to dimensions beyond three, leading to the deformation of purely topological, higher-dimensional CHSG by currents of conformal matter fields, but not by any fundamental HS Weyl curvatures.}. Subsequently, the coupling of conformal matter to 3D CHSG was studied by Nilsson \cite{Nilsson:2013tva, Nilsson} in the Poisson-bracket limit of the conformal HS algebra.
The somewhat surprising conclusion that the theory dual to 4D HSG is not free does not contradict the MZh theorem, since, being a gauge theory, it escapes at least one of the hypotheses the theorem is based on.
In Paper II, we shall examine these expansions further and provide a fully non-linear completion of matter-coupled CHSG. 

One of the main conclusions of \cite{Misha}, to be revisited and to some extent modified in a comparative study in Paper II, is that the 3D CHSG fields can be decoupled from matter currents by imposing specific boundary conditions for special configurations of the bulk HS fields \emph{only} in the Type A and Type B models \cite{Sezgin:2003pt}. 
However, strictly speaking, the approach of \cite{Misha} was intrinsically classical, on-shell and linear, though it was observed that the consistency of the CHSG equations beyond the linearized approximation implies that the 3D theory becomes non-linear. 
Moreover, being a rewriting of the 4D bulk equations, the analysis of \cite{Misha} does not make explicit the composite structure of the 3D currents in terms of 3D conformal matter fields; for more details, see Section 5 in Paper II.

\subsection{AKSZ-inspired holography proposal}

In this work, henceforth referred to as Paper I, being the first in a series of papers, we initiate an approach to holographic correspondences involving Vasiliev's 4D HSG based on its off-shell formulation \cite{Boulanger:2015kfa, Bonezzi:2016ttk} as an AKSZ sigma model of Frobenius--Chern--Simons (FCS) type with a \emph{fractional-spin gauge algebra} \cite{Boulanger:2015uha}; the proposed holographic dual consists of coloured, 3D conformal fields coupled to CHSG and topological colour gauge fields, giving rise to \emph{coloured conformal higher-spin gravity} (CCHSG).

Formulating the AKSZ sigma model on manifolds, referred to as its sources, with boundaries\footnote{Each boundary field configuration of the AKSZ sigma model is an entire unfolded, classical field history including  (asymptotic) boundaries modelled by suitable discontinuities.}, the boundary field equation is a flatness condition on the FCS superconnection, referred to henceforth as the parent field equation, admitting distinct, classically consistent truncations to HSG and CCHSG, to be spelt out in a companion Paper II \cite{paperII}.
Quantizing the sigma model canonically on a cylinder with dual HSG and CCHSG boundary conditions yields a vacuum state $|\Omega\rangle$ in the direct product of the spaces of HSG and CCHSG states, obeying \emph{overlap conditions} \cite{OLC}. Assuming ${\cal O}$ to be a local observable\footnote{In the non-commutative context, an observable ${\cal O}$ is referred to as being local if it is constructed using the basic algebra operations and the trace operation.}, i.e., a globally defined element in the BRST cohomology that can be evaluated on field configurations on separate foliates, the quantity ${\cal O}|\Omega\rangle$ is independent of whether  ${\cal O}$ acts in the HSG or CCHSG state spaces.
The resulting entanglement of the vacuum encodes a refined holographic correspondence between the HSG and CCHSG model, to be detailed in \cite{OLC}. 

Having stated our proposal, we stress that it provides an \emph{a priori} rationale for holographic dualities between two different theories living on spacetime manifolds of unequal dimensions.  Indeed, holographic relations between invariant functionals of the two dual theories become a direct consequence of the fact that the latter arise as classically consistent reductions of a single, common, AKSZ model.
Such a formulation of holography may be especially relevant for the HS/CFT correspondence in which
--- unlike holographic dualities involving a theory with broken HS symmetry like String Theory, where dual decoupling limits ultimately lead to conjectured equalities of partition functions of separately defined theories, to be verified \emph{a posteriori} --- no dual decoupling limit is available and, so far, the duality was only argued on the basis of symmetries.
More precisely, rather than directly comparing the partition functions of the dual theories, our strategy is to embed these theories classically as boundary states of an AKSZ sigma model on a cylinder in one higher dimension, represented by the entangled vacuum state.
Moreover, in the HSG context, and when compared to the holography proposal of \cite{Misha} based on dual foliations of unfolded HSG, our approach provides direct access to the fundamental conformal fields of the 3D CCHSG theory, while these fields are only accessed indirectly via their conformal currents in \cite{Misha}.

\subsection{Main results of Paper I}

In this paper, we formulate the parent field equation and construct the cohomologically non-trivial two-forms triggering the HSG and CCHSG defects to be studied in Paper II. The main results are:

\begin{itemize}

\item The construction of the representation of the fractional-spin algebra \cite{Boulanger:2013naa} underlying the parent model using holomorphic, symplectic oscillator realizations of complex metaplectic group algebras in Hermitian modules with split signatures.

\item The derivation of the two-by-two block structure of the fractional spin algebra through a geometric projection of a first-quantized two-particle system on a (fibered) correspondence space.

\item The identification of a discrete modular group in higher-spin gravity which connects twisted boundary conditions of the underlying first-quantized system, corresponding to distinct boundary conditions on the second-quantized fields. 

We shall illustrate these actions in the example of the outer and inner Klein operators, which are induced by symplectic reflections acting on the first-quantized, complexified twistor space, and that relate, for example, positive/negative energy modes in HSG and CCHSG, as well as particle/black-hole-like modes \cite{2017,COMST} and boundary-to-bulk propagators/singular solutions with vanishing scaling dimension, containing boundary Green's functions \cite{corfu19} in HSG  (as discussed in the comments to Eq. \eqref{2.86}). The elements of the modular group are also instrumental in singling out the structure groups of the defects, and we find that the modular transformations exchanging various boundary conditions of the HSG defect, which are generated by the standard inner Klein operators $\k_y$, $\bar\kappa_{\yb}$, have a counterpart on the CCHSG defect, generated by the operator $\kappa_{\!\mathscr{P}}$, whose composition with $\kappa_y$ generates the Fourier transform in the non-compact singleton module of relevance in our construction.

\item The construction of the expectation value of the dynamical two-form that singles out the CCHSG defect, and its interpretation as the curvature of a statistics gauge-field of a non-abelian anyon. 

\item The characterization of the two-form expectation values on the CCHSG and HSG defects via a second Chern class that can be uplifted to the parent model, ensuring that the two reductions are topologically non-trivial and mutually compatible; indeed, identifying the second Chern classes of the HSG and CCHSG defects yields a relation between HSG Weyl tensors and CCHSG currents compatible with the conjectured holographic duality. 

\end{itemize}

\noindent Before tending to details, we add further context to our holography proposal by outlining the notions of boundary defects and multi-dimensional partition functions of universal AKSZ sigma models in Sections \ref{Sec:1.6} and \ref{sec:aksz}, respectively, to be developed in Paper II and \cite{OLC}. We stress that the present paper's results are relevant for the research directions described in \ref{sec:aksz} but do not depend on these.

\subsection{Defects, gauge functions and fibre algebras}\label{Sec:1.6}

A key feature of the formalism is the treatment of the solution spaces of HSG and CCHSG, thought of as classical field theories, as spaces of boundary states of an AKSZ sigma model, i.e., as spaces of solutions to distinct, classically consistent truncations of the boundary field equation, which is a flatness condition on an FCS superconnection\footnote{\label{footnote4}
Given a homotopy associative algebra ($A_\infty$-algebra) $\boldsymbol{\cal A}$ with $n$-ary operations $M_n$, $n=1,2,\dots$, of degrees $2-n$, and a commuting manifold $\boldsymbol{M}$, the resulting space of flat superconnections is the subspace of the $A_\infty$-algebra $\boldsymbol{\cal E}(\boldsymbol{M};\boldsymbol{\cal A}):=\boldsymbol{\cal A}\otimes\Omega(\boldsymbol{M})$ consisting of elements $X\in \boldsymbol{\cal E}(\boldsymbol{M};\boldsymbol{\cal A})$ of total degree one with vanishing curvature $R^X:=dX+\sum_{n=1}^\infty M_n(X^{\stackrel{\wedge}{\otimes}n})$.
Upon quantization of $\boldsymbol{M}$ by deforming $\Omega(\boldsymbol{M})$ into a differential graded associative algebra $\Omega^{(\boldsymbol{\cal S})}(\boldsymbol{M})$ consisting of a space $\boldsymbol{\cal S}\subset \Omega(\boldsymbol{M})$ of symbols equipped with a differential $d$ and product $\star$ given by deformations of the de Rham differential and the wedge product, respectively, $\boldsymbol{\cal E}(\boldsymbol{M};\boldsymbol{\cal A})$ deforms as an $A_\infty$-algebra; correspondingly, the curvature deforms into $R^X=dX+\sum_{r=1}^\infty M_r(X^{\stackrel{\star}{\otimes}n})=0$.
More generally, $\boldsymbol{\cal E}(\boldsymbol{M})$ and $R^X$ deforms in concert with quantizations of $\boldsymbol{M}$ preserving homotopy associativity \cite{Gaberdiel:1997ia}.}, viz.,
\begin{align}\label{1.1}
dX+X\star X=0\ ,    
\end{align}
to be detailed in the bulk of this paper.
The classical boundary states are built from holonomies and cohomology elements, including integration constants in degree zero providing local degrees of freedom, glued together using transition functions from a subgroup of the parent structure group, to be introduced more explicitly in Section 5.2.
We thus think of the spaces of boundary states of the sigma model as topologically broken phases, or defects, characterized by the unbroken structure group.

The superconnection $X$ is an element of total degree one of a \emph{differential graded associative algebra} (DGA) given by the direct product of the DGA of \emph{horizontal forms} on a non-commutative, fibered \emph{correspondence space} and an internal DGA (possibly with vanishing differential). 
Taking the latter to be the three-graded ${\rm mat}_{1|1}$ (with trivial differential), yields a superconnection consisting of horizontal forms of degrees zero, one and two, i.e., forms on the base of the correspondence space of degrees zero, one and two valued in representations of its (ungraded) \emph{fibre algebra} in associated Hermitian modules.
The construction of classical solution spaces introduces a rich set of classical moduli in the form of \emph{gauge functions}, i.e.,  holonomies along open curves emanating from base points;  \emph{fibre-algebra representations} used to construct various cohomological elements; and \emph{structure groups} factored out at the level of classical observables.

The ensuing gauge function method \cite{Vasiliev:1990bu,Sezgin:2005pv,2011,2017,COMST,Iazeolla:2022dal} glues together locally defined configurations carrying various local degrees of freedom characterized by different types of asymptotic boundary conditions encoded into distinct representations of the fibre algebra, into globally defined configurations characterized by holonomies, Chern classes (including spacetime zero-form charges), and abelian $p$-form charges.
The method was originally designed for finite-dimensional CIS of potentials of strictly positive degrees  \cite{sullivan,DAuria:1982uck,vanNieuwenhuizen:1982zf,DAuria:1982mkx,Bryant} on commutative geometries.
It extends naturally to systems on non-commutative spaces since the de Rham differential and wedge product can be deformed in concert along differential Poisson structures into operations of non-commutative DGAs.
On non-commutative 
correspondence spaces, finite sets of horizontal forms (including forms in degree zero) pack infinite-dimensional spaces of forms on the base into representations of the fibre algebra (and holonomy group) realized in terms of operator algebras consisting of special functions of the non-commutative fibre coordinates.
The resulting cohomology elements on the base in various degrees, including \emph{zero-form integration constants}, store \emph{local} degrees of freedom.
Thus, the spacetime field configuration obeying various boundary conditions are encoded into cohomology elements belonging to specific representations of the fibre algebra.
In this approach to field theory, the local data contained in the cohomology elements is spread across spacetime charts by the star-product action of the gauge functions;
for examples of solution spaces containing scalars, Faraday tensors and generalized Weyl tensors in four dimensions, see \cite{Sezgin:2005pv,Iazeolla:2007wt,2011,GiombiYin2,2017,review,cosmo,BTZ,corfu19,COMST,Iazeolla:2022dal}; for fractional-spin fields in three dimensions, see \cite{Boulanger:2013naa,Boulanger:2015uha}; and for conformal scalars and spinors in three dimensions, see \cite{Iazeolla:2015tca,paper0}.

The gauge function method is facilitated by the local \emph{factorization} of horizontal forms into forms on the base valued in fibre-algebra representations \cite{2017,COMST}.
This in turn facilitates perturbative expansions around locally, constantly curved backgrounds\footnote{The parent gauge structure assigns the frame field an independent gauge parameter (as this leaves its BF-like action gauge invariant), which leads to configuration spaces including BTZ-like vacua with degenerate metrics \cite{BTZ,Iazeolla:2022dal}. 
Likewise, degenerate metrics are naturally incorporated into off-shell formulations of ordinary gravity as multi-dimensional AKSZ sigma-models with boundary equations of motion given by unfolded versions of Einstein's equations, unlike the original Einstein--Cartan formulation which treats the frame field as a covariantly constant section of constant rank.}.
The latter are encoded into \emph{vacuum fibre algebras}.
These act in \emph{Hermitian left-modules} whose endomorphism algebras can be realized (using Wigner--Ville-like maps) in terms of special fibre functions capturing the fluctuating local degrees of freedom.
The vacuum fibre algebras are direct sums of \emph{higher-spin algebras}, given by enveloping algebras of $\mso(2,3)\cong \msp(4;\Real)$ modulo annihilators of \emph{singletons} \cite{Konstein:1989ij,Vasiliev:2004cm,Bekaert:2005vh,fibre}, and \emph{internal matrix algebras}, given by enveloping algebras of finite-dimensional Chan--Paton-like factors, multiplied semi-directly with discrete, \emph{modular group algebras}\footnote{The modular group algebra elements are representations of discrete subgroups of $Mp(4;\Comp)$ by bounded, non-unitary operators acting in Hermitian modules arising upon factorization \`a la Flato--Fronsdal of linearized fluctuations around asymptotically, constantly curved backgrounds on spacetime leaves of the correspondence space.
On the one hand, the modular operators connect different spacetime boundary conditions, e.g., positive and negative energies as well as particle modes and Type D modes corresponding to linearized Coloumb-like and Schwarzschild-like solutions.
On the other hand, from the fibre algebra point-of-view, they are complexified Bogolyubov transformations that connect polarizations of the oscillator algebra described by elements from the upper \emph{and} lower Siegel half-planes, facilitated by the complexification of $Mp(4;\Real)$.
Interestingly, the modular group brings about a correspondence between the boundary conditions of first- and second-quantized systems: the former induce outer group algebra elements whose inner counterparts yields the modular algebra elements.}, given by complexified Bogolyubov transformations connecting various boundary conditions. 
The higher-spin and internal algebras arise as endomorphism algebras of direct sums of left-modules consisting of quantum states of \emph{multi-parton} systems \cite{Engquist:2005yt,Engquist:2007pr, Vasiliev:2018zer}.

In particular, the two-parton system admits a natural projection, to be spelled out below, providing a first-quantized, geometric origin for the two-by-two block structure of the \emph{fractional-spin algebra}, which was introduced by hand in the original formulation \cite{Boulanger:2013naa}.
Thus, as we shall see in Section \ref{sec:parent}, the parent superconnection $X$ can be represented via a graded $2\times 2$ matrix comprising two one-form connections $\mathbb{A}$ and $\widetilde{\mathbb{A}}$ on the diagonal, while a zero-form $\mathbb{B}$ and a two-form $\widetilde{\mathbb{B}}$ occupy the off-diagonal entries, viz.,
\be \label{1.2}
\left.X\phantom{)}\right\downarrow_{{\rm mat}_{1|1}} = \left[\begin{array}{c|c}\mathbb{A}&\mathbb{B}\\\hline\widetilde{\mathbb{B}}&\widetilde{\mathbb{A}}\end{array}\right]\ ,
\ee
where each block is valued in the fractional-spin algebra consisting of the endomorphisms of a Hermitian left-module of the vacuum algebra, given by the direct sum of an infinite-dimensional ``external'' spin space and a finite-dimensional ``internal'' colour space \cite{Boulanger:2013naa} (see also Section \ref{Sec:2.8}), viz.,
\be \label{1.3}
\left. (\mathbb{A}, \ \widetilde{\mathbb{A}}, \ \mathbb{B},\ \widetilde{\mathbb{B}}) \ \right\downarrow_{\boldsymbol{{\cal FS}}} \ \sim \ \left[\begin{array}{c|c}\ket{{\rm ext}}\bra{{\rm ext}} &\ket{{\rm ext}}\bra{{\rm int}}\\\hline\ket{{\rm int}}\bra{{\rm ext}}\ &\ket{{\rm int}}\bra{{\rm int}}\end{array}\right]  \ .\ee
To model holographic correspondences involving 4D, Lorentzian, global anti-de Sitter spacetimes, we take the external module to be the conformal singleton module (with endomorphism group related to the metaplectic group), and the internal module to be a finite-dimensional Hermitian module with split signature $(N,N)$ (with endomorphism group $U(N,N)$). 

As we shall see in Paper II, the decomposition of \ref{1.1} under \ref{1.2} and \ref{1.3} can be combined with further consistent truncations, with the HSG and CCHSG defects arising by taking the horizontal zero-form to be diagonal and off-diagonal, respectively, concerning \ref{1.3}.
As for the HSG defect, it arises as a proper subsystem of an intermediate truncation with full zero-form, which describes coloured, fractional-spin matter fields coupled to HSG and an internal colour sector (required for integrability), referred to as 4D  \emph{fractional-spin gravity} (FSG), whose linearization around $AdS_4$ is deferred to future work. 

The defects are thus built from three basic ingredients: i) background one-forms, valued in vacuum Lie algebras, and characterized by holonomies along open paths, referred to as vacuum gauge functions\footnote{In this work we shall not activate internal vacuum gauge functions.}, defined modulo the group of holonomies around closed loops, and represented \emph{unitarily}\footnote{
A complex group $G$ is represented unitarily in a Hermitian space ${\cal S}$ if the representation map $V: G\to {\rm End}({\cal S})$ obeys $V(g)^\dagger=V(\bar g^{-1})$; if $G$ contains a real subgroup $G_\Real$ and the Hermitian form is positive definite, then $V$ provides a unitary representation of $G_\Real$.}
in the Hermitian module; ii) integration constants (i.e., local data), valued in the endomorphism algebra of the Hermitian module, which decomposes under the (real) gauge function group into spaces of linearized modes subject to boundary conditions labelled by \emph{fibre polarizations} and connected by the modular transformations; and iii) cohomologically nontrivial horizontal $p$-forms (including the unit for $p=0$), which are part of the non-commutative background, and combine with (ii) into linearized fluctuations of horizontal $p$-forms.

In (iii), the elements with $p=2$ are built from two-forms $dd\phi$, which obey $ddd\phi= 0$, using polar coordinates $\phi$ of contractible circles appearing on various two-dimensional leaves of the base manifold\footnote{Letting $\phi$ be the azimuthal angle around the origin of $\mathbb{R}^2$ with Cartesian coordinates $(x,y)$, one has $d\phi= (-y dx+xdy)/(x^2+y^2)$ and $dd\phi=2\pi dx\wedge dy \delta(x)\delta(y)$ which in turn obeys $ddd\phi=0$.}. Dressing such two-forms with (ii) yields different types of conical defects: a) FSG, as well as HSG, encodes defects on symplectic leaves, built from group algebra elements given by \emph{inner Klein operators} and giving rise to deformed oscillator algebras \`a la Wigner; b) CCHSG encodes monodromies on two-dimensional Lagrangian sub-manifolds of four-dimensional symplectic leaves, built from \emph{twisted projectors} and giving rise to deformed oscillator algebras \`a la Leinaas and Myrheim \cite{Leinaas:1977fm}; c) monodromies akin to (b) on spacetime leaves, giving rise to HSG analogs of Lorentzian geometries used for entanglement computations in General Relativity \cite{Lewkowycz:2013nqa, Dong:2016fnf,Arias:2019pzy}; and d) mixed defects on two-dimensional planes with one spacetime direction and one non-commutative direction encoded into two-dimensional analogues of CCHSG and duals of near-horizon regions \cite{Carlip:2002be}. 
In all four cases, the perturbative expansion yields projected unfolded equations on spacetime leaves containing two-form cocycles describing the sourcing of linearized one-forms by linearized zero-forms. The main result of this paper is the identification of a cohomology class of type (b) triggering desirable two-form cocycles for CCHSG.

In (i), locally, constantly curved spacetime backgrounds for HSG and CCHSG\footnote{
One of the original motivations for this work is the formulation of a holographically dual description of 4D HSG expanded around the BTZ-like geometry of Type Ib \cite{Holst:1997tm}, given by the warped $ds^2_{AdS_3}\times_\xi S^1$ with $T^2\times \Real^2$ topology \cite{BTZ}, in terms of conformal scalars and spinors on its conformal boundary, given by the conformal, warped $ds^2_{T^2}\times_\xi S^1$.}
activate vacuum gauge functions in the \emph{metaplectic} double cover $Mp(4;\Real)$ of $Sp(4;\Real)$.
The corresponding vacuum algebras are extensions of $\msp(4;\Real)$ by a modular algebra generated by inner and outer Klein operators exchanging modes with positive and negative energies in the bulk and on the boundary.
The corresponding modules contain Hermitian singletons, i.e., direct sums of unitarizable singletons with positive energies and anti-singletons with negative energies connected by inner Klein operators.

\subsection{Multi-dimensional AKSZ partition functions}\label{sec:aksz}

The implementation of our holography proposal relies on a natural generalization of the AKSZ formalism.
So far, we have introduced the notion of a parent field equation as a CIS comprising the boundary field equations of an AKSZ sigma model formulated on a set of sources given by open, non-commutative manifolds of odd and even dimensions, respectively, using universal FCS and BF-like actions. 
The classical, boundary defects are free differential algebras (FDA) inside the DGAs of forms on the non-commutative boundaries, referred to as the integral manifolds of the parent field equations.

As for the sources included in the partition function, we are making the rudimentary assumption that they are \emph{fibrations}, the integral DGAs can be projected to subalgebras living on the bases; correspondingly, the integral FDAs can be projected to spaces of boundary states for AKSZ sigma models on the bases of the fibrations.
Each fibration yields a sub-partition function given by the sum over maps from its base into an unprojected target subject to boundary conditions descending from the projection of the FDA on the integral manifolds (weighted using the universal master action).
Thus, a given set of defects can be assigned a  \emph{multi-dimensional} partition function given by the sum of sub-partition functions over all possible fibrations interpolating the defects (which thus restricts the possible topologies).

The formalism assigns i) non-isomorphic operator algebras to the separate HSG and CCHSG boundary defects, with symbols given by classical functionals on the defect moduli spaces; and ii) entangled vacuum states to source manifolds with multiple boundary defects, obeying overlap conditions \cite{OLC} induced by sigma-model parent observables, e.g., off-shell topological invariants such as Chern classes.
The existence of saddle points requires the boundary states to be mutually compatible semi-classically, i.e., the sigma-model observables to coincide when evaluated on the classical boundary field configurations.

As mentioned earlier, the multi-dimensional partition function receives contributions from projected sources of various dimensions. In the present context, starting from a universal BF-like model, treating projections of the flat superconnection as configuration on the boundary of a disk, and coordinatizing the Weyl zero-form modules using harmonic expansions yields desired canonical commutations rules for creation and annihilation operators of massless particles (without referring to any Fronsdal action), which can furthermore be extended to various other types of modes associated to local degrees of freedom \cite{fibre,corfu19,COMST,OLC}.
Thus, applied to dual HSG and CCHSG boundary configurations admitting fibrations over one-dimensional circles, a leading contribution arises on the two-dimensional cylinder interpolating the dual operator algebras, resulting in entangled vacuum states encoding operator algebra morphisms interpretable holographically. This is the essence of our proposed formulation of holographic dualities.

Multi-dimensional AKSZ partition functions with desirable QFT properties are formulated naturally using horizontal forms on correspondence spaces whose fibre algebras contain unitarizable, particle representations of underlying spacetime symmetry algebras. 
The resulting sub-partition functions thus involve projecting the base while leaving the fibre unaffected, giving rise to a geometrical implementation of the basic idea.

\subsection{Outline of the paper}

The plan of the paper is as follows: 

\noindent Section \ref{sec:2} recalls the relevance of complex metaplectic group elements in HSG and introduces the operator algebras and representation theory underlying the parent model, stressing its origin in an underlying first-quantized two-parton system on a non-commutative correspondence space. 

\noindent In Section \ref{sec:fibrealg} we present the structure of the fibre algebra which will be relevant in constructing the parent field equations and its reductions. A prominent role will be played by special metaplectic group elements encoding endomorphism of a singleton module. The product of such elements maps to Dirac-style bra-ket computations with conformal singleton states and momentum eigenstates, and the so-realized non-compact singleton module can thus be equipped with a Hermitian form given by a regularized trace operation.

\noindent In Section \ref{sec:parent} we present the parent model and the superconnection, decomposing the latter under the three-graded ${\rm mat}_{1|1}$ and the two-parton algebra introduced earlier.

\noindent In Section \ref{sec:defects} we specify the parent model and the two-parton system underlying it to a class that is relevant for HS holography, resulting in a superconnection valued in the fractional spin algebra. 
We characterize the embeddings of 4D FSG and HSG, as well as 3D CCHSG, into the parent model by their two-form expectation values and related unbroken structure groups.

\noindent Finally, in Section \ref{sec:conclusionspart1}, we provide our conclusions, evaluate the second Chern classes, that capture the non-triviality of the two-form expectation values characterizing HSG and CCHSG, and provide an outlook to Paper II and related future works.

\noindent The paper is completed by Appendix \ref{App:emb}, which contains our conventions for realizing $\mso(2,3)\cong \msp(4;\Real)$ using $SL(2;\Comp)$- and $O(1,1)\times SL(2;\Real)$-covariant oscillators, and the internal matrix algebra using a finite-dimensional Fock space.

\subsection{Notation and nomenclature}

Manifolds are denoted by bold, capital, Roman letters, e.g. correspondence space $\boldsymbol{C}$, fibre $\boldsymbol{Y}$, base $\boldsymbol{M}$, commuting sub-base $\boldsymbol{X}$, and non-commutative sub-base $\boldsymbol{Z}$; the vector spaces of associative algebras by bold, calligraphic letters, e.g. the forms on the aforementioned spaces are denoted by $\boldsymbol{\cal C}$, $\boldsymbol{\cal Y}$, $\boldsymbol{\cal M}$ and $\boldsymbol{\cal X}$, related discrete map-algebras by $\boldsymbol{\cal K}$ and $\boldsymbol{\cal P}$, total fibre and base algebras by $\boldsymbol{\cal A}$ and $\boldsymbol{\cal B}$, metaplectic group subalgebras of the fibre algebra by $\boldsymbol{\cal G}$ and $\boldsymbol{\cal G}^{(\infty)}$, and their Frobenius subalgebras by $\boldsymbol{\cal F}$ and $\boldsymbol{\cal F}^{(\infty)}$, and the fractional spin algebra by $\boldsymbol{\cal FS}$; Lie algebras by lower-case, Gothic letters, e.g. $\mso(2,3)$, $\msp(4;\Real)$, $\msp(4;\Comp)$,  $\msl(2,\Comp)$, $\msl(2;\Real)$ and $\miso(1,2)$; inhomogenous, metaplectic left-orbits and their Hermitian subspaces by ${\cal O}$ and ${\cal S}$, respectively, and finite-dimensional left-modules of matrix algebras by ${\cal C}$; horizontal forms , alias master fields, by capital, Roman or Greek letters, e.g. the HSG master fields $A$ and $B$, the CCHSG master fields $W$, $C$ and $V$; the fractional-spin algebra valued master fields by $\mathbb{A}$, $\mathbb{B}$, $\widetilde{\mathbb{B}}$ and $\widetilde{\mathbb{A}}$; coordinates mostly by lower-case, Roman letters to denote  $x,y,...$ except $Sp(4)$-quartets $Y,Z,...$; twistor-space momenta in $\Real^2\cup i\Real^2$ by lower-case Greek letters $\lambda$, $\mu$, $\dots$\,, and twistor-space momenta in $\Real^2$ by lower-case Roman letters $k$, $l$, $\dots$\ ; partons of the two-parton system by $P,R,...=1,2$; internal, colour states by $I,J,..$; and automorphisms of algebras of functions given by pull-backs induced via maps, $\pi_i$ say, acting on the underlying manifolds, by $\pi^\ast$.

We refer to a one-sided module of $\mso(2,3)$ in which the conformal equation of motion holds as a \emph{singleton}, i.e., as a module in which the singleton annihilator \eqref{A.7} can be factored out.
We refer to a singleton equipped with an $\mso(2,3)$-invariant Hermitian form as a \emph{Hermitian singleton}.
A Hermitian singleton in which $Mp(4;\Real)$ is represented unitarily is referred to as a \emph{metaplectic singleton}.
We refer to the $Mp(4;\Real)$ irreps of a metaplectic singleton as \emph{unitarizable singletons}, which equips a metaplectic singleton with signature given by the signs appearing in the decomposition of its Hermitian form using the positive definite Hermitian forms of its unitarizable subspaces.
We refer to the unitarizable singletons given in compact basis of $\mso(2,3)$ --- i.e., the lowest-weight spaces ${\cal D}^+(1/2)$ and ${\cal D}^+(1)$ expanded using harmonic oscillator states (forming $\msu(2)$-tensors) --- as the \emph{compact singletons} (or harmonic singletons), and to the corresponding highest-weight spaces ${\cal D}^-(-1/2)$ and ${\cal D}^-(-1)$ as compact anti-singletons.
We refer to the corresponding Hermitian singletons (which are not metaplectic) given in conformal basis of $\mso(2,3)$, i.e., the highest-weight spaces ${\cal T}^+(-i/2)$ and ${\cal T}^+(-i)$ expanded using $\msl(2;\Real)$-tensors and the corresponding lowest-weight spaces ${\cal T}^-(i/2)$ and ${\cal T}^-(i)$, as the \emph{conformal singletons} and anti-singletons (or dual conformal singletons), respectively.
We refer to the unitarizable singletons expanded over eigenstates of real Hermitian momenta and dual momenta (i.e., coordinates) with real and purely imaginary eigenvalues (see Sec. 3.5) as the \emph{non-compact singletons} and anti-singletons, respectively.
Though not of use in this paper, the conformal and non-compact singletons are related by complex Bogolyubov transformations to harmonic and coherent states, respectively, in compact singletons; the corresponding modular transformations connecting harmonic expansions, are the topic of \cite{paper0}.

\section{Parent operator algebra}\label{sec:2}

In this Section, we introduce the non-commutative geometry underlying the parent model, which is based on an oscillator representation of the complexified metaplectic group providing a non-perturbative extension of the polynomial Weyl algebra with its Moyal product used in the perturbative Fronsdal approach \cite{meta}.
After exhibiting the need for the metaplectic extension of the Weyl algebra and its complexification in Section 2.1 \ref{sec:2.1}, we introduce the notion of non-commutative correspondence spaces equipped with holomorphic and K\"ahler symplectic structures in Section \ref{subsec:correspondencespace}.
Finally, in Section \ref{subsec:twopartonalgebra}, we construct a first-quantized two-parton system with boundary conditions that yield the two-by-two block structure of the fractional-spin algebra of the FSG and CCHSG defects of the flat superconnection.
The parton system also includes twisted boundary conditions giving rise to outer and inner Klein operators forming a modular group algebra connecting various boundary conditions of the second-quantized parent model.

\subsection{Metaplectic geometry}\label{sec:2.1}

In the context of higher-spin gravity, the interest in the metaplectic group algebra stems from the fact that gauge functions, inner Klein operators, and integration constants of physically relevant linearized as well as exact solutions involve operators given by Gaussians (or limits thereof) built from oscillators:
massless particle states \cite{fibre,2017,COMST}, boundary-to-bulk propagators and ${\cal D}$-functions \cite{Mishasiegel,Didenko:2012tv,corfu19}, HS black-holes (alias generalized Type D solutions) and black-branes \cite{Didenko:2009td,2011,Iazeolla:2012nf,Sundell:2016mxc,2017,review,COMST,corfu19,Didenko:2021vui,Didenko:2021vdb}, instantons, domain walls and FLRW-like solutions \cite{Sezgin:2005pv,Iazeolla:2007wt,Iazeolla:2015tca,cosmo} (see also \cite{Didenko:2023txr} for instantons in arbitrary spacetime dimensions). 
Independently of the ordering scheme in use, the Gaussian \emph{group} elements are represented by symbols with \emph{square-root} pre-factors depending on metaplectic group coordinates with characteristic \emph{$4\pi$-periodicities} on two-fold covers of the complex plane by Riemann sheets, that are crucial to achieving the associativity and reality conditions from which all other basic properties of the theory are derived; for further details, see Appendix B in \cite{meta}.

The metaplectic groups are double coverings of symplectic groups, i.e., groups of automorphisms of Heisenberg algebras; see, e.g., \cite{Folland,CSM,Guillemin:1990ew,Woit:2017vqo}.
The real, metaplectic group $Mp(4;\Real)$, alias the group of Bogolyubov transformations of a pair of harmonic oscillators, is the double covering of $Sp(4;\Real)$ consisting of unitary operators $U$ obeying 
\begin{align}\label{2.1}
U^{-1}\star Y_{\underline{\alpha}} \star U=S_{\underline{\alpha}}{}^{\underline{\alpha}} Y_{\underline{\beta}}\ ,\qquad S\in Sp(4;\Real)\ ,
\end{align} 
where $Y_{\underline{\alpha}}$, $\underline{\alpha}=1,\dots,4$, are real, canonical coordinates for the symplectic $\Real^4$ with translation-invariant structure, and the star denotes the representation of the operator product using symbols.
Since $\pm U$ correspond to the same matrix $S$, the inverse map $S\mapsto U(S)$ is a projective representation of $Sp(4;\Real)$ with cocycle 
\begin{eqnarray}\label{2.2}
&\sigma_\Real:Sp(4;\Real)\times Sp(4;\Real)\to \{\pm 1\}\ ,\qquad U(S_1) U(S_2)=\sigma_\Real(S_1,S_2) U(S_1 S_2)\ ,  &\\\label{2.3}
&\sigma_\Real(S_1,S_2 S_3)\sigma_\Real(S_2,S_3)=\sigma_\Real(S_1,S_2)\sigma_\Real(S_1S_2,S_3)\ .&
\end{eqnarray}
The complex, metaplectic group $Mp(4;\Comp)$ is a complexification of the real group given by a \emph{branched} double cover of $Sp(4;\Comp)$ arising by complexifying Eqs. \eqref{2.1}, \eqref{2.2} and \eqref{2.3}, viz.,
\begin{eqnarray}
&V^{-1}\star Y_{\underline{\alpha}} \star V=S_{\underline{\alpha}}{}^{\underline{\alpha}} Y_{\underline{\beta}}\ ,\qquad S\in Sp(4;\Comp)\ ,&\\
&V(S_1) V(S_2)= \sigma(S_1,S_2) V(S_1 S_2)\ ,&\\& \sigma(S_1,S_2S_3)\sigma(S_2,S_3)=\sigma(S_1,S_2)\sigma(S_1S_2,S_3)\ ,&
\end{eqnarray}
where i) $Y_{\underline{\alpha}}$ treated as canonical coordinates for the holomorphic, symplectic $\Comp^4$ with translation-invariant structure; ii) $V$ is non-unitary; and iii) the cocycle encodes the choice of a square-root branch cut inside $Sp(4;\Comp)$ for the inverse map $S\mapsto V(S)$, which is a holomorphic map that ramifies at the asymptotic boundary of $Sp(4;\Comp)$ \cite{meta}.
For this reason, $Mp(4;\Comp)$ is sometimes referred to as the holomorphic, complex metaplectic group\footnote{Bargmann's theorem implies that unitary representations of simply connected groups, such as $Sp(2n;\Comp)$, are proper, i.e., non-projective. It follows that $S\mapsto V(S)( V(S^{-1})^{\dagger_\Comp})$, where $\dagger_\Comp$ denotes the Hermitian conjugation operation on $\Comp^{2n}$ induced from the Hermitian conjugation operation of the algebra of complex functions on $\Real^{4n}$, is a unitary, proper representation of $Sp(2n;\Comp)$, which is hence \emph{not} faithful as a unitary representation of $Mp(2n;\Comp)$ \cite{meta}.}.

The oscillator realizations of the corresponding semi-direct product groups, i.e., the inhomogeneous, real and complex metaplectic groups $MpH(4;\Real)$ and $MpH(4;\Comp)$, respectively, give rise to group algebras $\Comp[MpH(4;\Real)]\subset \Comp[MpH(4;\Comp)]$ (modulo oscillator ideals) spanned by metaplectic group elements dressed by plane waves from the respective Heisenberg groups $H(4;\Real)$ and $H(4;\Comp)$. 
Their associative structures thus encode products of metaplectic group elements as well as plane waves; indeed, while the plane-wave subalgebra (at the metaplectic identity) can be converted into twisted convolutions closing in suitable classes of symbols (given by intersections of $L^p$-spaces), these classes do \emph{not} contain the metaplectic groups, which are given by Gaussians built from indefinite, bilinear forms\footnote{In the real case, the bilinear forms are purely imaginary.
In the complex case, the space of bilinear forms contains Siegel's upper half-plane, which is a module of the real group, but not a subgroup in itself; indeed, higher-spin gravity boundary conditions activate elements from both the upper and lower half-plane.}.
Thus, the associative structure of $\Comp[MpH(4;\Real)]\subset \Comp[MpH(4;\Comp)]$ (modulo the oscillator ideal) is \emph{not} represented faithfully by their formal Fourier expansions in $\Comp[H(4;\Real)]\subset \Comp[H(4;\Comp)]$ (modulo the oscillator ideal).
More precisely, the metaplectic group algebras consist of two subalgebras: i) invertible operators which represent proper group elements; and ii) non-invertible, idempotent elements, i.e., operators of the form $|\xi;\lambda\rangle\langle\xi';\lambda'|$, labelled by \emph{polarizations} $\xi$ and coherent-state parameters $\lambda$ of the oscillator algebra (from which discrete level indices can be obtained by taking $\lambda$-derivatives), which represent elements of the asymptotic boundary of the underlying symplectic matrix groups, and form a module of (i) that can be given an algebra structure by regularizing the matrix elements $\langle \xi;\lambda|\xi';\lambda'\rangle$ \cite{meta,paper0}; for momentum eigenstates, see Eq. \eqref{2.124}.

Holographic correspondences in 4D, global anti-de Sitter spacetime activate various subalgebras of $\Comp[MpH(4;\Comp)]$: the interior of $\Comp[Mp(4;\Real)]$ contains the vacuum holonomies and its asymptotic boundary contains operators that yield unfolded propagators from the Lorentzian spacetime boundary into the bulk \cite{GiombiYin2,Colombo:2010fu,Colombo:2012jx,Didenko:2012tv,corfu19}; the asymptotic boundary of $\Comp[Mp(4;\Comp)]$ contains operators that yield Wigner functions carrying states with compact weights, and its interior contains modular transformations, including inner Klein operators exchanging negative and positive energy modes. 

To exhibit the inner Kleinians, one may start from the symplectic $\Real^2\cong \Comp$ coordinatized by oscillators obeying $[a,a^{\dagger_{\Real^2}}]_\star=1$, where $\dagger_{\Real^2}$ denotes the Hermitian conjugation operation acting pointwise on the space of functions on $\Real^2$.
Bogolyubov transformations $U \in Mp(2;\Real)$ create a family of unitarily equivalent Fock-space vacua annhilated by $U^{-1} \star\, a \,\star\, U= \alpha a+\beta a^{\dagger_{\Real^2}}$ with $\alpha, \beta\in \Comp$ obeying $|\alpha|^2-|\beta|^2=1$.
The modular transformation that exchanges this family with a family of anti-vacua annihilated by $U^{-1} \star a^{\dagger_{\Real^2}} \star U$ is the inner realization, by a non-unitary element $\gamma\in Mp(2;\Comp)$, of a discrete map preserving the holomorphic, symplectic $\Comp^2$ containing the symplectic $\Real^2$ as a holomorphic, real slice.
This map can be obtained by coordinatizing $\Comp^2$ canonically using oscillators obeying $[a,b]=1$ and $[\bar a,\bar b]=-1$, where $a$ and $b$ are holomorphic and $\bar a:=a^{\dagger_{\Comp^2}}$ and $\bar b:=b^{\dagger_{\Comp^2}}$ are anti-holomorphic, with $\dagger_{\Comp^2}$ denoting the pointwise Hermitian conjugation operation on $\Comp^2$.
Letting $\dagger_{\Real^2}:= \dagger_{\Comp^2}\circ r^\ast_{\Comp^2}$ where $r_{\Comp^2}$ is the reflection map $(\bar a,\bar b)\circ r_{\Comp^2}:=(b,a)$, one may take $\gamma^{-1}\star a\star \gamma=b$ and $\gamma^{-1}\star b\star \gamma=-a$, i.e., $\gamma$ is a \emph{bounded} element in the holomorphic $Mp(2;\Comp)$ given by a \emph{fourth} root of the unity whose square $\gamma\star \gamma \in Mp(2;\Real)$ (viewed as a real, holomorphic subgroup) anti-commutes with $a$ and $b$.

Turning to the symplectic $\Real^4$, one may approach its complexification by viewing it as the holomorphic, symplectic $\Comp^2$ coordinatized by complex oscillators $a_i$ and $b^i:=(a_i)^{\dagger_{\Real^4}}$ obeying $[a_i,b^{j}]_\star=\delta_i^j$.
The corresponding vacua and anti-vacua form $Mp(4;\Real)$ left-orbits that can be related by two types of modular transformations playing the roles of first-quantized PT-operators of the CCHSG and FSG defects, respectively, as follows: i) the direct uplift of the group element $\gamma$ introduced above to an element $\kappa^{(Y)}_{\!\mathscr{P}}=\gamma_1\star \gamma_2\in Mp(4;\Comp)$, which is thus a fourth root of the unity acting diagonally on $\Comp^2 \cong \Comp\times \Comp$, and providing the inner realization of a discrete, holomorphic, symplectic map $\pi_{\!\mathscr{P}}:\Comp^4\to \Comp^4$ that implements PT-transformations in the $\frac12(1+((\pi_{\!\mathscr{P}})^\ast)^2)$-projection of the fibre algebra (assigned to the CCHSG); and ii) a root $\kappa_y\in Mp(4;\Comp)$ of the unit obeying $\kappa_y^{-1}\star a_i\star \kappa_y=b^j\epsilon_{ji}$ and $\kappa_y^{-1}\star a^i\star \kappa_y=\epsilon^{ij} a_j$, and providing the inner realization of a discrete, holomorphic, symplectic map $\pi_y:\Comp^4\to \Comp^4$ that preserves the holomorphic, symplectic, real slice $\Comp^2$, and implements PT-transformations in the $\frac12(1+(\pi_{y})^\ast(\bar\pi_{\bar y})^\ast)^2)$-projection of the fibre algebra (assigned to the FSG defect).

As will be detailed below, 
equipping the holomorphic oscillator algebra on $\Comp^4$ with a holomorphic star product based on a measure on a \emph{chiral} $\Real^4 \subset \Comp^4$ (denoted by $\widetilde \Real^4$ further below) the operator $\kappa_y$ and its holomorphic, Hermitian conjugate $\bar\kappa_{\bar y}:=(\kappa_y)^{\dagger_{\Real^4}}$, where $\dagger_{\Real^4}:=\dagger_{\Comp^4}\circ r^\ast_{\Comp^4}$, become represented by \emph{analytic} delta functions in Weyl order \cite{meta}; the operator $\kappa^{(Y)}_{\!\mathscr{P}}$ is is instead represented by a Gaussian that can be factorized in terms of $\kappa_y$ and an operator $\kappa^{(Y)}_{\!\mathscr{F}}\in Mp(4;\Real)$ implementing Fourier transformations on $\Real^4\cong T^\ast \Real^2$ using a rotated chiral slice.

Thus, to summarize, while (unitary) Bogolyubov transformations act within unitary representations associated to connected spaces of boundary conditions, their complexifications include transformations that exchange sectors of boundary conditions, and that can be implemented into first-quantized descriptions by complexifying underlying symplectic geometries.
Accordingly, the superconnection of the parent model is a \emph{holomorphic}, horizontal form on a complex, fibred, non-commutative differential Poisson manifold with fibres given by the translation-invariant, holomorphic, symplectic $\Comp^4$, whose quantization yields various DGAs containing classical moduli spaces exhibiting different boundary conditions.
Differential Poisson structures introduce two additional features with respect to symplectic structures: compatibility with differential form algebra and inclusion of degenerate brackets, of which the former is of direct relevance for the definition of the theory and the potential role of the latter remains to be investigated. 

In what follows, we proceed with the introduction of horizontal forms on manifolds of the aforementioned type and the Hermitian left module giving rise to the fractional-spin (fibre) algebra suitable for describing Lorentzian holography in the context of HSG \cite{bdhm,svr,silva,corfu19}.
The superconnection will then be defined in \ref{sec:parent}; in particular, the structure of the underlying differential Poisson manifold is spelled out Eqs. \eqref{5.18} and \eqref{3.61}.

\subsection{Correspondence space}\label{subsec:correspondencespace}

The parent model is formulated in terms of holomorphic, horizontal forms\footnote{The notion of horizontal form used in this paper, where horizontal forms are fibre zero-forms, differs from that used in some literature, where horizontal forms are fibre constants.} on a direct product space $\boldsymbol{C}\times \boldsymbol{C}$, where $\boldsymbol{C}$ is a complex\footnote{The HSG and CCHSG models obtained from the parent model are thus formulated in terms of holomorphic coordinates whose complex nature emerges at the level of chiral integration measures facilitating the analytic delta functions and the related independent ``holomorphic'' and ``anti-holomorphic'' involutions $\pi$ and $\bar\pi$; the holomorphic models admit standard real forms defined using holomorphic Hermitian conjugation operations relying on the existence of a reflection map on $\boldsymbol{C}$.} differential Poisson manifold equipped with a set of mutually commuting holomorphic vector fields $\vec s_{\underline{\alpha}}$ that generate inner derivatives compatible with the differential Poisson bracket \cite{Arias:2016agc}, resulting in a holomorphic fibre bundle 
\begin{align}\label{2.7}
\boldsymbol{Y}\to \boldsymbol{C} \stackrel{{\rm pr}_{\boldsymbol{C}}}{\longrightarrow} \boldsymbol{M}\ ,
\end{align} 
with vertical vector spaces spanned by $\vec s_{\underline{\alpha}}$, $\underline{\alpha}=1,\dots,k:={\rm dim}_\Comp(\boldsymbol{Y})$, which we refer to as a \emph{correspondence space}. 

\paragraph{One-parton algebra.}
First-quantizing $\boldsymbol{C}$ using a two-dimensional differential Poisson sigma model including boundary conditions twisted\footnote{\label{Foot:19}
Quantizing a differential Poisson manifold $\boldsymbol{P}$ with an action of a discrete group ${\cal K}$ of symmetries $\pi_i:\boldsymbol{P}\to \boldsymbol{P}$ using a two-dimensional sigma model (with target given by Lie groupoid $T[1]\boldsymbol{P}\stackrel{{\cal K}}{\to} T[1]\boldsymbol{P}$), open boundary segments are glued together using ${\cal K}$-twisted boundary conditions.
The resulting DGA $\Omega^{(\boldsymbol{\cal V})}({\cal K} \times \boldsymbol{P})$ has a module $\boldsymbol{\cal V}=\Comp_\alpha[{\cal K}]\otimes_{\cal K} \boldsymbol{\cal C}$, where $\boldsymbol{\cal C}$ is space of forms on $\boldsymbol{P}$ on which ${\cal K}$ acts and $\Comp_\alpha[{\cal K}]$ is a twisted version of the group algebra of ${\cal K}$, with product rule  $(e_{\pi_{i_1}}\otimes_{\cal K}\psi_1)\star (e_{\pi_{i_2}}\otimes_{\cal K} \psi_2)=\alpha(\pi_{i_1},\pi_{i_2})e_{\pi_{i_1}\pi_{i_2}}\otimes_{\cal K} (\pi_{i_2})^\ast(\psi_1)\star \psi_2$ where the cocycle $\alpha:{\cal K}\times {\cal K}\to \Comp$ obeys $\alpha(\pi_{i_1}\pi_{i_2},\pi_{i_3})\alpha(\pi_{i_1},\pi_{i_2})=\alpha(\pi_{i_1},\pi_{i_2}\pi_{i_3})\alpha(\pi_{i_2},\pi_{i_3})$.
If $\boldsymbol{\cal C}$ is unital, one writes $e_{\pi_i}\equiv e_{\pi_i}\otimes_{\cal K} 1$ and $e_{\pi_i}\star \psi\equiv e_{\pi_i}\otimes_{\cal K} \psi$ and $\boldsymbol{\cal V}\equiv \Comp_\alpha[{\cal K}]\star \boldsymbol{\cal C}$.} by a discrete group ${\cal K}$ of holomorphic differential Poisson maps $\pi_i:\boldsymbol{C}\to \boldsymbol{C}$, yields a DGA
\begin{align}
\Omega^{([\boldsymbol{\cal V}]_\Upsilon;\Upsilon)}_{\rm hor}({\cal K}\times \boldsymbol{C})=(\boldsymbol{{\cal K}}, [\boldsymbol{\cal C}]_\Upsilon;d,\star,\Upsilon,\dagger,{\rm STr})\ , \qquad [\boldsymbol{\cal V}]_\Upsilon:=\Comp_\alpha[{\cal K}]\otimes_{\cal K}[\boldsymbol{{\cal C}}]_\Upsilon\equiv \boldsymbol{{\cal K}}\star[\boldsymbol{{\cal C}}]_\Upsilon\ ,
\end{align}
comprising several structures as follows:  the algebra is obtained by factoring out an ideal from the DGA 
\begin{align}
\Omega^{(\boldsymbol{{\cal V}})}({\cal K}\times \boldsymbol{C})=(\boldsymbol{{\cal K}},\boldsymbol{{\cal C}};d,\star,\imath_{\underline{\alpha}},\Upsilon,\dagger,{\rm STr})\ ,\qquad \boldsymbol{{\cal V}}:=\Comp_\alpha[{\cal K}]\otimes_{\cal K}\boldsymbol{{\cal C}}\equiv\boldsymbol{{\cal K}}\star\boldsymbol{{\cal C}}\ ,
\end{align}
which in turn comprises\footnote{Adding a linear anti-automorphism $\tau$ to the list of DGA operations facilitates projections to minimal models \cite{Konstein:1989ij,review99,Sezgin:2003pt}.} 
\begin{enumerate}
\item[i)] a discrete group algebra $\boldsymbol{\cal K}:= \Comp_\alpha[{\cal K}]$ spanned by elements $e_{\pi_i}$, $\pi_i\in {\cal K}$, with product twisted by a cocycle $\alpha:{\cal K}_{\boldsymbol{M}}\times {\cal K}_{\boldsymbol{Y}}\to \Comp$; 
\item[ii)] a complex vector space $\boldsymbol{{\cal C}}$ of \emph{holomorphic} forms on $\boldsymbol{C}$ with a ${\cal K}$-action, referred to as the module of symbols of the algebra\footnote{The module of symbols is a coordinatization of an abstract algebra using a Wigner, or de-quantization, map. A given, abstract, algebra may be coordinatized using quite distinct classes of symbols, depending on the Wigner map; in particular, a given, abstract operator may be assigned a smooth and bounded symbol using one Wigner map, and a delta-function distribution using an another Wigner map. Thus, similarly to the treatment of differential geometric objects on an ordinary commuting manifold, while it is not possible to assign any intrinsic meaning to function classes, it is possible to assign meaning to equivalence classes of symbols by computing traces of operators. }; 
\item[iii)] a differential $d:\boldsymbol{{\cal V}}\to\boldsymbol{{\cal V}}$ of degree $+1$; 
\item[iv)] an associative product $\star: \boldsymbol{{\cal V}}\otimes \boldsymbol{{\cal V}}\to \boldsymbol{{\cal V}}$ of degree zero, written out as $\star(\psi_1\otimes\psi_2)\equiv \psi_1\star \psi_2$; 
\item[v)] a set of anti-differentials $\imath_{\underline{\alpha}}$ of degree $-1$;  
\item[vi)] a closed, central, ${\cal K}$-invariant, and non-degenerate holomorphic form $\Upsilon\in \boldsymbol{{\cal C}}$ of degree ${\rm dim}_\Comp(\boldsymbol{Y})$ inducing a complex vector space $[\boldsymbol{\cal C}]_{\Upsilon}$ spanned by non-trivial horizontal equivalence classes $[\psi]_\Upsilon$ via $\psi\sim \psi'$ iff $\Upsilon\star (\psi-\psi')=0$;  
\item[vii)] a holomorphic\footnote{A complex manifold $\boldsymbol{C}$ is equipped with a Hermitian conjugation operation $\dagger_\Comp$ inherited from the underlying real manifold, which interchanges holomorphic and anti-holomorphic functions. A complex manifold equipped with an involution  $r:\boldsymbol{C}\to \boldsymbol{C}$ that anti-commutes to the complex structure $J$, viz., $J(r^\ast \theta)=-r^\ast (J(\theta))$ for one-forms $\theta$, admits an additional, holomorphic, Hermitian conjugation map $\dagger:=r^\ast\circ \dagger$, which acts faithfully to on the space of holomorphic forms.}, $t$-twisted \footnote{Given a DGA $\boldsymbol{\cal A}$ with differential $d$, product $\star$ and degree map ${\rm deg}$, a $t$-twisted Hermitian conjugation operation, $t=0,1$, is an anti-linear, anti-involution $\dagger:\boldsymbol{\cal A}\to \boldsymbol{\cal A}$ obeying $(da)^\dagger=(-1)^{(t+1){\rm deg}(a)} d(a^\dagger)$ and $(a\star b)^\dagger=(-1)^{t{\rm deg}(a){\rm deg}(a)}b^\dagger\star a^\dagger$.} Hermitian conjugation operation $\dagger:\boldsymbol{{\cal V}}\to \boldsymbol{{\cal V}}$; and 

\item[viii)] holomorphic, graded trace operation\footnote{Given an anti-holomorphic volume-form $\overline \Omega$ on $\boldsymbol{C}$, a holomorphic, an alternative graded trace operation is given by $\int^\prime_{\boldsymbol{C}} \Upsilon\star \psi\star \overline \Omega$. } 
\begin{align}
{\rm STr}_{\boldsymbol{\cal V}}\, \psi:= \int_{\widetilde{\boldsymbol{C}}_\Real}^\prime {\rm Tr}_{\boldsymbol{\cal K}} \Upsilon\star \psi\ ,\qquad {\rm Tr}_{\boldsymbol{\cal K}} e_{\pi_i}=\delta_{\pi_i,{\rm id}}\ ,
\end{align}
using a real integration domain $\widetilde{\boldsymbol{C}}_\Real\subset \boldsymbol{C}$ of ${\rm dim}_\Real(\widetilde{\boldsymbol{C}}_\Real)={\rm dim}_\Comp(\boldsymbol{C}_\Comp)$, referred to as the \emph{chiral} integration domain, and the prime refers to anticipated regularization\footnote{The total integral factorizes into an integral over the fibre which is regularized by a subtraction scheme \cite{Boulanger:2015kfa,meta,paper0} specified in Eq. \eqref{2.67reg}, and base whose non-commutative factor is integrated out by assigning it top-form cohomology element; for the applications to the second Chern class, see Sections \ref{sec:2.3} and \ref{sec:conclusionspart1}. The remaining integration domain is a subspace of the commutative factor of the base, which may include asymptotic regions where holographic regularization schemes apply.} compatible with the monary DGA operations, viz., 
\begin{align}
{\rm STr}_{\boldsymbol{\cal V}}\circ d=0\ ,\qquad  {\rm STr}_{\boldsymbol{\cal V}}\circ \dagger=\dagger\circ {\rm STr}_{\boldsymbol{\cal V}}\ ,\qquad {\rm STr}_{\boldsymbol{\cal V}}\circ \pi^\ast_i={\rm STr}_{\boldsymbol{\cal V}}\ ,\quad \pi_i\in{\cal K}\ .
\end{align}
\end{enumerate}
The monary and binary operations of $\Omega^{(\boldsymbol{{\cal V}})}({\cal K}\times\boldsymbol{C})$ obey the compatibility conditions 
\begin{align}
& dd=0\ ,\quad [\imath_{\underline{\alpha}},\imath_{\underline{\beta}}]=0\ ,\quad\psi_1\star (\psi_2\star \psi_3)=(\psi_1\star \psi_2)\star \psi_2\ ,&\\
& d(\psi_1\star \psi_2)=(d\psi_1)\star\psi_2+(-1)^{{\rm deg}(\psi_1)}\psi_1 \star (d\psi_2)\ ,&\\
& \imath_{\underline{\alpha}}(\psi_1\star \psi_2)=(\imath_{\underline{\alpha}}\psi_1)\star\psi_2+(-1)^{{\rm deg}(\psi_1)}\psi_1 \star (\imath_{\underline{\alpha}}\psi_2)\ ,&
\end{align}
where $d$ and $\imath_{\underline\alpha}$ commute to the generators of $\boldsymbol{\cal K}$, 
and the joint kernels $\cap_{\underline{\alpha}}{\rm ker}(\imath_{\vec s_{\underline{\alpha}}})$ and $\cap_{\underline{\alpha}}{\rm ker}(\imath_{\underline{\alpha}})$ are assumed to be isomorphic as vector subspaces of $\boldsymbol{{\cal V}}$.
The non-degeneracy of $\Upsilon$ amounts to that $c:=\imath_{\underline{1}}\cdots \imath_{\underline{k}} \Upsilon$ is a non-trivial, central zero-form in $\cap_{\underline{\alpha}}{\rm ker}(\imath_{\underline{\alpha}})$, such that if $\psi \in \cap_{\underline{\alpha}}{\rm ker}(\imath_{\underline{\alpha}})$ obeys $\Upsilon \star \psi=0$, then $\psi=0$, that is,
\begin{align}
    [\boldsymbol{{\cal V}}]_\Upsilon\cong\cap_{\underline{\alpha}}{\rm ker}(\imath_{\vec s_{\underline{\alpha}}})\cong\cap_{\underline{\alpha}}{\rm ker}(\imath_{\underline{\alpha}})\ .
\end{align}
Thus, $\Omega^{([\boldsymbol{\cal V}]_\Upsilon;\Upsilon)}_{\rm hor}({\cal K}\times\boldsymbol{C})$ is a DGA with monary and binary operations\footnote{The joint kernel $\cap_{\underline{\alpha}}{\rm ker}(\imath_{\vec d_{\underline{\alpha}}})$ is a graded, associative subalgebra of $\boldsymbol{{\cal V}}$ without any natural differential.} 
\begin{align}
    d[\psi]_\Upsilon:=[d\psi]_\Upsilon\ ,\qquad [\psi_1]_\Upsilon\star [\psi_1]_\Upsilon:=[\psi_1\star \psi_2]_\Upsilon\ .
\end{align}
The DGA structures are assumed to be compatible with $\dagger$, viz.
\begin{align}
(d\psi)^\dagger=d(\psi^\dagger)\ ,\quad (d_{\underline{\alpha}}\psi)^\dagger=d_{\underline{\alpha}}(\psi^\dagger)\ ,\quad (\psi_1\star\psi_2)^\dagger=(-1)^{{\rm deg}(\psi_1){\rm deg}(\psi_2)}(\psi_2)^\dagger\star (\psi_1)^\dagger\ ,
\end{align}
and it is assumed that\footnote{Since the Hermitian conjugation operation $\dagger_\Comp$ commutes to $\pi^\ast$, it follows that $(e_\pi)^{\dagger_\Comp}=e_\pi$, while the holomorphic Hermitian conjugation operation $\dagger$ may act non-trivially on $\boldsymbol{\cal K}$.} 
\begin{align}
(\Upsilon)^\dagger=\Upsilon\ ,\qquad ([\psi]_\Upsilon)^\dagger:=[\psi^\dagger]_\Upsilon\ .
\end{align}
The horizontal DGA is equipped with a holomorphic, graded trace operation
\begin{align}
{\rm STr}_{[\boldsymbol{\cal V}]_\Upsilon}\, [\psi]_\Upsilon:= {\rm STr}_{\boldsymbol{\cal V}}\, \Upsilon\star \psi\equiv \int_{\widetilde{\boldsymbol{C}_{\Real}}}^\prime {\rm Tr}_{\boldsymbol{\cal K}}\Upsilon\star \psi\ .
\end{align}
Assuming a holomorphic section $s:\boldsymbol{M}\to \boldsymbol{C}$ that is a differential Poisson map yields a DGA projection $(s\circ {\rm pr}_{\boldsymbol{C}})^\ast:\Omega(\boldsymbol{C})\to \Omega(\boldsymbol{C})$ such that
\begin{align}
\Omega(\boldsymbol{M})\cong (s\circ {\rm pr}_{\boldsymbol{C}})^\ast \Omega(\boldsymbol{C})\ ,\qquad \Omega(\boldsymbol{Y})\cong (1-(s\circ {\rm pr}_{\boldsymbol{C}})^\ast)\Omega(\boldsymbol{C})\ ,
\end{align}
with unital fibre and base algebras. 
We also assume that
\begin{align}
{\cal K}={\cal K}_{\boldsymbol{M}}\times  {\cal K}_{\boldsymbol{Y}}\ ,\qquad {\cal K}_{\boldsymbol{Y}}\circ s\circ {\rm pr}_{\boldsymbol{C}}={\rm Id}_{\boldsymbol{Y}}\ ,\qquad {\cal K}_{\boldsymbol{M}}:={\cal K}/{\cal K}_{\boldsymbol{Y}}\ ,
\end{align}
with normal subgroup ${\cal K}_{\boldsymbol{Y}}$, inducing a semi-direct factorization 
\begin{align}
\boldsymbol{\cal K}=\boldsymbol{\cal K}_{\boldsymbol{M}}\star \boldsymbol{\cal K}_{\boldsymbol{Y}}\ ,\quad \boldsymbol{\cal K}_{\boldsymbol{M}}:=\Comp_\alpha[{\cal K}_{\boldsymbol{M}}]\ ,\quad \boldsymbol{\cal K}_{\boldsymbol{Y}}:=\Comp_\alpha[{\cal K}_{\boldsymbol{Y}}]\ ,\quad {\rm Tr}_{\boldsymbol{\cal K}}={\rm Tr}_{\boldsymbol{\cal K}_{\boldsymbol{Y}}}{\rm Tr}_{\boldsymbol{\cal K}_{\boldsymbol{M}}}\ ,
\end{align}
where the subalgebras are coupled via the off-diagonal cocycles $\alpha:{\cal K}_{\boldsymbol{M}}\times {\cal K}_{\boldsymbol{Y}}\to \Comp$ and $\alpha:{\cal K}_{\boldsymbol{Y}}\times {\cal K}_{\boldsymbol{Y}}\to \Comp$ (assumed to yield a Clifford algebra in the parent model).

\paragraph{Factorization.}
A factorization of the horizontal form algebra is a subspace $\boldsymbol{U}\subseteq \boldsymbol{C}$ in which 
\begin{align}
\Omega^{([\boldsymbol{{\cal V}}]_\Upsilon,\Upsilon)}_{\rm hor}({\cal K}\times\boldsymbol{C})|_{\boldsymbol{U}}\cong \boldsymbol{\cal K}\star\left(\Omega^{(\boldsymbol{{\cal Y}},\Upsilon)}_{[0]}(\boldsymbol{Y})\otimes\Omega^{(\boldsymbol{{\cal M}})}({\rm pr}_{\boldsymbol{C}}(\boldsymbol{U}))\right)\ ,
\end{align}
where $\Omega^{(\boldsymbol{{\cal M}})}({\rm pr}_{\boldsymbol{C}}(\boldsymbol{U}))$ is a DGA with module $\boldsymbol{\cal M}$ consisting of holomorphic forms on ${\rm pr}_{\boldsymbol{C}}(\boldsymbol{U})$, and $\Omega^{(\boldsymbol{{\cal Y}},\Upsilon)}_{[0]}(\boldsymbol{Y})$ is an associative algebra with module $\boldsymbol{\cal Y}$ consisting of a space of holomorphic functions on ${\boldsymbol{Y}}$.
Thus, the full module
\begin{align}
[\boldsymbol{\mathcal V}]_\Upsilon|_{\boldsymbol{U}}\equiv \boldsymbol{\mathcal{K}}\star [\boldsymbol{\mathcal C}]_\Upsilon|_{\boldsymbol{U}}=\boldsymbol{\cal A}\star \boldsymbol{\cal B}\ ,\qquad \boldsymbol{\cal A}:=\boldsymbol{\cal K}_{\boldsymbol{Y}}\star \boldsymbol{\cal Y}\ ,\qquad \boldsymbol{\cal B}:=\boldsymbol{\cal K}_{\boldsymbol{M}}\star \boldsymbol{\cal M}\ ,
\label{2.24}\end{align}
where $\boldsymbol{\cal A}$ and $\boldsymbol{\cal B}$ are mutually commuting, such that
if $\Psi_\lambda$ spans $\boldsymbol{\cal Y}$, then $[\psi]_\Upsilon\in [\boldsymbol{\mathcal V}]_\Upsilon$ can be represented by  $\psi|_{{\boldsymbol{U}}}=\sum_{i,\lambda} e_{\pi_i}\star \psi^{i,\lambda}\star \Psi_\lambda$ with $\psi^{i,\lambda} \in \boldsymbol{\cal M}$, and\footnote{In an abuse of notation, we denote $\Upsilon$ and its local representative by the same symbol.}
\begin{align}
{\rm STr}_{[\boldsymbol{\cal V}]_\Upsilon|_{\boldsymbol{U}}}\,[\psi]_\Upsilon=\sum_\lambda \left({\rm Tr}_{\boldsymbol{\cal M}}\,\psi^{{\rm id},\lambda}\right)\left( {\rm Tr}_{\boldsymbol{\cal Y}}\, \Upsilon\star \Psi^{\phantom{\rm id}}_{{\rm id},\lambda}\right)\ ,
\end{align}
using regularized, holomorphic trace operations
\begin{align}
{\rm Tr}_{\boldsymbol{\cal M}}\,\psi=\int_{\widetilde{{\rm pr}_{\boldsymbol{C}}(\boldsymbol{U})}_\Real}^{\prime}\psi\ ,\qquad {\rm Tr}_{\boldsymbol{\cal Y}}\, \Psi =\int_{\widetilde{\boldsymbol{Y}_\Real}}^{\prime} \Upsilon\star \Psi\ ~,\label{2.12}
\end{align}
with chiral integration domains $\widetilde{{\rm pr}_{\boldsymbol{C}}(\boldsymbol{U})}_\Real\subset {\rm pr}_{\boldsymbol{C}}(\boldsymbol{U})$ and $\widetilde{\boldsymbol{Y}_\Real}\subset \widetilde{\boldsymbol{Y}}$. 

\subsection{Intertwining two-parton algebra}\label{subsec:twopartonalgebra}

We assume that $\boldsymbol{C}\times \boldsymbol{C}$ is first-quantized subject to twisted boundary conditions dictated by projection maps
\begin{alignat*}{2}
{\rm pr}^\bullet_{11}:&\,\boldsymbol{C}\times \boldsymbol{C}\to \boldsymbol{C}\times \{p_\bullet\}\ ,\quad {\rm pr}^\bullet_{12}:&\,\boldsymbol{C}\times \boldsymbol{C}\to \{p_\bullet\}\times \boldsymbol{C}\ ,\\
&(p_1,p_2)\mapsto (p_1,p_\bullet) &(p_1,p_2)\mapsto(p_\bullet,p_1)\ ,\\
{\rm pr}^\bullet_{21}:& \,\boldsymbol{C}\times \boldsymbol{C}\to \boldsymbol{C}\times \{p_\bullet\}\ ,\quad {\rm pr}^\bullet_{22}: &\,\boldsymbol{C}\times \boldsymbol{C}\to \{p_\bullet\}\times \boldsymbol{C}\ ,\\
&(p_1,p_2)\mapsto (p_2,p_\bullet) &(p_1,p_2)\mapsto(p_\bullet,p_2)\ ,
\end{alignat*}
where $p_\bullet\in \boldsymbol{C}$ is a ${\cal K}$-invariant point, generating an associative algebra ${\cal P}$, viz.
\begin{eqnarray}
&{\rm pr}^\bullet_{11}\circ{\rm pr}^\bullet_{11}={\rm pr}^\bullet_{11}\ ,\qquad{\rm pr}^\bullet_{22}\circ{\rm pr}^\bullet_{22}={\rm pr}^\bullet_{22}\ ,&\\ 
&{\rm pr}^\bullet_{12}\circ{\rm pr}^\bullet_{11}={\rm pr}^\bullet_{22}\circ {\rm pr}^\bullet_{12}\ ,\qquad{\rm pr}^\bullet_{21}\circ{\rm pr}^\bullet_{22}={\rm pr}^\bullet_{11}\circ {\rm pr}^\bullet_{21}\ ,&\\
&{\rm pr}^\bullet_{11}\circ{\rm pr}^\bullet_{22}={\rm pr}^\bullet_{22}\circ {\rm pr}^\bullet_{11}={\rm pr}^\bullet_{12}\circ{\rm pr}^\bullet_{22}={\rm pr}^\bullet_{11}\circ {\rm pr}^\bullet_{12}={\rm pr}^\bullet_{21}\circ{\rm pr}^\bullet_{11}={\rm pr}^\bullet_{22}\circ {\rm pr}^\bullet_{12}=:{\rm pr}^\bullet\ .&
\end{eqnarray}
The projection maps factorize, viz.,
\begin{align} {\rm pr}^\bullet_{PP}=\sigma^\bullet_P\circ \pi^\bullet_P\ ,\qquad \pi^\bullet_P: \boldsymbol{C}\times \boldsymbol{C}\to \boldsymbol{C}\ ,\qquad \pi_P(p_1,p_2):=p_P\ ,
\end{align}
for $P=1,2$, using sections 
\begin{align}
\sigma^\bullet_P: \boldsymbol{C}\to \boldsymbol{C}\times \boldsymbol{C}\ ,\quad \sigma_P(p):=(\sigma_1(p),\sigma_2(p))\ ,\quad \sigma_P(p):=p\ ,\quad \sigma_{\check P}(p):=p_\bullet\ , 
\end{align}
where $\check 1:=2$, $\check 2:=1$; the sections combine with the projection maps into truncation maps, viz., 
\begin{align}
{\rm tr}^{\bullet}_P:={\rm pr}^\bullet_{PP}\circ \sigma^\bullet_P: \boldsymbol{C}\to \boldsymbol{C}\times \boldsymbol{C}\ , \qquad  ({\rm tr}^{\bullet}_P)^\ast: \Omega(\boldsymbol{C}\times \boldsymbol{C})\equiv \Omega_{(1)}(\boldsymbol{C})\otimes \Omega_{(2)}(\boldsymbol{C})\to \Omega_{(P)}\ ,
\end{align}
whose pull-backs thus delete a unit.
Taking the vertex operators to obey $({\rm pr}^\bullet)^\ast(\psi)=0$, yields a non-unital, projected two-parton DGA given by the semi-direct product\footnote{The projected element $({\rm pr}^{\bullet}_{PP})^\ast(\psi_{(\check P)})$ is proportional to the unit ${\rm Id}_{(\check P)}\in \Omega_{(\check P)}$, $P=1,2$, which is contracted to $\Comp$ by the contracting pull-back operation $(\sigma_P^\bullet)^\ast$; for the role of these maps in framed oscillator algebras of multi-parton systems, see \cite{Vasiliev:2018zer}.}
\begin{align}\label{2.29}
\Omega^{(\boldsymbol{{\cal P}}\star \boldsymbol{{\cal V}}^{\otimes 2})}( {\cal P}\times ({\cal K}\times \boldsymbol{C})^2)=  \boldsymbol{{\cal P}}\star (\Omega^{(\boldsymbol{\cal V})}( {\cal K}\times \boldsymbol{C}))^{\otimes 2}\equiv \boldsymbol{\cal P}\star \Omega^{(\boldsymbol{\cal V})}_{(1)}( {\cal K}\times \boldsymbol{C})\star \Omega_{(2)}^{(\boldsymbol{\cal V})}( {\cal K}\times \boldsymbol{C})\ ,
\end{align}
where $\boldsymbol{{\cal P}}$ is generated by $e^\bullet_{PQ}$ obeying
\begin{align}\label{2.32}
e^\bullet_{PQ}\star e^\bullet_{RS}:={}&\delta_{QR}e^\bullet_{PS}\ ,\\ \label{2.33}
e^\bullet_{PQ}\star \psi_{(1)}\star \psi_{(2)}:={}&e^\bullet_{PQ}\star ({\rm tr}^{\bullet}_Q)^\ast(\psi_{(1)}\star \psi_{(2)})\ ,\\
\psi_{(1)}\star \psi_{(2)}\star e^\bullet_{PQ}:={}&({\rm tr}^{\bullet}_P)^\ast(\psi_{(1)}\star \psi_{(2)})\star e_{PQ}\ ,\\
 e^\bullet_{{PQ}}\star \psi_{(Q)}:={}&\psi_{(P)}\star e_{PQ}\ ,
\end{align}
for $\psi_{(P)}\in \Omega_{(P)}^{(\boldsymbol{\cal V})}( {\cal K}\times \boldsymbol{C})$, and where the truncating contractors
\begin{align}
({\rm tr}^{\bullet}_P)^\ast(\psi_{(1)}\otimes \psi_{(2)}):={\rm str}^\bullet\left(\psi_{(\check P)}\right) \psi_{(P)}\ ,\quad {\rm str}^\bullet(\psi):=\psi|_{(e,p_\bullet)}\ ,
\end{align} 
using supertrace operations vanishing intrinsic degree acting in Weyl order by replacing symbols in $\Omega_{(P)}^{(\boldsymbol{\cal V})}( {\cal K}\times \boldsymbol{C})$ by ${\rm Id}_{(P)}$ times their value at $(e,p_\bullet)\in \boldsymbol{\cal K}\times \boldsymbol{C}$.
Thus, 
\begin{align}
\Omega^{(\boldsymbol{{\cal P}}\star \boldsymbol{{\cal V}}^{\otimes 2})}( {\cal P}\times ({\cal K}\times \boldsymbol{C})^2)=\bigoplus_{P,Q} e^\bullet_{PQ}\star \Omega_{(Q)}^{(\boldsymbol{\cal V})}( {\cal K}\times \boldsymbol{C})\ ,
\end{align}
containing the diagonal subalgebras $e^\bullet_{{PP}}\star \Omega_{(P)}^{(\boldsymbol{\cal V})}( {\cal K}\times \boldsymbol{C})\cong \Omega^{(\boldsymbol{\cal V})}( {\cal K}\times \boldsymbol{C})$.
The Hermitian conjugation and trace operations are extended to include $\boldsymbol{\cal P}$ by declaring
\begin{align}
(e^\bullet_{PQ})^\dagger=e^\bullet_{QP}\ ,\qquad {\rm Tr}_{\cal P}\, (e^\bullet_{PQ})=\delta_{PQ}\ .
\end{align}
The horizontal quotient DGA  
\begin{align}\label{2.32a}
\Omega^{(\boldsymbol{{\cal P}}\star([\boldsymbol{{\cal V}}]_\Upsilon)^{\otimes 2}, \Upsilon\otimes \Upsilon)}_{\rm hor}( {\cal P}\times ({\cal K}\times \boldsymbol{C})^2)={}& \boldsymbol{{\cal P}}\star (\Omega^{([\boldsymbol{\cal V}]_\Upsilon,\Upsilon)}_{\rm hor}( {\cal K}\times \boldsymbol{C}))^{\otimes 2}\\={}& \bigoplus_{P,Q} e^\bullet_{PQ}\star \Omega^{([\boldsymbol{\cal V}]_\Upsilon,\Upsilon)}_{(Q){\rm hor}}( {\cal K}\times \boldsymbol{C})\ ,    
\end{align}
containing the diagonal subalgebras $e^\bullet_{{PP}}\star \Omega^{([\boldsymbol{\cal V}]_\Upsilon,\Upsilon)}_{(Q){\rm hor}}( {\cal K}\times \boldsymbol{C})\cong \Omega^{([\boldsymbol{\cal V}]_\Upsilon,\Upsilon)}_{(Q){\rm hor}}( {\cal K}\times \boldsymbol{C})$
consisting of elements of the form $e^\bullet_{PP}\star ([\psi]_\Upsilon)_{(P)}$ with $[\psi]_\Upsilon\in \Omega^{([\boldsymbol{\cal V}]_\Upsilon,\Upsilon)}_{(Q){\rm hor}}( {\cal K}\times \boldsymbol{C})$.
In Section \ref{Sec:2.8}, the fractional-spin algebra is obtained as a subalgebra of the fibre component $\boldsymbol{\cal P}\star \boldsymbol{\cal A}$ arising by assigning the two partons separate left-modules given by left-orbits of two distinct group algebras contained in $\boldsymbol{\cal A}$.

\section{Metaplectic fibre algebra}\label{sec:fibrealg}

In this Section, we spell out the structure of the associative algebra of functions on the non-commutative fibre of the correspondence space, which will be relevant in constructing the parent field equations.
The algebra consists of endomorphisms of a left-orbit of the inhomogeneous, holomorphic  metaplectic group algebra acting on its asymptotic infinity. The latter has the structure of a singleton module equipped with a Hermitian form given by a regularized trace operation.

\subsection{Holomorphic symplectic structure and Klein operators}

\paragraph{Holomorphically real coordinates.}

The fibre of the correspondence space of the parent model is the holomorphic, symplectic 
\begin{align}
\boldsymbol{Y}=(\Comp^4,\omega)\ ,\qquad \Upsilon\equiv \Upsilon_{[4,0]}:=\frac12 \frac{\omega_{[2,0]}}{2\pi}\wedge \frac{\omega_{[2,0]}}{2\pi}\ ,
\end{align}
with translation-invariant structure $\omega$, 
coordinatized canonically by $Sp(4;\Comp)$-quartets $Y^{\underline{\alpha}}$ and their Hermitian conjugates $\widetilde{Y}^{\tilde{\underline{\alpha}}}$, viz., 
\begin{align}
\omega_{[2,0]}={}&\frac14dY^{\underline{\alpha}}\wedge dY_{\underline{\alpha}}\ ,\qquad [Y^{\underline{\alpha}}, Y^{\underline{\beta}}]_\star=2iC^{\underline{\alpha\beta}} \ ,\label{3.2}\\
\omega_{[0,2]}={}&\frac14d\widetilde Y^{\underline{\alpha}}\wedge d\widetilde Y_{\underline{\alpha}}\ ,\qquad[\widetilde{Y}^{\tilde{\underline{\alpha}}},\widetilde{Y}^{\tilde{\underline{\beta}}}]_\star=2iC^{\underline{\tilde{\alpha}\tilde{\beta}}} \ ,\\
\Upsilon_{[4,0]}={}& \frac{1}{(4\pi)^2} d^4Y\ ,\qquad d^4Y:= dY^{\underline{1}}\wedge dY^{\underline{2}}\wedge dY^{\underline{3}}\wedge dY^{\underline{4}}\ ,
\end{align}
i.e. $Y^{\underline{\alpha}}$ are holomorphically real (and the normalization is chosen in favor of unit contraction in Weyl order, viz. $Y^{\underline{\alpha}}\star Y^{\underline{\beta}}-Y^{\underline{\alpha}} Y^{\underline{\beta}}=iC^{\underline{\alpha\beta}}$).
The actions of the Hermitian conjugation operation $\dagger_\Comp$, defined using the algebra of complex functions on $\Real^8$, and the holomorphic Hermitian conjugation operation $\dagger$, on the canonical coordinates read
\begin{align}
(Y^{\underline{\alpha}})^{\dagger_\Comp}=\widetilde Y^{\underline{\alpha}}\ ,\quad (\widetilde Y^{\underline{\alpha}})^{\dagger_\Comp}= Y^{\underline{\alpha}} \ ,\quad (Y^{\underline{\alpha}})^\dagger= Y^{\underline{\alpha}}\ ,\quad (\widetilde Y^{\underline{\alpha}})^\dagger= \widetilde Y^{\underline{\alpha}}\ .
\end{align}
The fibre algebra of the parent model is thus represented in terms of symbols that are holomorphic functions and distributions on $\boldsymbol{Y}$ of which only a subset admit restrictions to the real slice 
\begin{align}
\boldsymbol{Y}_\Real:=(\Real^4,\omega_\Real)\subset \boldsymbol{Y}\ ,\qquad \widetilde{Y}^{\tilde{\underline{\alpha}}}|_{\boldsymbol{Y}_\Real}=Y^{{\underline{\alpha}}}|_{\boldsymbol{Y}_\Real}\ ; 
\end{align}
henceforth, we keep track only of the holomorphic coordinates of $\boldsymbol{Y}$.

\paragraph{Holomorphically complex and real doublets.}
The HSG and CCHSG branches activate cohomological two-forms that are distributions that do not admit any restriction to $\boldsymbol{Y}_\Real$.
On the HSG defect, these structures preserve an $SL(2;\Comp)\times \overline{SL}(2;\Comp)\subset Sp(4;\Comp)$ under which $Y^{\underline{\alpha}}$ splits into holomorphically complex doublets $(y^\alpha,\bar y^{\dot\alpha})$, obeying
\begin{align}
\omega_{[2,0]}={}& \frac14 dy^{{\alpha}}\wedge dy_{{\alpha}}+\frac14 d\bar y^{{\dot\alpha}}\wedge d\bar y_{{\dot\alpha}}\ ,\qquad [y^\alpha,y^\beta]_\star=2i\epsilon^{\alpha\beta}\ ,\qquad \yb^{\dot\alpha}:=(y^\alpha)^\dagger\ ,\label{2.49}\\\label{2.50}
\Upsilon_{[4,0]}={}&\frac{1}{(4\pi)^2}d^2y\wedge d^2\bar{y}\ ,\qquad d^2y:= dy^1\wedge dy^2\equiv -\frac12 dy^\alpha\wedge dy_\alpha\ .
\end{align}
The CCHSG structures instead preserve the manifest $SL(2;\Real)\times \Real\subset Sp(4;\Comp)$ symmetry of the conformal basis of $\mathfrak{sph}(4;\Real)$ defined in App. \ref{App:emb} in which $Y^{\underline{\alpha}}$ splits into a conjugate pair of holomorphically real doublets $y^{\xi,\alpha}$, $\xi=\pm$, normalized such that
\begin{align}
\omega_{[2,0]}={}& \frac12 dy^{+,{\alpha}}\wedge dy^-_{{\alpha}}\ ,\qquad [y^{\xi,\alpha},y^{\xi',\beta}]_\star=2i\delta^{\xi,\xi'}\epsilon^{\alpha\beta}\ ,\qquad (y^{\xi,\alpha})^\dagger=y^{\xi,\alpha}\ ,\label{3.9}\\ \label{2.52b}
\Upsilon_{[4,0]}={}&\frac{1}{(4\pi)^2}d^2y^+\wedge d^2y^-\ ,\qquad d^2y^\xi:=dy^{\xi,1}\wedge dy^{\xi,2}\equiv -\frac12 dy^{\xi,\alpha}\wedge dy^\xi_\alpha\ , 
\end{align}
where the conformal weights are determined by
\begin{align}
[D,y^\xi_\alpha]_\star:=\frac{i\xi}2 y^\xi_\alpha\ ,\qquad D:=\frac14 y^+ y^- ,
\end{align}
using the NW-SE convention for implicit spinor indices, in which $y^+ y^- \equiv y^{+\a} y^-_\a$.
We use conventions in which the two sets of coordinates are related as follows\footnote{To our knowledge, the basis \eq{2.67} for the oscillator realization of the 3D conformal group was first used in \cite{F&L}.}:
\begin{align}\label{2.67}
y^\xi_\alpha=\frac{e^{i\xi\pi/4}}{\sqrt{2}}(y_\alpha-i\xi \bar y_{\dot{\alpha}})\ ,\qquad y_\alpha=\frac{e^{-i\pi/4}}{\sqrt{2}}(y^+_\alpha+iy^-_\alpha)\ .
\end{align}

\paragraph{Outer Klein operators.}
We assume that ${\cal K}_{\boldsymbol{Y}}$ is generated by the  holomorphic\footnote{From $\dagger_\Comp\circ \pi^\ast_y=\pi^\ast_y\circ \dagger_\Comp$ idem $\bar\pi_{\bar y}$, it follows that $(\tilde y^{\dot\alpha},\tilde{\bar y}^{\alpha})\circ \pi_y=(-\tilde y^{\dot\alpha},\tilde{\bar y}^{\alpha})$ and $(\tilde y^{\dot\alpha},\tilde{\bar y}^{\alpha})\circ \bar\pi_{\bar y}=(\tilde y^{\dot\alpha},-\tilde{\bar y}^{\alpha})$.}, symplectic involutions $\pi_y,\bar\pi_{\bar y},\pi_{\!\mathscr{P}}^{_{_{ (Y)}}}:\boldsymbol{C}\to \boldsymbol{C}$ given by  
\begin{align}\label{piyyb}
(y^\alpha,\bar y^{\dot\alpha})\circ \pi_y:=(-y^\alpha,\bar y^{\dot\alpha})\ ,\qquad (y^\alpha,\bar y^{\dot\alpha})\circ \bar\pi_{\bar y}:=(y^\alpha,-\bar y^{\dot\alpha})\ ,\qquad y^{\xi,\alpha}\circ \pi_{\!\mathscr{P}}^{_{_{ (Y)}}} :=-i\xi y^{\xi,\alpha}\ ,
\end{align}  
inducing an untwisted outer group algebra $\boldsymbol{{\cal K}}_{\boldsymbol{Y}}=\Comp[{\cal K}_{\boldsymbol{Y}}]$ with generators 
\begin{align}
k_y\equiv  e_{\pi_y}\ ,\qquad \bar k_{\bar y}\equiv e_{\bar \pi_{\bar y}}\ ,\qquad k_{\!\mathscr{P}}^{_{_{ (Y)}}}\equiv e_{\pi_{\!\mathscr{P}}^{_{_{ (Y)}}}}\ ,
\end{align} 
obeying 
\begin{eqnarray}
&k_y\star k_y=1\ ,\qquad k_y\star \bar k_{\bar y}=\bar k_{\bar y}\star k_y\ ,\qquad \bar k_{\bar y}\star \bar k_{\bar y}=1\ ,&
\label{2.57}\\
&k_{\!\mathscr{P}}^{_{_{ (Y)}}}\star k_{\!\mathscr{P}}^{_{_{ (Y)}}}=k_y\star \bar k_{\bar y}\ ,\qquad k_y\star k_{\!\mathscr{P}}^{_{_{ (Y)}}}=k_{\!\mathscr{P}}^{_{_{ (Y)}}}\star \bar k_{\bar y}=:k^{_{_{ (Y)}}}_{\!\mathscr{F}}\ ,&\\&
k_y\star k_{\!\mathscr{F}}^{_{_{ (Y)}}}=k_{\!\mathscr{F}}^{_{_{ (Y)}}}\star \bar k_{\bar y}\ ,\qquad k_{\!\mathscr{F}}^{_{_{ (Y)}}}\star k_{\!\mathscr{F}}^{_{_{ (Y)}}}=1\ ,&\\
&(k_y)^\dagger=\bar k_{\bar y}\ ,\qquad (\bar k_{\bar y})^\dagger=k_y\ ,\qquad 
\left(k_{\!\mathscr{P}}^{_{_{ (Y)}}}\right)^\dagger=k_{\!\mathscr{P}}^{_{_{ (Y)}}}\ ,\qquad 
\left(k_{\!\mathscr{F}}^{_{_{ (Y)}}}\right)^\dagger=k_{\!\mathscr{F}}^{_{_{ (Y)}}}\ ,&\\
&{\rm Tr}_{\boldsymbol{\cal K}_{\boldsymbol{Y}}}\, (k_y)^{m} \star (\bar{k}_{\bar y})^{\bar m}\star (k_{\!\mathscr{P}}^{_{_{ (Y)}}})^n = \delta_{m,0}\delta_{\bar m,0}\delta_{n,0}\ ,\qquad m,\bar m, n=0,1\ ;&\label{2.61}
\end{eqnarray}
in conformal basis, one has 
\begin{align}
y^\xi_\alpha\circ \pi_y=-i\xi y^{-\xi}_\alpha\ ,\qquad  y^\xi_\alpha\circ \pi_{\!\mathscr{F}}^{_{_{ (Y)}}}=y^{-\xi}_\alpha\ . \end{align}
The semi-direct product $\boldsymbol{{\cal K}}_{\boldsymbol{Y}}\star \boldsymbol{\cal Y}$ is then formed using the associative composition rule
\begin{align}
\Psi\star e_{\pi_i}= e_{\pi_i}\star \pi_i^\ast(\Psi)\ ,\qquad \pi_i\in {\cal K}_{\boldsymbol{Y}}\ ,\qquad \Psi\in \boldsymbol{\cal Y}\ , 
\end{align}
together with the product rule for $\boldsymbol{\cal Y}$.

\subsection{Holomorphic metaplectic group algebra}

\paragraph{Bulk algebra and boundary module.}

Non-commutative geometries with desirable higher-spin gravity properties, arise in correspondence spaces with fibre modules 
\begin{align}
\boldsymbol{\cal A}=\boldsymbol{\cal K}_{\boldsymbol{Y}}\star \boldsymbol{\cal Y}\ ,\qquad \boldsymbol{\cal Y}= \boldsymbol{\cal G} \cup \boldsymbol{\cal G}^{(\infty)}\ ,
\end{align} 
where $\boldsymbol{\cal G}$ is the module of the holomorphic oscillator realization of the group algebra $\Comp[MpH(4;\Comp)]$ of the complex, inhomogeneous, metaplectic group, and $\boldsymbol{\cal G}^{(\infty)}$ is a two-sided $\boldsymbol{\cal G}$-module consisting of symbols representing the ramification points of $Mp(4;\Comp)$ viewed as a branched double cover of $Sp(4;\Comp)]$, which sit at the asymptotic boundary of $Sp(4;\Comp)$ \cite{meta}. 
The associative algebra structure, viz.,
\begin{eqnarray}
&\boldsymbol{\cal G}\star \boldsymbol{\cal G}=\boldsymbol{\cal G}\ ,\qquad \boldsymbol{\cal G}^{(\infty)}\star \boldsymbol{\cal G}^{(\infty)}=\boldsymbol{\cal G}^{(\infty)}\ ,&\\ & \boldsymbol{\cal G}\star \boldsymbol{\cal G}^{(\infty)}=\boldsymbol{\cal G}^{(\infty)}=\boldsymbol{\cal G}^{(\infty)}\star \boldsymbol{\cal G}\ ,& \end{eqnarray}
comprises that of the bulk group algebra $\boldsymbol{\cal G}$; its left- and right-actions on the boundary module $\boldsymbol{\cal G}^{(\infty)}$; and the latter's associative structure which needs regular prescriptions \cite{meta}. 
The group algebra contains an inner realization
\begin{align}
\boldsymbol{\cal K}_{\boldsymbol{\cal G}}:=\Comp[\kappa_y,\bar\kappa_{\bar y},\kappa_{\!\mathscr{P}}^{_{_{ (Y)}}}]\subset \boldsymbol{\cal G}\ ,
\end{align}
of $\boldsymbol{\cal K}_{\boldsymbol{Y}}$ generated by holomorphic oscillator realizations $\kappa_y$, $\bar\kappa_{\bar y}$ and $\kappa_{\!\mathscr{P}}^{_{_{ (Y)}}}$ of three properly complex group elements in $Mp(4;\Comp)$ obeying 
which obey 
\begin{align}\label{2.44}
\kappa_{y} \star \Psi\star \kappa_{y}=(\pi_y)^\ast(\Psi)\ , \quad \bar\kappa_{\bar y}\star \Psi\star \bar\kappa_{\bar y}=(\bar\pi_{\bar y})^\ast(\Psi)\ ,\quad (\kappa_{\!\mathscr{P}}^{_{_{ (Y)}}})^{-1}\star \Psi\star \kappa_{\!\mathscr{P}}^{_{_{ (Y)}}}=(\pi_{\!\mathscr{P}}^{_{_{ (Y)}}})^\ast(\Psi)\ ,
\end{align}  
for $\Psi \in \boldsymbol{\mathcal{A}}$~.
\paragraph{$SL(2,\Comp)$-covariant basis.} The associative product rule for $\boldsymbol{\cal Y}$ is represented by the analytical continuation\footnote{The convolution product is well-defined for symbols in $L^1(\boldsymbol{Y})\cap L^\infty(\boldsymbol{Y})$, which provides a representation of the group algebra $\Comp[H(4;\Real)]$ of the real Heisenberg group, including Hilbert-space endomorphisms in the image of Wigner--Ville maps. The resulting non-commutative geometries capture the evolution of linear quantum states, i.e., density matrices $\Psi^\dagger\star \Psi$ built from first-quantized Koopman--von Neumann wave functions $\Psi$, on symplectic backgrounds with constant Hamiltonian structures. 
Identifying the latter with vacuum expectation values of the one-form component of the superconnection, and $\Psi$ with its zero-form fluctuations, the full flatness condition describes how linear quantum states backreact to symplectic-Hamiltonian backgrounds producing non-commutative geometries built from holomorphic, metaplectic group algebras in images of Wigner--Ville maps applied to indefinite Hermitian modules (containing the Hilbert spaces as proper subspaces).} in $Mp(4;\Comp)$ parameters of the holomorphic, twisted convolution\footnote{The twisted convolution with symbols from $L^1(\Real^4)\cap L^\infty(\Real^4)$ represents the Heisenberg algebra in Weyl order. Geometrically, the convolution product is an integral over ordered, symplectic triangles $(y;\bar y;\xi',\bar\xi';\eta',\bar\eta')$ with $(y,\bar y)$ held fixed using a unit-strength, symplectic measure $\Upsilon|_{\xi',\bar \xi'}\wedge \Upsilon|_{\eta',\bar \eta'}$ and kernel $\exp{i{\rm Area}(y;\bar y;\xi',\bar\xi';\eta',\bar\eta')}$, which reproduces \eqref{2.37} upon setting $(\xi',\eta')=2(\xi,\eta)$ and $(\bar\xi',\bar\eta')=2(\bar\xi,\bar\eta)$.}
\begin{align}\label{2.37}
\Psi_1\star \Psi_2 := 16\int\displaylimits_{\widetilde{\boldsymbol{Y}_\Real}\times \widetilde{\boldsymbol{Y}_\Real}}\Upsilon|_{\xi,\bar \xi}\wedge \Upsilon|_{\eta,\bar \eta}\,e^{i\left(\eta\xi + {\bar \eta}{\bar \xi}\right)}~T^\ast_{\xi,\bar{\xi}} \Psi_1 ~T^\ast_{\eta,\bar{\eta}}\Psi_2~,\quad \Psi_1, \Psi_2\in \boldsymbol{\cal Y}\ ,
\end{align}
with $(y,\bar{y})\circ T_{\xi,\bar \xi} := (y + \xi, \bar{y}+\bar{\xi})$, and chiral domain $\widetilde{\boldsymbol{Y}_\Real}$ is embedded into $\boldsymbol{Y}$ such that 
\begin{align}\label{2.38}
(\xi^\alpha,\bar \xi^{\dot\alpha})|_{\widetilde{\boldsymbol{Y}_\Real}}  \in \Real^2\times \Real^2\ ,\qquad \Upsilon|_{\xi,\bar \xi}=\frac{d^2\xi\wedge d^2 \bar{\xi}}{(4\pi)^2}\ .
\end{align}
i.e., $(\xi^\alpha)^\dagger=\xi^\alpha$ and $(\bar \xi^{\dot\alpha})^\dagger=\bar \xi^{\dot\alpha}$ idem $(\eta^\alpha,\bar \eta^{\dot\alpha})$.
In this basis, the inner Kleinians $\kappa_y$ and $\bar\kappa_{\bar y}$ are given by the analytic delta functions
\begin{align}
\kappa_y:=2\pi\delta^2_\Comp(y)\ ,\qquad \bar\kappa_{\bar y}:=2\pi\delta^2_\Comp(\bar y)\ ,\qquad (\kappa_y)^\dagger=\bar\kappa_{\bar y}\ ,
\end{align}
defined on the two-sheeted Riemann surface of the square-root function, viz., 
\begin{align}
\delta^2_\Comp(M y)=\frac{1}{{\rm det}(M)}\delta^2_\Comp(y)\ ,\qquad \delta^2_\Comp(\bar M\bar y)=\frac{1}{{\rm det}(\bar{M})}\delta^2_\Comp(\bar y)\ ,\label{deltaC}
\end{align}
for $M,\bar M\in GL(2;\Comp)$; for details, see \cite{meta}.

\paragraph{Regularized chiral trace.} A subspace 
\begin{align}
\boldsymbol{\cal F}^{(\infty)}\subset \boldsymbol{\cal K}_{\boldsymbol{Y}}\star\boldsymbol{\cal G}^{(\infty)}\ ,
\end{align}
with a unique decomposition
\begin{align}\label{2.39}
& \Psi ={} \sum_{\sigma,\sigma'=\pm}\Pi^{(\sigma)}_K\star\left(\Psi^{(\sigma,\sigma')}_{0}+\Psi^{(\sigma,\sigma')}_1\star\kappa_y\right)\star \Pi^{(\sigma')}_K~,\qquad \Psi\in \boldsymbol{\cal F}^{(\infty)}\ ,\\ & \hspace{2cm} \Pi^{(\pm)}_K :={} \frac12 (1\pm K)\ ,\qquad K:=\kappa_y\star\bar{\kappa}_{\bar y}\ ,
\end{align}
into supertraceable components $\Psi^{(\sigma,\sigma')}_{\ell}$, $\ell=0,1$, that is respected by the associative product, i.e., 
\begin{align}
(\Psi\star \Xi)^{(\sigma,\sigma')}_0={}& \sum_{\sigma''=\pm}\left(  \Psi^{(\sigma,\sigma'')}_0\star \Xi^{(\sigma'',\sigma')}_0  + \Psi^{(\sigma,\sigma'')}_1\star \pi^\ast_y(\Xi^{(\sigma'',\sigma')}_1)\right)\ , \\
(\Psi\star \Xi)^{(\sigma,\sigma')}_1={}& \sum_{\sigma''=\pm} \left( \Psi^{(\sigma,\sigma'')}_0\star \Xi^{(\sigma'',\sigma')}_1  + \Psi^{(\sigma,\sigma'')}_1\star \pi^\ast_y(\Xi^{(\sigma'',\sigma')}_0)\right)\ , 
\end{align}
admits a regularized chiral trace operation, viz.,\footnote{Formally, the trace operation turns $\boldsymbol{\cal F}^{(\infty)}$ into a, possibly infinite-dimensional, symmetric, Frobenius algebra.}
\begin{align}
{\rm Tr}_{\boldsymbol{\cal F}^{(\infty)}}\,\Psi :={}& {\rm Tr}_{\boldsymbol{\cal K}_{\boldsymbol{Y}}}\int\displaylimits_{\widetilde{\boldsymbol{Y}_\Real}} \Upsilon\rvert_{(y,\bar{y})} K\star\left(\Psi^{(+,+)}_{0}-\Psi^{(-,-)}_{0}\right)\nonumber\\
={}& {\rm Tr}_{\boldsymbol{\cal K}_{\boldsymbol{Y}}}\int\displaylimits_{\widetilde{\boldsymbol{Y}_\Real}} \Upsilon\rvert_{(y,\bar{y})} K\left(\Psi^{(+,+)}_{0}-\Psi^{(-,-)}_{0}\right)\equiv {\rm Tr}_{\boldsymbol{\cal K}_{\boldsymbol{Y}}}{\rm STr}_{\boldsymbol{\cal Y}}\left(\Psi^{(+,+)}_{0}-\Psi^{(-,-)}_{0}\right)\ ,\label{2.67reg}
\end{align}
which induces Hermitian modules for non-commutative geometries in higher-spin gravity. We shall provide further details and applications of this trace operation in a paper in preparation \cite{paper0}. 

\paragraph{$SL(2,\Real)\times \Real$-covariant basis.}
In the conformal basis, the twisted convolution formula \eqref{2.37} reads 
\begin{align}\label{2.79}
\Psi_1\star \Psi_2 = 16\int\displaylimits_{\widetilde{\boldsymbol{Y}_\Real}\times \widetilde{\boldsymbol{Y}_\Real}}\Upsilon|_{\xi^+,\xi^-}\wedge \Upsilon|_{\eta^+, \eta^-}\,e^{i\left(\eta^+ \xi^- + {\eta^-}{\xi^+}\right)}~T^\ast_{\xi^+,\xi^-} \Psi_1 ~T^\ast_{\eta^+,{\eta^-}}\Psi_2~,
\end{align}
which thus extends to an associative product for $\boldsymbol{\cal Y}$ as described adjacently to \eqref{2.37}, and where the chiral integration measure
\begin{align}
\Upsilon|_{\xi^+,\xi^-}=\frac{d^2\xi^+ \wedge d^2\xi^-}{(4\pi)^2}\ ,
\end{align}
and integration domain $\widetilde{\boldsymbol{Y}_\Real}$ given by two-dimensional real planes embedded into $\Comp^2$ coordinatized by $\xi^\pm_\alpha(\xi,\bar \xi)$ in accordance with \eqref{2.67} and \eqref{2.38}.
From the conformal coordinatization of the inner Klein operators, viz.,
\begin{align}
\kappa_y=4\pi i\, \delta^2_\Comp(y^+ + iy^-) \ ,\quad \bar\kappa_{\bar y}=4\pi i\, \delta^2_\Comp(y^- + iy^+)\ ,
\end{align}
it follows that the chiral domain $\widetilde{\boldsymbol{Y}_\Real}$ can be rotated into a new domain $\widetilde{\boldsymbol{Y}_\Real}'$ where
\begin{align}
(\xi^+_\alpha,\xi^-_\alpha)|_{\widetilde{\boldsymbol{Y}_\Real}'}\in(\Real^2)^2\ ,
\end{align}
to be used for symbols of operators acting in Hermitian modules given by orbits of conformal reference states.

\subsection{Hermitian modules}

\paragraph{Left orbits.} Fluctuations around asymptotically, constantly curved backgrounds consisting of localizable degrees of freedom\footnote{Certain domain walls and FLRW-like solutions \cite{Aros:2017ror} arise from integration constants inside $\boldsymbol{\cal G}$.} arise from integration constants in subspaces of $\boldsymbol{\cal A}$ given by endomorphisms of left-orbits\footnote{The inclusion of $\boldsymbol{\cal K}_{\boldsymbol{Y}}$ into the orbits facilitates massive deformations \cite{Prokushkin:1998vn}.}
\begin{align}\label{2.67orb}
  {\cal O}(\xi_0):=\boldsymbol{\cal K}_{\boldsymbol{Y}}\star\boldsymbol{\cal K}_{\boldsymbol{\cal G}}\star \boldsymbol{\cal G}_\Real\star\Psi_{\xi_0|\cdot}\ ,\qquad \Psi_{\xi_0|\cdot}\in \boldsymbol{\cal K}_{\boldsymbol{Y}}\star\boldsymbol{\cal G}^{(\infty)}\ ,\qquad \boldsymbol{\cal G}_\Real:=\Comp[MpH(4;\Real)]\ ,
\end{align}
i.e., spans of elements $|\xi_0;u\rangle:=  u\star \Psi_{\xi_0|\cdot}$ obtained by acting with $u\in \boldsymbol{\cal K}_{\boldsymbol{Y}}\star\boldsymbol{\cal K}_{\boldsymbol{\cal G}}\star \boldsymbol{\cal G}_\Real$ on reference elements $|\xi_0;1\rangle\equiv \Psi_{\xi_0|\cdot}$  with active \emph{left-polarization}\footnote{The algebra $\boldsymbol{\cal G}^{(\infty)}$ contains elements $\Omega_{\xi|\xi'}$ with fixed left- and right-polarizations, i.e., $Y_{\xi}^{\underline{\alpha}}\star \Omega_{\xi|\xi'}=0=\Omega_{\xi|\xi'}\star Y_{\xi'}^{\underline{\alpha}}$, where $Y_{\xi}^{\underline{\alpha}}:=(\Pi_{\xi})^{\underline{\alpha\beta}}Y_{\underline\beta}$ using projectors $\Pi_{\xi}:=\frac12(1+S_{\xi})$ with $S_{\xi}\in Sp(4;\Comp)\cap \msp(4;\Comp)$ being fixed points of the Cayley map $C:Sp(4;\Comp)\to \msp(4;\Comp)\cup \infty$, viz., $C(S):=(S-1)/(S+1)$, i.e., $C(S_\xi)=S_\xi$, that is, $(S_\xi)^2=-1$. If $S_\xi+S_{\xi'}=0$, then $P_\xi\equiv \Omega_{\xi|\xi'}=4\exp(-4K_\xi)$, where $K_\xi=\frac18 Y S_\xi Y$, is idempotent; if $S_\xi+S_{\xi'}$ is invertible, then $\Omega_{\xi|\xi'}=P_\xi\star P_{\xi'}$; and if $S_\xi+S_{\xi'}$ is non-vanishing and non-invertible, then $\Omega_{\xi|\xi'}$ is a codimension-two delta function \cite{paper0}.} $\xi_0$ (defined modulo $Sp(4;\Real)$) and \emph{muted} right-polarization; thus, as stressed notationally, the elements of ${\cal O}(\xi_0)$ are states of a left-module for $\boldsymbol{\cal K}_{\boldsymbol{Y}}\star\boldsymbol{\cal K}_{\boldsymbol{\cal G}}\star \boldsymbol{\cal G}_\Real$.
The algebra of (left-)endomorphisms of ${\cal O}(\xi_0)$ is represented as a subalgebra 
\begin{align}
\boldsymbol{\cal A}^{(\infty)}(\xi_0) \subset  \boldsymbol{\cal K}_{\boldsymbol{Y}}\star\boldsymbol{\cal G}^{(\infty)}\ ,  
\end{align} 
given by the image of the holomorphic Wigner--Ville\footnote{The Wigner--Ville map represents the associative algebra of endomorphism of the Hilbert space of wave functions on a Lagrangian submanifold of the real, symplectic $\boldsymbol{Y}_\Real=(\Real^4,\omega_\Real)$ in terms of symbols given by distributions on $\boldsymbol{Y}_\Real$.
Extending the latter space by Gaussians phase factors (including delta sequences) yields a projective representation of $SpH(4;\Real)$ lifting to a proper representation of $MpH(4;\Real)$, whose complexification yields the holomorphic representation of $MpH(4;\Comp)$ underlying \eqref{2.74}.} map 
\begin{align}\label{2.74}
\varphi_{\rm WV}: {\cal O}(\xi_0)\otimes ({\cal O}(\xi_0))^\dagger \to{}&  \boldsymbol{\cal A}^{(\infty)}(\xi_0)\ ,\\ 
\varphi_{\rm WV}\left(|\xi_0;u\rangle \otimes (|\xi_0;v\rangle)^\dagger |\right):={}&  u \star \Omega_{\xi_0|\bar\xi_0}\star v^\dagger\ ,
\end{align}
where the reference element 
\begin{align}\label{2.65}
\Omega_{\xi_0|\bar{\xi}_0}\equiv \varphi_{\rm WV}\left(|\xi_0;1\rangle \otimes(|\xi_0;1 \rangle)^\dagger\right)\equiv\varphi_{\rm WV}\left(\Psi_{\xi_0|\cdot} \otimes(\Psi_{\xi_0|\cdot})^\dagger\right)   
\end{align}
is Hermitian and left- \emph{and} right-polarized as $\Psi_{\xi_0|\cdot}$ and $\overline{\Psi}_{\cdot|\bar \xi_0}\equiv (\Psi_{\xi_0|\cdot})^\dagger$, respectively, i.e.,\footnote{The isomorphism in \eqref{2.52} amounts to a first-order partial differential equation for the symbol of $\Omega_{\xi_0|\bar\xi_0}$ which has a unique solution (that may be a delta-function distribution).
Alternatively, assuming that the muted right-polarization of $\Psi_{\xi_0|\cdot}$ is projected with a hermitian, compact projector ${\cal P}$, i.e., $\Psi_{\xi_0|\cdot}=\Psi_{\xi_0|\cdot}\star {\cal P}$ with ${\cal P}\star {\cal P}={\cal P}$, ${\cal P}^\dagger={\cal P}$ and ${\rm Tr}_{\boldsymbol{\cal F}^{(\infty)}}=1$, one may choose $\varphi_{\rm WV}(|\xi_0;u\rangle \otimes(|\xi_0;v \rangle)^\dagger)= |\xi_0;u\rangle \star (|\xi_0;v \rangle)^\dagger$.}
\begin{align}\label{2.52}
\left(\Omega_{\xi_0|\bar\xi_0}\right)^\dagger=\Omega_{\xi_0|\bar \xi_0}\ ,\qquad \Omega_{\xi_0|\bar\xi_0}\stackrel{\rm pol}{\cong} \Psi_{\xi_0|\cdot}\otimes \overline{\Psi}_{\cdot|\bar \xi_0}\ .
\end{align}
\paragraph{Hermitian singletons.}
The corresponding Hermitian form\footnote{$(\cdot, \cdot)_{{\cal O}(\xi_0)}$ is invariant under renormalizations of ${\rm Tr}_{\boldsymbol{\cal A}^{(\infty)}(\xi_0)}$ and proportional to the normalization of $\Omega_{\xi_0|\bar\xi_0}$.}
\begin{align}
\langle \bar\xi_0;u|\xi_0;v\rangle\equiv \left( |\xi_0;u \rangle, |\xi_0;v \rangle\right)_{{\cal O}(\xi_0)}:=\langle\bar\xi_0;1|\xi_0;1\rangle\frac{{\rm Tr}_{\boldsymbol{\cal A}^{(\infty)}(\xi_0)}\, \varphi_{\rm WV}(|\xi_0;u\rangle \otimes\langle \bar\xi_0;v|)}{{\rm Tr}_{\boldsymbol{\cal A}^{(\infty)}(\xi_0)}\, \Omega_{\xi_0|\bar \xi_0}}\ ,\label{2.54}
\end{align}
is bounded on a subspace
\begin{align}\label{2.99}
{\cal S}(\xi_0)\subset {\cal O}(\xi_0)\ ,\quad |\xi_0;u\rangle\,,\ |\xi_0;v\rangle \in {\cal S}(\xi_0) \Rightarrow |\langle \bar\xi_0;u|\xi_0;v\rangle|<\infty\ ,
\end{align}
which we refer to as the \emph{Hermitian singleton} with polarization $\xi_0$ and attached modular group $\boldsymbol{\cal K}_{\boldsymbol{M}}\star\boldsymbol{\cal K}_{\boldsymbol{\cal G}}$.
Its endomorphism algebra 
\begin{align}
\boldsymbol{\cal F}^{(\infty)}(\xi_0):= \varphi_{\rm WV}\left({\cal S}(\xi_0)\otimes ({\cal S}(\xi_0))^\dagger\right)\ ,  
\end{align}
is a symmetric Frobenius algebra subject to Dirac-style bra-ket calculus, viz., 
\begin{align}\varphi_{\rm WV}(|\xi_0;u\rangle \otimes \langle\bar\xi_0;v |)\star \varphi_{\rm WV}(|\xi_0;u'\rangle \otimes\langle \bar\xi_0;v'|)=
\langle \bar\xi_0;v|\xi_0;u'\rangle\, \varphi_{\rm WV}(|\xi_0;u\rangle \otimes\langle \bar\xi_0;v' |)\ ,
\end{align}
using\footnote{From \eqref{2.52}, it follows that $\Omega_{\xi_0|\bar\xi_0}\star v^\dagger\star u'\star  \Omega_{\xi_0|\bar\xi_0}={\cal C}(v,u')\Omega_{\xi_0|\bar\xi_0}$, whose trace yields ${\cal C}(v,u'){\rm Tr}_{\boldsymbol{\cal A}^{(\infty)}(\xi_0)}\,\Omega_{\xi_0|\bar\xi_0}={\rm Tr}_{\boldsymbol{\cal A}^{(\infty)}(\xi_0)}\,(\Omega_{\xi_0|\bar\xi_0}\star v^\dagger\star u'\star  \Omega_{\xi_0|\bar\xi_0})={\rm Tr}_{\boldsymbol{\cal A}^{(\infty)}(\xi_0)}\,( v^\dagger\star u'\star  \Omega_{\xi_0|\bar\xi_0}\star\Omega_{\xi_0|\bar\xi_0})$, i.e., ${\cal C}(v,u')={\cal N}_{\xi_0}{\rm Tr}_{\boldsymbol{\cal A}^{(\infty)}(\xi_0)}\,( v^\dagger\star u'\star  \Omega_{\xi_0|\bar\xi_0})/ {\rm Tr}_{\boldsymbol{\cal A}^{(\infty)}(\xi_0)}\,(\Omega_{\xi_0|\bar\xi_0})$ where $\Omega_{\xi_0|\bar\xi_0}\star\Omega_{\xi_0|\bar\xi_0}=:{\cal N}_{\xi_0}\Omega_{\xi_0|\bar\xi_0}$. Thus, ${\cal N}_{\xi_0}={\cal C}(1,1)$, hence ${\cal C}(v,u')\equiv\langle \bar\xi_0;v|\xi_0;u'\rangle$ by \eqref{2.54}.} $\Omega_{\xi_0|\bar\xi_0}\star v^\dagger\star u'\star  \Omega_{\xi_0|\bar\xi_0}=\langle \bar\xi_0;v|\xi_0;u'\rangle\Omega_{\xi|-\bar\xi}$.
Henceforth, the symbol $\varphi_{\rm WV}$ will be suppressed, i.e., we identify
\begin{align}
{\cal O}(\xi_0)\equiv{}&\boldsymbol{\cal A}^{(\infty)}(\xi_0)\star\Psi_{\xi_0|\cdot}\ , \qquad \boldsymbol{\cal A}^{(\infty)}(\xi_0)\equiv {\cal O}(\xi_0)\otimes ({\cal O}(\xi_0))^\dagger \subset  \boldsymbol{\cal K}_{\boldsymbol{Y}}\star \boldsymbol{\cal G}^{(\infty)}\ , \\
{\cal S}(\xi_0)\equiv{}&\boldsymbol{\cal F}^{(\infty)}(\xi_0)\star\Psi_{\xi_0|\cdot}\ , \qquad \boldsymbol{\cal F}^{(\infty)}(\xi_0)\equiv {\cal S}(\xi_0)\otimes ({\cal S}(\xi_0))^\dagger \subset  \boldsymbol{\cal K}_{\boldsymbol{Y}}\star \boldsymbol{\cal G}^{(\infty)} \ ;
\end{align}
correspondingly, the identity operator ${\rm Id}_{\boldsymbol{\cal F}^{(\infty)}(\xi_0)}$ and trace operation ${\rm Tr}_{\boldsymbol{\cal F}^{(\infty)}(\xi_0)}$ can be expanded over bases using standard conventions for a Hermitian space.

\paragraph{Unitarizable, non-compact singletons.}
The Hermitian singletons decompose under $\boldsymbol{\cal G}_\Real$ into a spectrum of unitarizable, irreducible singletons with polarizations $\xi_{0;\lambda}$, viz.,\footnote{The unitarizable singletons provide Hilbert spaces for underlying, first-quantized conformal particles  \cite{Gunaydin:1999jb,Bars}.}
\bea
& \left.{\cal S}(\xi_0)\right\downarrow_{\boldsymbol{\cal G]}_\Real}={}\bigoplus_{\lambda} \boldsymbol{\cal K}_{\boldsymbol{Y}}\star {\cal S}^{(\xi_0)}(\xi_{0;\lambda})\ ,\qquad {\cal S}^{(\xi_0)}(\xi_{0;\lambda})=\boldsymbol{\cal G}_\Real\star \Psi_{\xi_{0;\lambda}|\cdot}\ ,& \\ 
&\kappa_{\pi_i}\star {\cal S}^{(\xi_0)}(\xi_{0;\lambda})={}{\cal S}^{(\xi_0)}(\xi_{0;\pi_i(\lambda)})\ ,\qquad \pi_i\in {\cal K}_{\boldsymbol{Y}}\ ,&\label{2.86}
\eea
in which $\boldsymbol{\cal G}_\Real$ is represented unitarily\footnote{
Applying the formalism to a one-dimensional harmonic oscillator with Hamiltonian $H:=\frac12(-(d/dx)^2+x^2)$, quantizes $H$ in a Hilbert space with Hermitian form $\langle u_2| u_1\rangle :={\rm STr} (u_1(x, d/dx) \star 2\exp(-2H) \star  (u_2(x,d/dx))^\dagger)$ (using Weyl order) in which $x^\dagger:=x$ and $(d/dx)^\dagger := - d/dx$ (rather than as a self-adjoint operator in a subspace of $L^2(\Real)$).
The resulting unitary $MpH(2;\Real)$-module admits a natural extension to an indefinite Hermitian $MpH(2;\Comp)$-module containing also the anti-vacuum suitable for creating deformations of the non-commutative geometry. Further details will be collected in \cite{paper0}.}; extending the modular group ${\cal K}$, thereby extending the modular algebra in \eqref{2.71}, enlarges the spectrum of unitarizable singletons contained in the Hermitian singleton.

\paragraph{Generalized Flato-Fronsdal factorization.}
The Frobenius algebra thus factorizes \`a la Flato-Fronsdal \cite{FF} , viz.,
\begin{align}
\boldsymbol{\cal F}^{(\infty)}(\xi_0) =\bigoplus_{\lambda,\lambda'}{\cal S}^{(\xi_0)}(\xi_{0;\lambda})\otimes ({\cal S}^{(\xi_0)}(\xi_{0;\lambda'}))^\dagger\ , 
\end{align}
where each block gives rise to a space of unfolded, linearized, massless fields obeying distinct boundary conditions \cite{fibre,2011,2017,BTZ,COMST,corfu19} related by modular transformations; for further examples, see \cite{paper0}.

Taking $\xi_0$ to be the compact polarization referred to the energy generator $E\equiv P_0$ of the $AdS_4$ isometry algebra $\mso(2,3)$ (see Appendix \ref{App:emb}), for which the reference state is the lowest-weight state of the compact singleton ${\cal D}^+(1/2)$, the blocks correspond to the $AdS_4$ massless particles/anti-particles irreps in the case of identical left and right polarizations, and singular black-hole-like solutions for opposite left-right polarizations \cite{fibre,2011,2017,COMST}; taking instead $\xi_0$ to be the non-compact polarization referred to the dilation generator $D$  corresponding to the conformal reference states (which can be taken to coincide with a spatial $AdS_4$ transvection, see \ref{App:emb}), to be spelled out below, the blocks correspond to bulk-to-boundary propagators and their counterpart ``at infinity'' (obtained via inversion) for identical left and right polarizations, and singular solutions with vanishing scaling dimension, containing boundary Green's functions in the case of opposite left-right polarizations \cite{corfu19}. 

\subsection{Projected two-parton modules} The diagonal projection $(e^\bullet_{11}\star \boldsymbol{\cal Y}_{(1)})\oplus (e^\bullet_{22}\star \boldsymbol{\cal Y}_{(2)})$ of the fibre module of the horizontal, projected two-parton DGA \eqref{2.32a} contains pairs of one-parton left-orbits, viz., 
\begin{align}\label{3.56}
{\cal O}_{\rm 2-p}(\xi_1,\xi_2):=(e^\bullet_{11}\star {\cal O}_{(1)}(\xi_1))\oplus ( e^\bullet_{22}\star {\cal O}_{(2)}(\xi_2))\ ,
\end{align}
with ${\cal O}_{(P)}(\xi_P))$ as in \eqref{2.67orb}.
Assuming a mutual, muted, right-projection by a Hermitian, rank-one projector, viz.,
\begin{align}
\Psi_{\xi_P|\cdot}=\Psi_{\xi_P|\cdot}\star {\cal P}_{\cdot|\cdot}\ ,\qquad {\cal P}_{\cdot|\cdot}\star {\cal P}_{\cdot|\cdot}={\cal P}_{\cdot|\cdot}\ ,\qquad ({\cal P}_{\cdot|\cdot})^\dagger={\cal P}_{\cdot|\cdot}\ ,\qquad P=1,2\ ,
\end{align}
the holomorphic, one-parton, Wigner--Ville map \eqref{2.65} can be extended to a two-parton map 
\begin{eqnarray}
\varphi_{\rm WV}^{\rm 2-p}: {\cal O}_{\rm 2-p}(\xi_1,\xi_2)\otimes\left({\cal O}_{\rm 2-p}(\xi_1,\xi_2)\right)^\dagger\to\boldsymbol{\cal A}_{\rm 2-p}^{(\infty)}(\xi_1,\xi_2)\ ,
\end{eqnarray}
by declaring 
\begin{align}
\varphi^{\rm 2-p}_{\rm WV}\left(\Psi_{(P)}\star e^\bullet_{PP}\otimes (\Xi_{(Q)}\star e^\bullet_{QQ})^\dagger\right):= \Psi_{(P)}\star e^\bullet_{PQ}\star (\Xi_{(Q)})^\dagger= e^\bullet_{PQ}\star \Psi_{(Q)}\star (\Xi_{(Q)})^\dagger\ ,
\end{align}
for $P,Q=1,2$. The resulting endomorphism algebra
\begin{align}
\boldsymbol{\cal A}_{\rm 2-p}^{(\infty)}(\xi_1,\xi_2):={}&\varphi_{\rm WV}\left({\cal O}_{\rm 2-p}(\xi_1,\xi_2)\otimes ({\cal O}_{\rm 2-p}(\xi_1,\xi_2)^\dagger\right)\\={}&\sum_{P,Q} {\cal O}_{(P)}(\xi_P)\star e^\bullet_{PQ}\star {\cal O}_{(Q)}^\dagger(\bar\xi_Q)\\={}&\sum_{P,Q} e^\bullet_{PQ}\star {\cal O}_{(Q)}(\xi_P)\star {\cal O}_{(Q)}^\dagger(\bar\xi_Q)\equiv\left[\begin{array}{c|c} {\cal O}(\xi_1)\star {\cal O}^\dagger(\bar\xi_1)&{\cal O}(\xi_1)\star {\cal O}^\dagger(\bar\xi_2)\\\hline{\cal O}(\xi_2)\star {\cal O}^\dagger(\bar\xi_1)&{\cal O}(\xi_2)\star {\cal O}^\dagger(\bar\xi_2)\end{array}\right]\ ,
\end{align}
thus consists of the blocks
\begin{align}
{\cal O}(\xi_P)\star {\cal O}^\dagger(\bar\xi_Q)= \boldsymbol{\cal A}^{(\infty)}(\xi_P)\star   \Omega_{\xi_P|\bar\xi_Q}\star \boldsymbol{\cal A}^{(\infty)}(\xi_Q)\ ,\quad \Omega_{\xi_P|\bar\xi_Q}:=\Psi_{\xi_P|\cdot}\star \overline\Psi_{\cdot|\bar\xi_Q}\ ,\quad P,Q=1,2\ ;
\end{align}
restricting the orbits to Hermitian modules ${\cal S}(\xi_P)$, yields the Frobenius subalgebra
\begin{align}\label{3.64}
\boldsymbol{\cal F}_{\rm 2-p}^{(\infty)}(\xi_1,\xi_2):=\left[\begin{array}{c|c} {\cal S}(\xi_1)\star {\cal S}^\dagger(\bar\xi_1)&{\cal S}(\xi_1)\star {\cal S}^\dagger(\bar\xi_2)\\\hline{\cal S}(\xi_2)\star {\cal S}^\dagger(\bar\xi_1)&{\cal S}(\xi_2)\star {\cal S}^\dagger(\bar\xi_2)\end{array}\right]\subset \boldsymbol{\cal A}_{\rm 2-p}^{(\infty)}(\xi_1,\xi_2)\ ,
\end{align}
with identity operators and traces given by sums over the bases using standard conventions.

\subsection{Conformal and non-compact singletons}
\paragraph{Left-orbits.} Decomposing the 3D conformal algebra $\mathfrak{so}(2,3)$ under $SL(2;\mathbb{R})_{\rm Lor}\times SO(1,1)_{\rm Dil}$, its generators can be realized using $y^{\xi}_\alpha$ as 
\bea & \displaystyle T_{\alpha\beta}=\frac12\,y^+_\alpha y^+_\beta\ , \qquad K_{\alpha\beta}=-\frac12\,y^-_\alpha y^-_\beta & \label{PK}\\
& \displaystyle  M_{\alpha\beta}=\frac12\,y^+_{(\alpha} y^-_{\beta)}\ ,\qquad D= \frac14 \,y^{+\alpha} y^-_{\alpha} \ ,& \label{MD}\eea
where $M_{\alpha\beta}$ and $D$ generate $\mathfrak{sl}(2;\mathbb{R})_{\rm Lor}\oplus \mathfrak{so}(1,1)_{\rm Dil}$, with $[D,y^\pm_\a]_\star=\pm\frac{i}{2}\,y^\pm_\a$, and $T_{\alpha\beta}$ and $K_{\alpha\beta}$, respectively, are the translations and special conformal translations; for details, see Appendix \ref{App:emb}. 
The conformal reference states
\begin{align}
|\xi; 1\rangle\equiv |(-i\xi /2)\rangle\ ,\qquad \xi=\pm 1\ ,\label{2.113}
\end{align}
are elements in $\boldsymbol{\cal G}^{(\infty)}$ (with muted right-polarization) obeying\footnote{The polarization matrices of the conformal reference states of the singleton are given by $S_\xi=i \xi i n^a \Gamma_{0'a}$, $\xi=\pm$, where $n^a$ is a Euclidean unit vector in $Mink_4$; we choose $S_\xi=i\xi   \Gamma_{0'2}$, i.e., $K_\xi=\xi P_2\equiv \xi D$, see Appendix \ref{App:emb}.} 
\begin{eqnarray}
&y^\xi_\alpha\star|(-i\xi/2) \rangle=0\ ,\qquad (D-i\xi/2)\star |(i\xi /2) \rangle=0\ ,&\\&
\kappa_{y}\star |(i\xi /2) \rangle= i\xi|(-i\xi /2)
\rangle\ ,\qquad \bar\kappa_{\bar y}\star |(i\xi /2) \rangle=-i\xi |(-i\xi /2) \rangle\ ,&
\end{eqnarray} 
i.e., they are the conformal singleton highest-weight ($\xi=+1$)  and conformal anti-singleton lowest-weight ($\xi=-1$) states, respectively.

\paragraph{Conformal singletons.}
The left-orbit\footnote{The left-orbit can alternatively be generated from compact reference states $|\pm 1/2\rangle$ with polarization matrices $S_\pm=\pm \Gamma_{0'0}$ related to the conformal ones via a modular transformation $\gamma\in \boldsymbol{\cal G}$, viz., $|\pm 1/2\rangle =\gamma\star |(\pm i/2)\rangle$; the details of this relation will be spelled out in \cite{paper0}.} 
\begin{align}\label{2.71}
{\cal O}(i/2)\cong {\cal O}(-i/2)\cong \boldsymbol{\cal K}_{\boldsymbol{Y}}\star \boldsymbol{\cal K}_{\cal G}\star\boldsymbol{\cal G}_\Real \star \Psi_{\pm i/2|\cdot}\ ,
\end{align}
contains the non-unitarizable subspaces\footnote{The enveloping algebra of $\mso(2,3)$ is contained in $\boldsymbol{M}_\Real$ in the vicinity of the identity element. In the Hermitian form provided by the regularized trace, ${\cal T}^-(-i/2)$ is paired with ${\cal T}^+(i/2)$; for further details on this point, see \cite{paper0}.}
\begin{align}\label{confbasis}
{\cal T}^\pm(\pm i/2):={\rm Env}(\mso(2,3))\star \Psi_{\pm i/2|\cdot}\ ,
\end{align}
decompose under $SO(1,1)_D \times SL(2,\Real)$ into Lorentz tensors with distinct conformal weights; ; see App. \ref{App:emb}.  With the choice of translations made in \eqref{PK}, the spaces ${\cal T}^-(- i/2)$ and ${\cal T}^+(i/2)$ contain the tensors required for unfolding conformal scalar and spinor fields and their duals, respectively, on 3D, locally, conformally flat spaces.
\paragraph{Non-compact singletons.}
The left orbit \eqref{2.71} contains a metaplectic   singleton 
\begin{align}\label{2.117}
{\cal S}(-i/2)\downarrow_{G^{(0)}}\cong {\cal S}(i/2)\downarrow_{G^{(0)}}\cong \bigoplus_{\epsilon,\sigma=\pm}{\cal S}^{(\sigma;\epsilon)}(\pm i/2) \ ,  
\end{align}
i.e., a subspace with bounded Hermitian form decomposing into unitarizable singletons in which\footnote{We choose van der Waerden symbols such that $(\gamma_0)_{\alpha\beta}=-\delta_{\alpha\beta}$, i.e., we choose a gauge in which the conformal particle on $T^\ast \Real^{2,3}$ reduces to a free, non-relativistic particle in two-dimensional, Euclidean space \cite{Bars}.}  
\begin{align}\label{2.118}
{\rm Spec}_{{\cal S}^{(\sigma;\epsilon)}(\pm i/2)}(\epsilon T_0)>0\ ,\qquad   \Pi^{(\sigma)}_K\star{\cal S}^{(\sigma';\epsilon)}(\pm i/2)=\delta^{\sigma,\sigma'} {\cal S}^{(\sigma;\epsilon)}(\pm i/2)\ ,
\end{align} 
i.e., ${\cal S}^{(\sigma;\epsilon)}(\pm i/2)$ yield 3D modes with positive ($\epsilon=+$) and negative ($\epsilon=-$) Poincar\'e energy suitable for imposing boundary conditions in conformal ${\rm Mink}_3$ on conformal scalars ($\sigma=-$) and spinors ($\sigma=+$); correspondingly, ${\cal S}^{(\sigma;\epsilon)}(-i/2)\otimes ({\cal S}^{(\sigma';\epsilon')}(-i/2))^\dagger$ yield 4D modes suitable for expanding boundary-to-bulk propagators in Poincar\'e patches subject to various boundary conditions \cite{corfu19}.
The Hermitian subspace ${\cal S}^{(\sigma;\epsilon)}(-i\xi/2)$ consists of $L^2$-normalizable states
\begin{align}
&{\cal S}^{(\sigma;\epsilon)}(-i\xi/2)\ni |\xi;\epsilon;\sigma;\phi\rangle:=\int_{\zeta_\epsilon\Real^2} \frac{d^2\lambda}{4\pi} \phi^{(\epsilon)}_{\sigma}(\lambda)|\xi;\epsilon;\lambda\rangle\ ,\quad\phi^{(\epsilon)}_{\sigma}(-\lambda)=(-1)^{\sigma+1} \phi^{(\epsilon)}_{\sigma}(\lambda)\ ,\label{mom}\\
&\zeta_{\epsilon} := e^{i(1-\epsilon)\pi/4}\ ,\quad \phi^{(\epsilon)}_{\sigma}\in L^2(\zeta_\epsilon\Real^2)\ ,\quad \Pi^{(\sigma)}_K\star|\xi;\epsilon;\sigma';\phi\rangle=\delta_{\sigma,\sigma'}|\xi;\epsilon;\sigma;\phi\rangle\ ,
\end{align}
expanded over \emph{momentum eigenstates}\footnote{In an abuse of nomenclature, we refer here collectively to both $y^+$- and $y^-$-eigenstates as ``momentum eigenstates'', though, in view of the commutation relations \eq{3.9}, we should refer to them as momentum and coordinate eigenstates.} 
\begin{align}\label{coherent}
|\xi;\epsilon;\lambda\rangle:=\exp(\frac{i}2 \lambda y^{-\xi})\star |(-i\xi/2)\rangle\ ,  \quad (y^\xi_\alpha-\lambda_\alpha)|\xi;\epsilon;\lambda\rangle=0\ ,\quad \lambda\in \zeta_\epsilon\Real^2\ ,
\end{align}
which span ${\cal S}^{(\sigma;\epsilon)}(-i\xi /2)$ modulo the Fourier transformations 
\begin{align}
|-\xi;\epsilon;\lambda\rangle=\int\displaylimits_{\zeta_\epsilon \Real^2} \frac{d^2 \mu}{4\pi} e^{\frac{i}2 \mu\lambda} |\xi;\epsilon;\mu\rangle \ ,
\end{align}
that can be factored out due the normalizability of the wave functions.
The elements\footnote{The $SO(1,1)$ subgroup of $Mp(4;\Real)$ represented by $\exp_\star (i x D)$ generates real scale transformations of $y^\xi_\alpha$.} 
\begin{align}
\kappa_{\!\mathscr{P}}:=\exp_\star \left(\pi D\right)\in \boldsymbol{\cal G}\setminus \boldsymbol{\cal G}_\Real\ , \quad\kappa_{\!\mathscr{F}}:=-\kappa_y\star \kappa_{\!\mathscr{P}}\in \boldsymbol{\cal G}_\Real\ , 
\end{align}
behave as follows under star multiplication and Hermitian conjugation:
\begin{eqnarray}
&\kappa_{\!\mathscr{P}}\star \kappa_{\!\mathscr{P}}=K\ ,\quad \kappa_y\star \kappa_{\!\mathscr{P}}=\kappa_{\!\mathscr{P}}\star \bar\kappa_{\bar y}\ ,\quad \kappa_y\star \kappa_{\!\mathscr{F}}=\kappa_{\!\mathscr{F}}\star \bar\kappa_{\bar y}\ ,\quad \kappa_{\!\mathscr{F}}\star \kappa_{\!\mathscr{F}}=1\ ,&\\
&
(\kappa_{\!\mathscr{P}})^\dagger=\kappa_{\!\mathscr{P}}\ ,\qquad 
(\kappa_{\!\mathscr{F}})^\dagger=\kappa_{\!\mathscr{F}}\ .&
\end{eqnarray}
From $\kappa_{\!\mathscr{P}}|i\xi/2\rangle=i\xi |i\xi/2\rangle$ and
\begin{align}
\kappa_y\star y^\xi_\alpha\star (\kappa_y)^{-1}=-i\xi y^{-\xi}_\alpha\ ,\quad \kappa_{\!\mathscr{P}}\star y^\xi_\alpha\star (\kappa_{\!\mathscr{P}})^{-1}= i\xi y^\xi_\alpha\ ,\qquad  \kappa_{\!\mathscr{F}}\star y^\xi_\alpha\star (\kappa_{\!\mathscr{F}})^{-1}=y^{-\xi}_\alpha\ ,     
\end{align}
it follows that
\begin{align}
\kappa_y\star |\xi; \epsilon;\lambda\rangle={}& -i\xi |-\xi;-\epsilon;i\xi \lambda\rangle\ ,\\
\kappa_{\!\mathscr{P}}\star |\xi; \epsilon;\lambda\rangle={}&-i\xi|\xi;-\epsilon;-i\xi \lambda\rangle\ ,\\
\kappa_{\!\mathscr{F}}\star |\xi; \epsilon;\lambda\rangle={}&|-\xi;\epsilon; \lambda\rangle\ ,
\end{align}
i.e., $\kappa_{\!\mathscr{F}}$ is a real, metaplectic group element represented by Fourier transformation, viz.\footnote{The Fourier transform on the ``doubly-symplectic'' $T^\ast\Real^2$ squares to the identity, viz., 
${\cal F}\circ {\cal F}={\rm Id}_{L^2(\Real^2)}$, while the operation on $T^\ast\Real$ squares to the parity transformation on the base.}
\begin{align}
\kappa_{\!\mathscr{F}}\star |\xi;\epsilon;\sigma;\phi\rangle=|\xi;\epsilon;\sigma;{\cal F}\phi\rangle\ ,\quad ({\cal F}\phi)(\lambda):=\int\displaylimits_{\zeta_\epsilon \Real^2} \frac{d^2 \mu}{4\pi} e^{\frac{i}2 \mu\lambda}\phi(\mu)\ ,
\end{align}
and $\kappa_{\!\mathscr{P}}$ is a complex, metaplectic group element implementing the 3D parity transformation, viz.,
\begin{align}
\kappa_{\!\mathscr{P}}\star (D,M_{mn},T_m,K_m)\star (\kappa_{\!\mathscr{P}})^{-1}=(D,M_{mn},-T_m,-K_m)\ .
\end{align}

\paragraph{Hermitian form.}
The holomorphic Wigner--Ville map is given by 
\begin{align}
|\xi;\epsilon;\lambda\rangle \langle\xi';\epsilon';\lambda'|=\exp{\frac{i}2 \lambda y^{-\xi}}\star |(-i\xi /2)\rangle \langle(i\xi'/2)|\star  \exp{-\frac{i}2 \overline{\lambda'} y^{-\xi'}}\ ,
\end{align}
where $\langle(i\xi'/2)|\equiv (|(-i\xi'/2)|)^\dagger$, and
\begin{align}
|(+i/2)\rangle \langle(+i/2)|=4\exp(iy^+ y^-)\ ,\qquad |(i/2)\rangle \langle(-i/2)|=4\pi  \delta^2_\Comp(y^-)\ ,\\
|(-i/2)\rangle \langle(i/2)|=4\pi  \delta^2_\Comp(y^+)\ ,\qquad |(-i/2)\rangle \langle(-i/2)|=4\exp(-iy^+ y^-)\ .
\end{align}
As will be shown in \cite{paper0}, combining the holomorphic Wigner--Ville map and regularized trace operation equips ${\cal S}(-i/2)$ with the Hermitian form 
\begin{align}\label{2.124}
\langle \xi;\epsilon;\lambda|\xi';\epsilon';\lambda'\rangle= 4\d_{\e\e'}\left(\d_{\xi\xi'}4\pi \d^2_\Comp(\l'+\bar\l)+\d_{\xi,-\xi'}e^{i\bar\lambda \lambda'/2}\right)\ ,
\end{align}
which is positive on ${\cal S}^{(+;+)}(-i/2)$ and, using the analytic property \eq{deltaC} of the complex delta function, negative on ${\cal S}^{(+;-)}(-i/2)$ and with respect to which positive-energy and negative-energy states are orthogonal.

\subsection{Finite group-algebra orbits}

The classical moduli space of the parent model includes Chan--Paton-like states spanning finite-dimensional Hermitian vector spaces 
\begin{align}\label{2.78}
{\cal C}(N_+,N_-)\equiv \bigoplus_{\hat I} \Comp \otimes |e^{\hat I}\rangle :=\boldsymbol{\cal F}^{(\infty)}_{\rm CP}(N_+,N_-)\star \Psi_{\xi_{\rm CP}|\cdot}\ ,
\end{align} 
of signature $(N_+,N_-)$ with Wigner--Ville maps defined as for the infinite-dimensional module \eqref{2.67orb}, are generated from normalizable reference states $\Psi_{\xi_{\rm CP}|\cdot}\in \boldsymbol{\cal G}^{(\infty)}$ by subalgebras
\begin{align}
\qquad \boldsymbol{\cal F}^{(\infty)}_{\rm CP}(\xi_0;N_+,N_-)=\bigoplus_{\hat I,\hat J} \Comp\otimes |e^{\hat I}\rangle \langle e_{\hat J}|\subset\boldsymbol{\cal K}_{\boldsymbol{Y}}\star \boldsymbol{\cal K}_{\boldsymbol{\cal G}}\star\boldsymbol{\cal G}^{(\infty)}\ ,  
\end{align} 
of the asymptotic group algebra, where 
\begin{align}
\langle e_{\hat I}|:= (|e^{\hat J}\rangle)^\dagger \eta_{\hat J\hat I}\ ,\quad \eta_{\hat I\hat K}\eta^{\hat J\hat K}=\delta^{\hat J}_{\hat I}\ ,\quad \eta^{\hat I\hat J}:=\left( |e^{\hat I}\rangle, |e^{\hat J}\rangle\right)_{{\cal O}(\xi_{\rm CP})} \ ;
\end{align}
we refer to ${\cal C}(N_+,N_-)$ as being modular if it represents $\boldsymbol{\cal K}_{\boldsymbol{Y}}\star \boldsymbol{\cal K}_{\boldsymbol{\cal G}}$; the FSG and CCHSG defects may break the modular symmetries by restricting ${\cal C}(N_+,N_-)$ to a subspace with definite norm.

\section{Parent field equations} \label{sec:parent}

In this Section, we assemble the projected, horizontal, two-parton algebra of Section 2 and a three-graded matrix algebra into a DGA equipped with a structure group containing a flat superconnection serving as parent field on-shell for FSG and CCHSG defects to be defined in the next Section.  

\subsection{Differential graded associative algebra}\label{sec:2.3}

\paragraph{Global formulation.}

The boundary configurations of the parent model arise \emph{off-shell} in the projected, direct-product DGA
\begin{align}\label{2.152b}
\boldsymbol{\cal E}^H_{\rm hor}({\cal K}\times \boldsymbol{C};\boldsymbol{\cal N}):= \left[{\rm Pr}_{\boldsymbol{{\cal E}}}\left(\boldsymbol{\cal N}\star \Omega_{\rm hor}^{([\boldsymbol{\cal V}]_\Upsilon,\Upsilon)}({\cal K}\times \boldsymbol{C})\right)\right]^H\ ,\qquad \boldsymbol{\cal N}:= {\rm mat}_{1|1}\star \boldsymbol{\cal P}\ ,
\end{align} 
where the three-graded 
\begin{align}
&{\rm mat}_{1|1}=\bigoplus_{i,j=1,2} \Comp\otimes m_{ij}\ ,\qquad {\rm deg}(m_{ij})=j-i\ ,\\&m_{ij}\star m_{kl}=\delta_{jk}m_{il}\ ,\qquad {\rm STr}_{{\rm mat}_{1|1}}\,m_{ij}=(-1)^{i+1}\delta_{ij}\ ,\qquad (m_{ij})^\dagger=i^{|i-j|t}m_{ji}\ ,
\end{align} 
is generated by the ghost system of an additional $\Real$-gauging of the two-parton system.
To this end, the latter is first restricted to local trivializations over $\boldsymbol{U}\subseteq \boldsymbol{C}$, which releases locally defined modules $\boldsymbol{\cal P}\star \boldsymbol{\cal A}\star \boldsymbol{\cal B}$ that can be tensored with ${\rm mat}_{1|1}$ using Koszul signs determined by the total degree 
\begin{align}
{\rm deg}_{\boldsymbol{{\cal E}}}={\rm deg}_{{\rm mat}_{1|1}}+{\rm deg}_{\boldsymbol{\cal B}} \ .
\end{align}
The resulting locally defined DGAs are subjected to an irreducibility projection
\begin{align}\label{2.152}
{\rm Pr}_{\boldsymbol{{\cal E}}} (\psi):={}& \prod_{r}\Pi^{(+)}_{\Gamma_r}\star  \psi \star \Pi^{(+)}_{\Gamma_r}\ ,\quad \Pi^{(\pm)}_{\Gamma_r}:={}\frac12( 1\pm \Gamma_r)\ ,\quad \Gamma_r\in {\rm mat}_{1|1}\star \boldsymbol{\cal K}\ ,\\
\Gamma_{r}\star \Gamma_{r'}={}&\Gamma_{r'}\star \Gamma_{r}\ ,\quad \Gamma_r\star \Gamma_r=1\ ,\quad (\Gamma_r)^\dagger=\Gamma_r\ ,
\end{align}
for $\psi\in \boldsymbol{\cal N}\star \boldsymbol{\cal A}\star \boldsymbol{\cal B}$, after which the irreducible subalgebras are glued together into \eqref{2.152b} using transition functions from a structure group $H$ generated by elements of a differential, graded, Lie sub-subalgebra
\begin{align}
\mathfrak{L}({\cal K}\times \boldsymbol{U};\boldsymbol{\cal N}):= \mathfrak{P}\left(\boldsymbol{\cal E}_{\rm hor}({\cal K}\times\boldsymbol{U};\boldsymbol{\cal N})\right)\subset \boldsymbol{\cal E}_{\rm hor}({\cal K}\times\boldsymbol{U};\boldsymbol{\cal N})\ ,    
\end{align}
with graded bracket
\begin{align}
[\psi_1, \psi_2]_\star:=\psi_1\star \psi_2-(-1)^{\boldsymbol{{\cal E}}\,(\psi_1)\boldsymbol{{\cal E}}\,(\psi_2)}\psi_2\star \psi_1\ ,   
\end{align}
and where $\mathfrak{P}:\boldsymbol{\cal E}_{\rm hor}({\cal K}\times\boldsymbol{U};\boldsymbol{\cal N})\to \boldsymbol{\cal E}_{\rm hor}({\cal K}\times\boldsymbol{U};\boldsymbol{\cal N}) $ is a projector obeying
\begin{align}
\mathfrak{P}(d\psi)=d(\mathfrak{P}(\psi))\ ,\qquad \mathfrak{P}\left([\psi_1, \psi_2]_\star\right)= [\mathfrak{P}(\psi_1), \mathfrak{P}(\psi_2)]_\star\ .
\end{align} 
Assuming the base to be a fibration
\begin{align}\label{2.161}
\boldsymbol{M}\stackrel{{\rm pr}_{\boldsymbol{M}}}{\longrightarrow} \boldsymbol{X}\ ,
\end{align}
with commutative base and non-commutative fibres (which need not be symplectic), the transition functions are assumed to be homotopic to a set of transition functions on $\boldsymbol{X}$.

\paragraph{Fibrations and integrations.}

The fibration of $\boldsymbol{M}$ over $\boldsymbol{X}$ in \eqref{2.161} combines with the bundle structure \eqref{2.7} into a fibration of $\boldsymbol{C}$ over $\boldsymbol{X}$ with non-commutative fibres given by fibre bundles with fibre $\boldsymbol{Y}$, viz.
\begin{align}\label{2.173}
\boldsymbol{C}\stackrel{{\rm pr}'_{\boldsymbol{C}}}{\longrightarrow} \boldsymbol{X}\ ,\qquad \boldsymbol{Y}\to \boldsymbol{T}_x\to \boldsymbol{Z}_x\ ,\qquad x\in \boldsymbol{X}\ ,
\end{align}
where $\boldsymbol{Z}_x:=({\rm pr}_{\boldsymbol{M}})^{-1}(x)$ and $\boldsymbol{T}_x:= ({\rm pr}'_{\boldsymbol{C}})^{-1}(x)$.
On a restriction of \eqref{2.173} with bundle structure, viz.
\begin{align}
\boldsymbol{T}\to \boldsymbol{U} \stackrel{{\rm pr}'_{\boldsymbol{C}}}{\longrightarrow} \boldsymbol{X}'\subseteq \boldsymbol{X}\,
\end{align}
the supertrace operation factorizes, viz.,
\begin{align}
{\rm STr}_{\boldsymbol{\cal E}^H_{\rm hor}({\cal K}\times\boldsymbol{U};\boldsymbol{\cal N})}\,\psi= \int_{\widetilde{\boldsymbol{X}'_\Real}}^{\prime}\oint_{\widetilde{\boldsymbol{Z}_\Real}}^{\prime} {\rm STr}_{\boldsymbol{\cal N}}\,{\rm Tr}_{\boldsymbol{\cal K}}\,{\rm Tr}_{\boldsymbol{\cal Y}}\,\psi\ ,\label{4.13}
\end{align}
where $\widetilde{\boldsymbol{X}'_\Real}$ and $\widetilde{\boldsymbol{Z}_\Real}$ are orientable, chiral domains and the primes refer to suitable regularizations.

\paragraph{Irreducibility projection and structure group.}

We assume that ${\cal K}_{\boldsymbol Z}\times \boldsymbol{Z}$ is a compacted version of ${\cal K}_{\boldsymbol Y}\times \boldsymbol{Y}$, with ${\cal K}_{\boldsymbol Z}$ generated by $\{\pi_z,\bar \pi_{\bar z},\pi_{\!\mathscr{P}}^{_{_{ (Z)}}}\}$, inducing an algebra $\boldsymbol{\cal K}_{\boldsymbol{Z}}$ of outer Klein operators generated by $\{k_z,\bar k_{\bar z},k_{\!\mathscr{P}}^{_{_{ (Z)}}}\}$ obeying counterparts of \eqref{2.57}-\eqref{2.61} and the twisted relations
\begin{align}
k_z\star k_y=-k_y\star k_z\ ,\quad k_z\star \bar k_{\bar y}=\bar k_{\bar y}\star k_z\ ,\quad k_{\!\mathscr{P}}^{_{_{ (Z)}}}\star k_{\!\mathscr{P}}^{_{_{ (Y)}}}=-k_{\!\mathscr{P}}^{_{_{ (Y)}}}\star k_{\!\mathscr{P}}^{_{_{ (Z)}}}\ 
\end{align}
The irreducibility projection \eqref{2.152} is imposed using\footnote{The fractional-spin/massive deformation requires a relaxed projection releasing an outer Clifford algebra with holomorphic and anti-holomorphic generators \cite{wip}.} 
\begin{align}
\Gamma_1:={}& k\star \bar k\ ,\quad  k:=ik_y\star k_z\ ,\quad \bar k=i\bar k_{\bar y}\star \bar k_{\bar z}\ ,\\
\Gamma_2:={}& k_{\!\mathscr{P}}\star k_{\!\mathscr{P}}\,\qquad k_{\!\mathscr{P}}:=i k_{\!\mathscr{P}}^{_{_{ (Y)}}}\star k_{\!\mathscr{P}}^{_{_{ (Z)}}}\ .
\end{align}
The non-trivial actions of the structure group projector are taken to be
\begin{align}
\mathfrak{P}(m_{ij})= \delta_{ij} m_{ii}\ ,\quad \mathfrak{P}(e^\bullet_{PQ})=\delta_{PQ}e^\bullet_{PP}\ ,
\end{align}
where $i,j=1,2$ and $P,Q=1,2$.

\subsection{Flat superconnection}

The boundary defects of the multi-dimensional model consist of projections of a superconnection
\begin{align}
X\in \boldsymbol{\cal E}^H_{\rm hor}({\cal K}\times\boldsymbol{C};\boldsymbol{\cal N})\ ,\qquad {\rm deg}_{\boldsymbol{\cal E}}\,X=1\ ,
\end{align}
obeying the flatness condition
\begin{align}
dX+X\star X= 0\ ,\label{32b}    
\end{align}
and the reality condition\footnote{It follows that \begin{align}(dX+X\star X)^\dagger={}&(-1)^{t+1} d(X^\dagger)+(-1)^t X^\dagger\star X^\dagger&\nonumber\\={}&(-1)^{t+1} d(\Gamma\star X\star \Gamma)+(-1)^t \Gamma\star X\star \Gamma\star \Gamma\star X^\star \Gamma\ =\ (-1)^t \Gamma\star (dX+X\star X)\star \Gamma\ .\nonumber \end{align}} 
\begin{align}
X^\dagger = \Gamma\star X \star \Gamma\ ,\qquad \Gamma\star \Gamma = 1\ ,\qquad \Gamma^\dagger \star \Gamma = \Gamma\star \Gamma^\dagger = i^t\ ,
\end{align}
using an odd conjugation operator $\Gamma \in {\rm mat}_{1|1}$; we take $\Gamma=m_{12}+m_{21}$, which obeys $\Gamma^\dagger=i^t \Gamma$.

\paragraph{Structure group and Chern classes.}
Letting  
\begin{align}
X^{\mathfrak{L}}:= \mathfrak{P}(X)\ ,\qquad E:=(1-\mathfrak{P})(X)\ ,
\end{align} 
the equations of motion decompose into 
\begin{align}
dX^{\mathfrak{L}}+X^{\mathfrak{L}}\star X^{\mathfrak{L}}+\mathfrak{P}(E\star E)={}& 0\ , \\
dE+ X^{\mathfrak{L}}\star E+E\star X^{\mathfrak{L}} +(1-\mathfrak{P})(E\star E)={}& 0\ .
\end{align} 
The structure group connection is assumed to be embedded into $X^{\mathfrak{L}}$ such that $E\in \boldsymbol{\cal E}^H_{\rm hor}({\cal K}\times \boldsymbol{C};\boldsymbol{\cal N})$ decomposes into $H$-irreducible sections, viz., 
\begin{align}
E=\sum_{\rho} {\rm Pr}_\rho (E)\ ,\qquad [\mathfrak{P}(\epsilon),{\rm Pr}_\rho(E)]_\star ={\rm Pr}_\rho([\mathfrak{P}(\epsilon),{\rm Pr}_\rho(E)]_\star)\ ,
\end{align}
for a set of mutually compatible DGA projectors ${\rm Pr}_\rho$ commuting to $d$.
Using the integration measures \eq{4.13},
closed sub-manifolds $\boldsymbol{X}'\subseteq \boldsymbol{X}$ are assigned Chern classes of shifted connections, viz.,
\begin{align}\label{4.25}
C_{n;\{t_\rho\}}(\boldsymbol{X}'\times \boldsymbol{Z}):={}&{\rm STr}_{\boldsymbol{\cal E}^H({\cal K}\times  \boldsymbol{U};\boldsymbol{\cal N})}\,(dX^{\mathfrak{L}}_t+X^{\mathfrak{L}}_t\star X^{\mathfrak{L}}_t)^{\star n}\ ,\qquad \quad n=\frac12{\rm dim}(\boldsymbol{X}'\times \boldsymbol{Z})\ ,\\
X^{\mathfrak{L}}_t:={}& X^{\mathfrak{L}}+\sum_\rho t^\rho {\rm Pr}_\rho(E)\ ,\qquad t^\rho\in \Real\ ,
\end{align}
which are classical observables with a direct off-shell resolution.

\paragraph{Component fields.}

The decomposition of the superconnection under ${\rm mat}_{1|1}$ into  forms in $\boldsymbol{\cal P}\star \Omega_{\rm hor}({\cal K}\times \boldsymbol{C})$ , viz., 
\begin{align}
X=m_{11}\star\mathbb{A}+m_{22} \star\widetilde{\mathbb{A}}+\zeta_t m_{12}\star \mathbb{B}+\bar\zeta_t m_{21}\star \widetilde{\mathbb{B}}\ ,\qquad {\rm deg}_{\boldsymbol{\cal B}}(\mathbb{A},\widetilde{\mathbb{A}};\mathbb{B},\widetilde{\mathbb{B}})=(1,1;0,2)\ ,   
\end{align} 
where $\zeta_t:=e^{it\pi/4}$,
yields the Cartan integrable system
\begin{align}
d\mathbb{A}+\mathbb{A}\star \mathbb{A}+\mathbb{B}\star \widetilde{\mathbb{B}}= 0\ ,\qquad d\widetilde {\mathbb{A}}+\widetilde{\mathbb{A}}\star \widetilde{\mathbb{A}}+\widetilde{\mathbb{B}}\star {\mathbb{B}}= 0\ ,\\
d\mathbb{B}+\mathbb{A}\star \mathbb{B}-\mathbb{B}\star \widetilde{\mathbb{A}}= 0\ ,\qquad d\widetilde {\mathbb{B}}+\widetilde{\mathbb{A}}\star \widetilde{\mathbb{B}}-\widetilde{\mathbb{B}}\star {\mathbb{A}}= 0\ ,
\end{align}
where the signs arise from odd forms passing over odd ${\rm mat}_{1|1}$-generators.
The component fields obey the reality conditions
\begin{align}
\mathbb{A}^\dagger = -\widetilde{\mathbb{A}}~,\qquad \mathbb{B}^\dagger = \mathbb{B}~,\qquad \widetilde{\mathbb{B}}^\dagger = (-1)^t \widetilde{\mathbb{B}}~.
\end{align}
In what follows, we choose twisted reality conditions, i.e., $t=1$, which conform with the standard conventions used for differential form calculus.
We denote the decomposition of $X$ under ${\rm mat}_{1|1}$ and the further decomposition under $\boldsymbol{\cal N}$ into horizontal forms on ${\cal K}\times \boldsymbol{C}$ as follows:
\begin{align}
X=\left[\begin{array}{c|c}\mathbb{A}&\mathbb{B}\\\hline\widetilde{\mathbb{B}}&\widetilde{\mathbb{A}}\end{array}\right]=\left[\begin{array}{cc|cc} A&\Psi&B&\Sigma\\\overline{\Psi}&U&\overline{\Sigma}&M\\\hline\widetilde{B}&\widetilde{\Sigma}&\widetilde{A}&\widetilde{\Psi}\\\widetilde{\overline{\Sigma}}&\widetilde{M}&\widetilde{\overline{\Psi}}&\widetilde{U}\end{array}\right]\ ,   
\end{align} 
obeying the reality conditions
\begin{eqnarray} &A^\dagger=-\widetilde A\ ,\qquad U^\dagger=-\widetilde{U}\ ,\qquad \Psi^\dagger=-\widetilde{\overline{\Psi}}\ ,\qquad \widetilde{\Psi}^\dagger=-\overline{\Psi}\ ,&\\ &B^\dagger=B\ ,\qquad M^\dagger=M\ ,\qquad \Sigma^\dagger=\overline{\Sigma}\ ,&\\& \widetilde{B}^\dagger=-\widetilde{B}\ ,\qquad \widetilde{M}^\dagger=-\widetilde{M}\ ,\qquad
\widetilde{\Sigma}^\dagger=-\widetilde{\overline{\Sigma}}\ .&\end{eqnarray}

\section{Classical higher-spin gravity defects}\label{sec:defects}

In this Section, we characterize the FSG and CCHSG defects of the flat superconnection containing higher-spin gravity coupled to matter in the presence of internal, colour gauge fields.
The defects are classical moduli spaces of classically consistent truncations obtained by first projecting $\boldsymbol{\cal N}$, the horizontal DGA, and the structure group, and then activating holonomies represented in fractional-spin algebras, two-form cohomologies and reduced structure groups.

\subsection{DGA projections}

\paragraph{On-shell projections.}

Whereas DGAs on commutative manifolds can be projected to embedded manifolds using pull-back operations induced by embedding maps that are defined locally in the immediate neighbourhood of the embedded surface, the non-commutative nature of the correspondence space implies that projections of the horizontal DGA require global structures induced by fibrations of the base equipped with sections.
Thus, a class of defects of the parent model are spaces
\begin{align}
{\cal C}_\Xi\equiv 
{\cal C}^{{H_\Xi}}({\cal K}_\Xi\times \boldsymbol{C}_\Xi;\boldsymbol{\cal N}_{\Xi})\subset \boldsymbol{\cal E}_{\rm hor}^{{H_\Xi}}({\cal K}_\Xi\times \boldsymbol{C}_\Xi;\boldsymbol{\cal N}_{\Xi}) \cong {\rm Pr}_\Xi^{_{(H)}}  ({\rm Pr}_\Xi^{_{(\boldsymbol{C})}})^\ast\left( {\rm Pr}_\Xi^{_{(\boldsymbol{\cal N})}}\boldsymbol{\cal E}_{\rm hor}^H({\cal K}\times \boldsymbol{C};\boldsymbol{\cal N})\right)\ , 
\end{align}
consisting of flat superconnections in DGAs $\boldsymbol{\cal E}_{\rm hor}^{{H_\Xi}}({\cal K}_\Xi\times \boldsymbol{C}_\Xi;\boldsymbol{\cal N}_{\Xi})$ obtained by applying three separate projection maps to the universal DGA: i) a graded, associative algebra projection 
\begin{align}
{\rm Pr}_\Xi^{_{(\boldsymbol{\cal N})}}:\boldsymbol{\cal N}\to \boldsymbol{\cal N}\,\qquad {\rm Pr}_\Xi^{_{(\boldsymbol{\cal N})}}(\boldsymbol{\cal N})\cong \boldsymbol{\cal N}_\Xi\ ;  
\end{align} 
ii) a horizontal DGA projector 
\begin{align}
({\rm Pr}_\Xi^{_{(\boldsymbol{C})}})^\ast:\Omega_{\rm hor}({\cal K}\times\boldsymbol{C})\to \Omega_{\rm hor}({\cal K}\times\boldsymbol{C})\ ,\qquad ({\rm Pr}_\Xi^{_{(\boldsymbol{C})}})^\ast(\Omega_{\rm hor}({\cal K}\times\boldsymbol{C}))\cong \Omega_{\rm hor}({\cal K}_\Xi\times\boldsymbol{C}_\Xi)\ ,     
\end{align}
where ${\cal K}_\Xi={\cal K}_{\boldsymbol{M}_\Xi}\times {\cal K}_{\boldsymbol{Y}}$, induced by a projector 
\begin{align}
{\rm Pr}_\Xi^{_{(\boldsymbol{C})}}=s_\Xi\circ {\rm pr}_\Xi:\boldsymbol{C}\to \boldsymbol{C}\ ,    
\end{align}
arising from a bundle fibration $\boldsymbol{C}\stackrel{{\rm pr}_\Xi}{\longrightarrow} \boldsymbol{C}_\Xi$ induced from a fibration  $\boldsymbol{M}\stackrel{{\rm pr}^{(\boldsymbol{M})}_\Xi}{\longrightarrow} \boldsymbol{M}_\Xi$ of the base, i.e., ${\rm pr}_{\boldsymbol{C}_\Xi}\circ {\rm pr}_\Xi={\rm pr}^{(\boldsymbol{M})}_\Xi\circ  {\rm pr}_{\boldsymbol{C}}$, equipped with a section $s_\Xi:\boldsymbol{C}_\Xi \to \boldsymbol{C}$; 
and iii) a group projection 
\begin{align}
{\rm Pr}_\Xi^{_{(H)}}:H\to H\ ,\qquad {\rm Pr}_\Xi^{_{(H)}}(H)\cong H_\Xi\ .    
\end{align}
The projected DGA is then restricted to the shell ${\cal C}_\Xi$ by choosing a representation for the operator algebra, i.e., a set of holonomies, cohomology elements, integration constants and transition functions, which correspond to imposing boundary conditions on the horizontal forms.
This activates a set of classical moduli parameters, which thus coordinatize ${\cal C}_{\Xi}$, that can be second-quantized using the multi-dimensional AKSZ formalism.

\paragraph{Off-shell projections.}

As will be elaborated on further in \cite{OLC}, the defect ${\cal C}_\Xi$ is quantized by treating projections of $\boldsymbol{M}_\Xi$ as boundaries of a set of ``virtual'' sources $\widehat{\boldsymbol{M}}_\tau$.
The resulting multi-dimensional AKSZ model has a configuration space consisting of field configurations in sub-DGAs 
\begin{align}\label{5.6}
\boldsymbol{\cal E}^{\widehat{H}_\tau}_{\rm hor}(\widehat{\cal K}_\tau\times \widehat{\boldsymbol{C}}_\tau;\boldsymbol{\cal N}) \cong {\rm Pr}^{_{(\widehat{H})}}_\tau({\rm Pr}^{_{(\widehat{\boldsymbol{C}})}}_\tau)^\ast \left(\boldsymbol{\cal E}_{\rm hor}^{\widehat{H}}(\widehat{\cal K}\times \widehat{\boldsymbol{C}};\boldsymbol{\cal N})\right)\ , 
\end{align}
of a universal DGA obtained using group projectors ${\rm Pr}^{_{(\widehat{H})}}_\tau$ and horizontal DGA projectors $({\rm Pr}^{_{(\widehat{\boldsymbol{C}})}}_\tau)^\ast$ induced by fibre-preserving bundle maps $\widehat{\boldsymbol{C}}\stackrel{{\rm pr}_\tau}{\longrightarrow} \widehat{{\boldsymbol{C}}}_\tau$
induced from fibrations $\widehat{\boldsymbol{M}}\stackrel{{\rm pr}^{(\widehat{\boldsymbol{M}})}_\tau}{\longrightarrow}\widehat{\boldsymbol{M}}_\tau$ equipped with sections $s_\tau:\widehat{\boldsymbol{C}}_\tau\to \widehat{\boldsymbol{C}}$.
The source manifold $\widehat{\boldsymbol{M}}_\tau$ is open and the off-shell projections are assumed to have well-defined restrictions to $\boldsymbol{M}_\tau:=\partial\widehat{\boldsymbol{M}}_\tau$.
If ${\rm Pr}^{_{(\widehat{H})}}_\tau$ and ${\rm Pr}^{_{(\widehat{\boldsymbol{C}})}}_\tau$ are compatible with ${\rm Pr}^{_{(H)}}_\Xi$ and ${\rm Pr}^{_{(\boldsymbol{C})}}_\Xi$, then the sub-partition function on $\widehat{\boldsymbol{M}}_\tau$ can be computed with boundary field configuration ${\rm Pr}^{_{(\widehat{H})}}_\tau({\rm Pr}^{_{(\widehat{\boldsymbol{C}})}}_\tau)^\ast (X_\Xi)$ for $X_\Xi\in {\cal C}_\Xi$, thus adding a contribution to the quantization of ${\cal C}_\Xi$.

\paragraph{Fronsdal vs AKSZ formulations.}

Vasiliev's equations bridge two stand-alone formulations of HSG: 

Following the AKSZ approach, the defects ${\cal C}_\Xi$ are constructed using the gauge function method from metaplectic group algebra elements which introduces second-quantizable, classical moduli parameters. Correspondingly, $H_\Xi$ is chosen in concordance with the basic DGA operations used in defining globally defined functionals on- and off-shell.

To make contact with the Fronsdal approach, the flat DGA element $X_\Xi$ is homotopy contracted perturbatively on $\boldsymbol{Z}$ before imposing any boundary conditions on $\boldsymbol{X}$.
This yields a flat superconnection $X^\#_\Xi$ in an $A_\infty$-algebra $\boldsymbol{\cal E}^{H^\#_\Xi}_{\rm hor}({\cal K}^\#_\Xi\times \boldsymbol{C}^\#_\Xi;\boldsymbol{\cal N}^\#_\Xi)$ consisting of a shell ${\cal C}^\#_\Xi$ of horizontal forms on a fibre bundle $Y\to \boldsymbol{C}^\#_\Xi\to \boldsymbol{X}^\#_\Xi$ consisting of $H^\#_\Xi$-tensors, i.e., linear representations of $H^\#_\Xi$, composed using a set of manifestly $H^\#_\Xi$-covariant, $n$-ary products $m^\#_n:\left(\boldsymbol{\cal V}_\#\right)^{\otimes n}\to \boldsymbol{\cal V}_\#$, $n=1,2,\dots$.
Before imposing any boundary conditions, 
the shell ${\cal C}^\#_\Xi$ is defined formally since the tensorial bases do not span strict operator algebras without any further assumption on the class of symbols (such as, e.g., the non-covariant metaplectic group algebra elements). 

In the case of HSG, the perturbatively defined structure group is taken to be $SL(2,\Comp)$, and the resulting formally defined system can be linearized around constantly curved spacetime backgrounds, which gives rise to cocycles describing unfolded Fronsdal fields on-shell as stated by the Central on Mass-Shell Theorem (COMST).

As we shall see, CCHSG admits a similar treatment with structure group given by $SL(2,\Real)\times \Real$.

\subsection{Moduli parameters}

The gauge function method yields classical solution spaces given by local configurations
\be X_{i}= L_i^{-1}\star (d+X'_i)\star L_i\ ,\qquad d_{\boldsymbol{Z}}X'_i+X'_i\star X'_i= 0\ ,\qquad d_{\boldsymbol{X}} X'_i=0\ ,\ee
where $L_i$ are gauge functions on $\boldsymbol{U}_i=\boldsymbol{X}_i\times \boldsymbol{Z}$, and $X'_i$, referred to as \emph{local data}, comprise deformed oscillators on $\boldsymbol{Y}\times \boldsymbol{Z}$, glued together using transition functions $T_i^j=(T_j^i)^{-1}$ defined on overlaps of $\boldsymbol{X}$ (see e.g. \cite{Iazeolla:2022dal}), viz.
\begin{align}
L_i=H_i^{\prime j}\star L_j\star T^j_i\ ,\qquad X'_i=H_i^{\prime j}\star (d_{\boldsymbol{Z}}+X'_j)\star H_j^{\prime i}\ ,\qquad d_{\boldsymbol{X}}  H_i^{\prime j}=0\ .  
\end{align}
The resulting defects contain backgrounds parametrized by i) zero-form expectation values that preserve the background symmetry (e.g., mass parameters or sizes of domain walls); ii) flat one-forms with open-curve holonomy groups $G^{(0)}$ comprising asymptotic gauge functions $L^{(0)}$ (of possibly locally degenerate metrics) and closed-loop holonomy groups ${\rm Hol}^{(0)}$ (given by path-ordered products of the locally defined constant group elements $H_i^{\prime j}$); and iii) spaces $Z^{(0)}$ of closed and central cohomology elements (including monodromies of conical singularities which trigger cocycles gluing infinite-dimensional spaces of integration constants to fluctuations in gauge potentials); and iv) reduced structure groups.
Linearized fluctuations around the background arise from switching on integration constants\footnote{Examples of classical moduli spaces with  integration constants lying inside $\boldsymbol{\cal G}$ are given by certain space-like and time-like domain walls \cite{Aros:2017ror}.} in endomorphism algebras \eqref{3.64} of projected, two-parton left-orbits \eqref{3.56} consisting of Hermitian singletons; requiring these to be regular\footnote{In a regular configuration, the Lorentz-tensorial component fields arising upon power-series expansions of the symbols in $\boldsymbol{\cal A}$ and $\boldsymbol{\cal B}$ in canonical coordinates may exhibit classical singularities, while the underlying operators in $\boldsymbol{\cal V}$ remain bounded; for examples involving resolutions of classical gauge field singularities, see \cite{2011,2017,BTZ,Iazeolla:2022dal}.} and single-valued, respectively, requires
\begin{align}
G^{(0)}\star \boldsymbol{{\cal S}}(\xi_P)\subseteq\boldsymbol{{\cal S}}(\xi_P)\ ,\qquad {\rm Ad}_{{\rm Hol}^{(0)}}\,\Psi=\Psi\ ,\qquad \Psi\in \boldsymbol{\cal F}_{\rm 2-p}^{(\infty)}(\xi_1,\xi_2)\ . 
\end{align}
In what follows, we shall consider backgrounds in which $L^{(0)}$ describes 4D, asymptotically, locally anti-de Sitter spacetimes and 3D, conformally flat spacetimes, which requires the zero-form expectation value to vanish\footnote{Subjecting the defect configurations to asymptotic boundary conditions on $\boldsymbol{X}$, the gauge functions and integration constants can be corrected perturbatively  such that the asymptotically free fields do not receive any higher-order corrections \cite{COMST}.}.

\subsection{Holonomies and fractional-spin algebra }\label{Sec:2.8}

The FSG and CCHSG defects have open-curve holonomy groups
\begin{align}
G^{(0)}= Mp(4;\Real)\times U(N_+,N_-)\ ,
\end{align}
which trigger the integration constants to belong to fractional-spin algebras.
The resulting class of parent models arises from two-parton systems with fibre algebras given by endomorphisms of projected left-orbits
\begin{align}
 {\cal S}_{\rm 2-p}(\xi_0;N_+,N_-):=(e^\bullet_{11}\star {\cal S}(\xi_0))\oplus(e^\bullet_{22}\star {\cal C}(N_+,N_-))\ ,   
\end{align}
consisting of a Hermitian singleton, as in Eq. \eqref{2.99}, and a finite-dimensional, Hermitian space of Chan--Paton-like factors as in Eq. \eqref{2.78}; the resulting fractional-spin\footnote{The singletons have fractional $SL(2;\Comp)$-spins.
The zero-form can be assigned a vacuum expectation value $\nu\in \Real$ that deforms the fractional spins and breaks $O(1,1)_D\times SL(2,\Real)_{\rm Lor}$ to $SL(2,\Real)_{\rm Lor}$; correspondingly, the superconnection contains matter fields on the FSG branch with properly fractional spins, and massive scalars and spinors on the corresponding broken CCHSG branch \cite{wip}.} algebra \cite{Boulanger:2013naa, Boulanger:2015uha}\footnote{The projected, two-parton algebra realizes of the two-by-two block structure introduced by hand in the original construction \cite{Boulanger:2013naa}. 
The adapted trace operation \eqref{2.67reg} refines the supertrace operation for the fractional-spin algebra introduced in \cite{Boulanger:2013naa}. 
The activation of the full two-parton algebra yields the N=2 tensor models of  \cite{Vasiliev:2018zer}.} 
\begin{align}\label{2.143}
\boldsymbol{\cal FS}:= 
\boldsymbol{\cal F}_{\rm 2-p}^{(\infty)}(\xi_0;N_+,N_-)\equiv \left[\begin{array}{c|c} {\cal S}(\xi_0)\otimes {\cal S}^\dagger(\xi_0) &{\cal S}(\xi_0)\otimes {\cal C}^\dagger(N_+,N_-)\\\hline{\cal C}(N_+,N_-)\otimes {\cal S}^\dagger(\xi_0)&{\cal C}(N_+,N_-)\otimes {\cal C}^\dagger(N_+,N_-)\end{array}\right]\ ,    
\end{align} 
with Hermitian conjugation operation
\be \left[\begin{array}{cc} \Psi_{11} & \Psi_{12}\\ \Psi_{21}&\Psi_{22}\end{array}\right]^\dagger= \left[\begin{array}{cc} (\Psi_{11})^\dagger & (\Psi_{21})^\dagger \\ (\Psi_{12})^\dagger &(\Psi_{22})^\dagger \end{array}\right]\ .\ee 
For definiteness, we assume that
\begin{align}
\Psi_{\xi_{\rm CP}|\cdot}={}&{\cal P}_{\cdot|\cdot}={\cal P}_{1/2|1/2}=4\exp(-4E)\ ,\qquad N_+=N_-=:N\ ,\\
{\cal C}(N,N)={}&\bigoplus_{I=1}^N \Comp\otimes\left(|e_+^I\rangle\oplus   |e_-^I\rangle\right)\ ,\quad |e_\pm^I\rangle=\kappa_y\star |e_\mp^I\rangle\ ,
\end{align}
consisting of $N$ states $|e_+^I\rangle$ in the Fock space and $N$ images $|e_-^I\rangle=\kappa_y\star|e_+^I\rangle$ in the anti-Fock space with maximally split Hermitian form and resulting identity operator 
\begin{align}
\left(| e^I_\epsilon \rangle, |e^{I'}_{\epsilon'}\rangle\right)_{{\cal C}(N,N)}:=\epsilon \delta_{\epsilon,\epsilon'}\delta^{II'}\ ,\qquad 
{\rm Id}_{{\cal C}(N,N)}=\sum_{\epsilon=\pm}\sum_{I=1}^N \ket{e^I_\epsilon}\bra{e_{I}^\epsilon}\ ,\quad \bra{e_{I}^\epsilon}=\epsilon (|e^I_\epsilon\rangle)^\dagger\ ;
\end{align}
the resulting space of Chan--Paton-like factors can be reduced to a positive definite space, i.e., a Hilbert space, at the expense of breaking the modular element $\kappa_y$.

\subsection{FSG and CCHSG defect topology}\label{sec:FSGandCHSG}

The FSG and CCHSG defects have correspondence spaces given by fibre bundles characterized by reduced DGA structure groups\footnote{As shown in \cite{paperII}, $U(N,N)$ can be included into the structure group by treating the $\boldsymbol{Z}$-components of the one-forms as statistical gauge fields for non-abelian Anyons.} and chiral domains\footnote{The chiral domains of the FSG and CCHSG defects are related by a rotation of integration contours which do not affect star products and traces of metaplectic group algebra elements.
These analytical continuations may be obstructed, however, by symbols in $\boldsymbol{Z}$ arising at higher orders in classical perturbation theory.} as follows:
\begin{align}
\mbox{FSG}\, (\Xi=\mathbb{C})\  :& \ H_\Comp=\frac12(1+\pi^\ast)(H)\stackrel{\#}{\longrightarrow} SL(2;\mathbb{C})_{\rm Lor}\times U(N,N)\ ,\\
&\ \boldsymbol{T}^{(\mathbb{C})}_8\to \boldsymbol{C}^{(\mathbb{C})}_{12}\stackrel{{\rm pr}_{\boldsymbol{C}}'}{\longrightarrow} \boldsymbol{X}^{(\Comp)}_4\ ,\label{5.18}\\
&\ \widetilde{\boldsymbol{Y}_\Real}\subset \boldsymbol{Y}\to \boldsymbol{T}^{(\mathbb{C})}_8\to \boldsymbol{Z}^{(\Comp)}_4\cong  S^2_{\Comp}\times \overline{S}^2_{\Comp}\supset \widetilde{\boldsymbol{Z}_\Real}\\
\mbox{CCHSG}\,(\Xi=\mathbb{R})\ :&\ H_\Real=\frac12(1+\pi^\ast_{\!\mathscr{P}})(H)\stackrel{\#}{\longrightarrow}  SL(2;\mathbb{R})_{\rm Lor}\times O(1,1)_{\rm Dil}\times U(N,N)\ ,\\
&\  \boldsymbol{T}^{(\mathbb{R})}_8\to \boldsymbol{C}^{(\mathbb{R})}_{12}\stackrel{{\rm pr}_{\boldsymbol{C}}'}{\longrightarrow} \boldsymbol{X}_4^{(\Real)}\ ,\label{3.61}\\ 
&\ \widetilde{\boldsymbol{Y}_\Real}'\subset \boldsymbol{Y}\to \boldsymbol{T}^{(\mathbb{R})}_8\to \boldsymbol{Z}^{(\Real)}_4\cong S^2_+ \times S^2_-\supset \widetilde{\boldsymbol{Z}_\Real}'\ ,\label{3.62}
\end{align} 
where i) $\boldsymbol{X}^{(\Xi)}_4$ are non-commutative, compact, orientable K\"ahler manifolds of complex dimension four whose module of holomorphic symbols form a commuting algebra equipped with a holomorphic Hermitian conjugation operation; ii) $\boldsymbol{Z}^{(\Xi)}_4$ are non-commutative, compact\footnote{By $\boldsymbol{Z}^{(\Xi)}_4$ being compact, it is meant that algebra of differential forms on $\boldsymbol{Z}^{(\Xi)}_4$ has a well-defined trace operation projecting onto top forms (used in defining Chern classes) described using holomorphic inner Klein operators realized using the complexified metaplectic group (rather than Moyal--Weyl star products). In this sense, the HSG branch of the model is equivalent to Vasiliev's system locally but not globally, meaning on the entire noncommutative geometry including boundary conditions on $\boldsymbol{Z}^{(\Xi)}_4$.}, orientable, holomorphic, symplectic manifolds of complex dimension four whose modules of holomorphic symbols represent complex, inhomogeneous metaplectic group algebras; iii) $H_\Xi$ are reduced DGA structure groups projected out using associative algebra automorphisms; and iv) the assignments of chiral domains facilitate the construction of background cohomology groups
\begin{align}
H(S^2_{\Comp}\times \overline{S}^2_\Comp)={}&\Comp\otimes(1\oplus j_z)\star (1\oplus \bar j_{\bar z}) \ ,\qquad \bar j_{\bar z}=(j_z)^\dagger\ ,\label{78b}\\  \label{3.6} 
H(S^2_+\times S^2_-) ={}& \Comp\otimes(1\oplus j_{z^+})\star (1\oplus j_{z^-})\ , \qquad (j_{z^\pm})^\dagger=j_{z^\pm}\ ,
\end{align}
where $j_{z}$ and $\bar j_{\bar z}$, respectively, are twisted-central elements arising from conical singularities on two-dimensional, holomorphic, symplectic leaves $S^2_\Comp$ and $\overline{S}^2_\Comp$, respectively, related by holomorphic Hermitian conjugation, and $j_{z^\pm}$ are non-central elements arising from conical singularities on holomorphic, Lagrangian sub-manifolds $S^2_\pm$ left invariant by holomorphic Hermitian conjugation.

\paragraph{FSG cohomology.}

We assign  $\boldsymbol{Z}^{(\Comp)}_4$ canonical coordinates $z_\alpha$  obeying 
\begin{align}
[ z_\alpha,z_\beta]_\star= -2i\epsilon_{\alpha\beta}\ ,\qquad k_z\star z_\alpha+z_\alpha\star k_z=0\ ,\qquad {\bar z}_{\dot\alpha}:=-(z_\alpha)^\dagger\ ,\label{realyz}    
\end{align}
assume an ordering scheme that reduces to Weyl order on $\boldsymbol{Y}$ and $\boldsymbol{Z}^{(\Comp)}_4$, and assign $Z^{(\Comp)}_4$ the chiral integration domain $\widetilde{Z_\Real}\cong \widetilde{Y_\Real}$.
The holomorphic two-form in \eq{78b} is represented by 
\begin{align}
j_z=-\frac{i}{4} dz^\alpha\wedge dz_\alpha \kappa_z\ , \qquad \kappa_z:=2\pi\delta^2_\Comp(z)\ , \label{jC}    
\end{align}
which is manifestly $SL(2,\mathbb{C})_{\rm Lor}$-invariant and twisted-central, viz.,
\begin{align}
[j_z \star k_z, \psi]_\star = 0\ ,\qquad \psi\in \Omega(\boldsymbol{Z}^{(\Comp)}_4)\ .   
\end{align}
It follows that the complex two-form\footnote{In $I_{\Comp}$, the factor $\kappa_y\star k$ acts faithfully on the internal space ${\cal C}(N,N)$.}
\begin{align}\label{3.15a}
\mathbb{I}_{\Comp}:=I_{\Comp}\star {\rm Id}_{\boldsymbol{\cal FS}}\ ,\qquad I_{\Comp}:= \frac12(1+k\star \bar k)\star j_z\star \kappa_y\star k\ ,\qquad {\rm Id}_{\boldsymbol{\cal FS}}=\left[\begin{array}{cc} {\rm Id}_{{\cal S}(\xi_0)}&0\\0&{\rm Id}_{{\cal C}(N,N)}\end{array}\right]\ ,
\end{align}
is central (already) in $\boldsymbol{\cal E}_{\rm hor}^{H}({\cal K}\times \boldsymbol{C};\boldsymbol{\cal N})$ and cohomologically non-trivial, i.e., $d\mathbb{I}_{\Comp}=0$ and 
\begin{align}
\mathbb{I}_{\Comp}\star \overline {\mathbb{I}}_{\Comp} \propto  d^4Z \,\kappa_z\star \bar\kappa_{\bar z}\star K\ ,\qquad \overline {\mathbb{I}}_{\Comp}:=(\mathbb{I}_{\Comp})^\dagger\ .
\end{align}

\paragraph{CCHSG cohomology.}

We assign  $\boldsymbol{Z}^{(\Real)}_4$ canonical coordinates $z^\xi_\alpha$, $\xi=\pm$, obeying 
\begin{align}
[z^\xi_\alpha,z^{\xi'}_\beta]_\star= -2i\epsilon_{\alpha\beta}\d^{\xi,-\xi'}
\ ,\qquad (z^\xi_\alpha)^\dagger=-z^\xi_\alpha\ , 
\end{align}
assume an ordering scheme that reduces to Weyl order on $\boldsymbol{Y}$ and $\boldsymbol{Z}^{(\Real)}_4$, and assign $Z^{(\Real)}_4$ the chiral integration domain $\widetilde{Z_\Real}'\cong \widetilde{Y_\Real}'$. 
The two-forms in \eqref{3.6} are built from manifestly $SL(2,\mathbb{R})_{\rm Lor}\times O(1,1)_{\rm Dil}$-covariant twisted projectors, viz.,
\begin{align}
I^\pm_{\Real}=j_{z^\pm}\ ,\qquad j_{z^\pm}:=\frac{i}8 dz^{\pm\alpha}\wedge dz^\pm_\alpha \, \tilde\kappa_{z^\pm}\ ,\quad \tilde\kappa_{z^\pm}:=4\pi\delta^2_\Comp(z^\pm)\ ,\label{IR} 
\end{align}
whose de Rham non-triviality follows from\footnote{If $\xi=\xi'$, then the supports of the delta functions are point-split by the prescription 
\be  (dz^\xi)^2 \d^2_\Comp(z^\xi)\star (dz^\xi)^2 \d^2_\Comp(z^\xi) = \lim_{\e\to 0}  (dz^\xi)^2 \d^2_\Comp(z^\xi)\star  (dz_\e^\xi)^2 \d^2_\Comp(z_\e^\xi)\ ,\qquad z^\pm_\e := \frac{1}{\sqrt{1+\e^2}}(z^\pm + \e z^\mp) \ ,  \nonumber\ee
resulting in 
\be \lim_{\e\to 0}  (dz^\xi)^2 \d^2_\Comp(z^\xi)\star  (dz_\e^\xi)^2 \d^2_\Comp(z_\e^\xi) = -\lim_{\e\to 0}\frac{1}{(2\pi)^2}\, \frac{4\e^2}{1+\e^2}d^2z^+ d^2z^{-}\, \frac{1+\e^2}{\e^2} \,\exp(i\xi z^+z^-)\ , \nonumber\ee
which yields \eq{VolIR} (while no regularization is required if $\xi=-\xi'$). This regularization, which is the only one that we use in this paper and in Paper II, assigns a finite volume to $\boldsymbol{Z}^{(\Real)}_4\cong S^2_\Real\times S^2_\Real$. On the contrary, we are taking $\boldsymbol{Y} = T^\ast \Real^2$, and, correspondingly, we do not regularize star-products of symbols on $\boldsymbol{Y}$; indeed, the divergent nature of $\delta(y^-) \star \delta(y^-)$ is desirable given the structure of 3D CFT correlation functions \cite{Didenko:2012tv,Colombo:2010fu,Colombo:2012jx}; see also comments in Section 4.7 in Paper II \cite{paperII}).} 
\begin{align}
I^\xi_\Real\star I^{\xi'}_\Real= -\frac{1}{4}\,d^2z^+ d^2z^- e^{i\xi z^+ z^-}  \ ,\qquad \xi,\xi'=\pm 1\ ,\label{VolIR}
\end{align}
which essentially assigns $S^2_+\times S^2_-$ a finite volume.
Representing $SL(2,\mathbb{R})_{\rm Lor}\times O(1,1)_{\rm Dil}$ on symbols on $\boldsymbol{Z}^{(\Real)}_4$ using
\begin{align}
M^{(z)}_{\alpha\beta}:= -\frac12 z^+_{(\alpha}z^-_{\beta)}\ , \ ,\quad D^{(z)}:=-\frac14 z^+ z^-\ ,\quad [D^{(z)},z^\pm_\a]_\star=\pm\tfrac{i}2\,z^\pm_\a\ ,    
\end{align}
the lowest-weight projector
\begin{align}
{\cal P}^{(z)}_{\pm i/2|\pm i/2}:=4\exp(\pm 4iD^{(z)})\ ,\qquad {\cal P}^{(z)}_{\pm i/2|\pm i/2}\star {\cal P}^{(z)}_{\pm i/2|\pm i/2}={\cal P}^{(z)}_{\pm i/2|\pm i/2}\ ,   
\end{align}
obeys $M^{(z)}_{\a\b}\star {\cal P}^{(z)}_{\pm i/2|\pm i/2}=0={\cal P}^{(z)}_{\pm i/2|\pm i/2}\star M^{(z)}_{\a\b}$ and 
\begin{align}
D^{(z)}\star{\cal P}^{(z)}_{\pm i/2|\pm i/2} = \pm\frac{i}2 {\cal P}^{(z)}_{\pm i/2|\pm i/2} = {\cal P}^{(z)}_{\pm i/2|\pm i/2} \star D^{(z)}\ .    
\end{align}
In terms of ${\cal P}^{(z)}_{\pm i/2|\pm i/2}$, one has 
\begin{align}\label{IR2}
\widetilde{{\cal P}}^{(z^\pm)}_{\pm i/2|\mp i/2}=i{\cal P}^{(z)}_{\pm i/2|\pm i/2}\star \kappa_z=\tilde\kappa_{z^\pm}\ ;\end{align}
thus, comparing \eq{jC} and \eq{IR}, it follows that the cohomology elements of the CCHSG defect possess lower symmetry than those of the FSG defect due to the insertion of the ${\cal P}^{(z)}_{\pm i/2|\pm i/2}$ which indeed breaks $SL(2,\Comp)_{\rm Lor}$ down to $SL(2,\Real)_{\rm Lor}\times SO(1,1)_{\rm Dil}$.

\section{Conclusions of Part I}\label{sec:conclusionspart1}

Motivated by the quest for a framework capable of including Vasiliev's higher-spin gravity as well as holographic dual descriptions, and of giving an a priori rationale to holographic relations, in this paper we have initiated an AKSZ approach to holographic dualities based on the Frobenius-Chern-Simons action for Vasiliev's higher-spin gravity (HSG). In particular, we propose that HSG and its dual theory are embedded into a common parent theory, as consistent reductions (or defects) of the latter, as we shall show in the companion paper \cite{paperII}. Thus, the basic object controlling holography will be the multi-dimensional partition function of the parent AKSZ theory, consisting of sub-partition functions given by sums over sub-configuration spaces. The latter are, in turn, projections of the universal configuration space of the parent field theory that are compatible with boundary conditions given by on-shell projections of classical moduli spaces (alias, defects).

The reductions of the parent action thus equip the defects with generalized, symplectic structures encoded into the generalized Hamiltonian action as terms that are of quadratic or higher order in AKSZ momenta, resulting in multi-dimensional correlation functions encoding generalized, second-quantized star products.
In particular, the Fourier modes of the Fronsdal fields, which are read off from the Weyl zero-forms and dualized into gauge fields using cocycles on defects with 4D spacetime leaves, are deformed into creation and annihilation operators for massless particle states through contractions read off from the sub-partition function obtained by simultaneously reducing the parent action down to a two-dimensional gauged Poisson sigma model on a disk, with boundary conditions given the reduced on-shell values for the Weyl zero-forms, idem the quantization of the matter fields of the dual defects \cite{OLC}. 
Besides providing a unified approach to holography, the advantage of this approach to the quantization of higher-spin gravity is that it applies naturally to the non-commutative geometries underlying the theory. 

More specifically, in this paper we have constructed the appropriate non-commutative geometry from the quantization of two conformal particles --- alias, string partons --- subject to a boundary condition that pins one of the two particles to a base point, which gives rise to an ungraded two-by-two matrix algebra.
Moreover, the target space is assumed to have the structure of a correspondence space, i.e., its algebra of differential forms admits a set of derivations in negative degree.
The latter closes with the de Rham differential into a set of Lie derivatives assigning the correspondence space the structure of a fibre bundle equipped with an algebra of horizontal forms given by forms on the base valued in an ungraded algebra of special functions on the fibre.
The resulting horizontal algebra is then tensored with a three-graded two-by-two matrix algebra, to form a superconnection consisting of sixteen horizontal forms, i.e., four zero-forms, eight one-forms and four two-forms.

The correspondence space is quantized by taking the horizontal forms to be symbols of a holomorphic metaplectic group algebra, i.e., the correspondence space is equipped with a holomorphic, symplectic structure equipped with an involution, and the horizontal forms are taken to be holomorphic forms obeying a holomorphic reality condition (imposed using the composition of the canonical Hermitian conjugation map and the involution). 
As a result, the metaplectic group algebra contains a discrete set of modular transformations used in imposing boundary conditions on the horizontal forms; moreover, these elements combine with corresponding outer Klein operators of the first-quantized system and forms on the base into closed and central two-forms that support the defects. As we have seen in this paper, the modular transformations exchanging various boundary conditions of the HSG defect, which are generated by the standard inner Klein operators $\k_y$, $\bar\kappa_{\yb}$, have a counterpart on the CCHSG defect, generated by the operator $\kappa_{\!\mathscr{P}}$, whose composition with $\kappa_y$ acts by Fourier transforming the wave-functions of the Hermitian singleton module in the non-compact momentum basis.

Then, we turned to singling out the relevant defects. The latter are created by first switching on holonomies that break the symmetry under the exchange of parton indices, given by the direct product of metaplectic and $U(N,N)$ group elements assigned to the separate partons.
One corresponds to 4D fractional-spin fields carrying local degrees of freedom coupled to Vasiliev's higher-spin gravity and $U(N,N)$ gauge fields (defining a 4D fractional-spin gravity, which can be truncated to Vasiliev's higher-spin gravity); while the other describes 3D conformal matter fields coupled to topological, conformal higher-spin gravity and $U(N,N)$ gauge fields (coloured conformal higher-spin gravity), as we shall study in detail in Paper II. 

As discussed in Section 5, the defects arise from projections of the underlying DGA that involve structure group reductions, projections of the base of the correspondence space, i.e. the fibred manifold on which the master fields live, and projections of the graded four-by-four algebra comprising the projector algebra of the two-parton system introduced in Section \ref{sec:2}, and an additional three-graded ${\rm mat}_{1|1}$ (which one may speculate arises naturally by adding a set of chiral spin-zero ghosts to the gauged WZW model of \cite{Engquist:2007pr} to make the chiral $W_{\mathfrak{sp}(2)+\infty}$ gauging critical).
Despite their distinct characters, clear signs that the two defects are dual to each other appear already at the level of Chern classes, which can be spelt out already at this stage without entering into any details of the defect dynamics. 
To this end, we quote here the relevant parts of the HSG and CCHSG equations of motion, viz.
\begin{align}
{\rm HSG}:{}&\quad dA+A\star A+B\star I_\Comp^{(\theta_0)}=0\ ,\\
{\rm CCHSG}:{}&\quad dW+W\star W+C\star \overline{C}\star I^-_\Real=0\ ,
\end{align}
where $B$ contains the bulk Weyl zero-form and $C$ contains the boundary conformal matter fields.
Assuming that the AKSZ formulation assigns a cylinder with dual boundary conditions a morphism connecting isomorphic factors of the HSG and CCHSG operator algebras, it is natural to expect this map to intertwine $B$ and $C\star \overline C$.
Indeed, this can be corroborated by matching the second Chern classes \eqref{4.25} for $A$ and $W$ on $\boldsymbol{Z}\times \{x\}$ where $x\in \boldsymbol{X}$, which are given by 
\begin{align}
{\rm HSG}:{}&C_2(\boldsymbol{Z}\times \{x\})\propto\oint_{\widetilde{\boldsymbol{Z}_\Real}}^{\prime} {\rm Tr}_{\boldsymbol{\cal K}}\,{\rm Tr}_{\boldsymbol{\cal Y}}\, I^{(\theta_0)}_\Comp\star I^{(\theta_0)}_\Comp\star B\star B \ ,\\
{\rm CCHSG}:{}&C_2(\boldsymbol{Z}\times \{x\})\propto\oint_{\widetilde{\boldsymbol{Z}_\Real}'}^{\prime}{\rm Tr}_{\boldsymbol{\cal K}}\,{\rm Tr}_{\boldsymbol{\cal Y}}\, I^-_\Real\star I^-_\Real\star (C\star \overline{C})^{\star 2}\ ,
\end{align}
as will be the topic of \cite{OLC}.

The fractional-spin algebra arises upon performing the classical reductions in the presence of a background holonomy element. The resulting algebra has a two-by-two block decomposition \cite{Boulanger:2013naa}: the first diagonal block consists of the endomorphism algebra of the Hermitian singleton which is a subalgebra of the group algebra of the centrally-extended, inhomogenous, complex metaplectic group $MpH(4;\Comp)$ arising at its asymptotic boundary; the second diagonal block is the internal colour algebra; and the off-diagonal blocks are coloured, Hermitian singletons. 

More broadly, the Hermitian modules provide representation spaces for non-commutative geometries making up parent field configurations subject to various boundary conditions.
Their quantum numbers constitute a set of classical moduli parameters analogous to brane charges in supergravity.
Indeed, their introduction triggers extensions of pure HSG/CHSG models by matter fields that are non-perturbative from the underlying first-quantized point-of-view, while admitting perturbative descriptions as classical parent field configurations (in asymptotic regions where curvatures fall off on-shell).
In this sense, we view the CCHSG defects as analogues of supergravity $D$-brane/anti-$D$-brane systems in which topological world-volume metrics have acquired higher-spin partners, and, correspondingly, the FSG defects as analogues of supergravities in which fractional-spin one-forms play the role of potentials.

\paragraph{Acknowledgements.}
We would like to thank the Referee for many valuable suggestions that helped us improve the presentation. We have benefitted from discussions with L. Andrianopoli, R. Aros, M. Bianchi, N. Boulanger, S. Deger, V. E. Didenko, J. Lang, S. Lysov, Y. Neiman, B.E.W. Nilsson, C. Reyes, E. Skvortsov, D. Sorokin, M. Trigiante, M. Tsulaia, M. Valenzuela, B. Vallilo, M. A. Vasiliev and J. Zanelli. PS is grateful for the support during various stages of the project of the Centro de Ciencias Exactas at Universidad del Bio-Bio; the Centro de Estudios Cientificos at Universidad San Sebastian; the Service de Physique de l’Univers, Champs et Gravitation at
Universit\'e de Mons; the Department of Mathematics at Bogazici University; and the Quantum Gravity Unit of the Okinawa Institute of Science and Technology. FD and PS would like to thank the Institute of Mathematics of the Czech Academy of Sciences for hospitality during the final stage of this project. The work of PS is partially supported by the funding from the European Research Council (ERC) under Grant No. 101002551 and by the Tubitak Bideb-2221 fellowship program. The work of FD is supported by {\sc Beca Doctorado nacional} 
(ANID) 2021 Scholarship No. 21211335, ANID/ACT210100 Anillo Grant ``{\sc Holography and its applications to High Energy Physics, Quantum Gravity and Condensed Matter Systems}'' and FONDECYT Regular grant No. 1210500.

\begin{appendix}

\section{Bases and oscillator realizations of $\mso(2,3)$}\label{App:emb}

\subsection{Compact vs. conformal basis of $\mathfrak{so}(2,3)$}

In the conventions of \cite{fibre}, the $\mso(2,3)$ generators $M_{AB}=-M_{BA}$, $A,B=0',0,1,2,3$, are taken to obey 
\be [M_{AB},M_{CD}]_\star =\ 4i\y_{[C|[B}M_{A]|D]}\ ,\qquad
(M_{AB})^\dagger\ =\ M_{AB}\ ,\label{sogena}\ee
where $\eta_{AB}={\rm diag}(--+++)$.
The generators of the Lorentz subalgebra $\mso(1,3)$ are taken to be $M_{ab}$, $a,b=0,1,2,3$; the transvections 
\begin{align}
    P_a:=M_{0'a}~
\end{align}
in units where the cosmological constant  $\L=-3$, obey 
\be 
[M_{ab},P_c]_\star\ =\ 2i\y_{c[b}P_{a]}\ ,\qquad [P_a,P_b]_\star\ =\
i M_{ab}\ ,\label{sogenb}\ee
where $\eta_{ab}={\rm diag}(-+++)$.

In order to exhibit the maximal compact subalgebra $\mso(2)_E\oplus\mso(3)_M$ generated by the energy generator $E=M_{0'0}=P_0$ and the spatial rotation generators $M_{rs}$ with $r,s=1,2,3$, we arrange the remaining generators into energy-raising and lowering operators
\be L^\pm_r =M_{0r}\mp iM_{0'r}\ =\ M_{0r}\mp iP_r\ ,\label{Lplusminus}\ee
leading to the following $E$-graded decomposition of the
commutation rules \eq{sogena}:
\begin{align} 
\label{el} 
[E,L^{\pm}_r]_\star  ={}& \pm L^{\pm}_r\ , && [L^-_r,L^+_s]_\star = 2iM_{rs}+2\d_{rs}E \ ,\\
[M_{rs},M_{tu}]_\star ={}& 4i\d_{[t|[s}M_{r]|u]} \ ,&& [M_{rs},L^\pm_t]_\star =
2i\d_{t[s}L^\pm_{r]}\ .\label{ml}
\end{align}
The generators $(E,M_{rs},L^\pm_r)$ are referred to as the \emph{compact basis}, or \emph{compact split} of $\mso(2,3)$. Representations in which $E$ is bounded from below and above, respectively, referred to as lowest- and highest-weight representations, arise from specific functions in the enveloping algebra of $\mso(2,3)$ modulo ideals. In particular, the ultra-short unitary irreducible singleton and anti-singleton representations arise by factoring out the ideal generated by
\begin{align}\label{A.7}
    V_{AB} :={}& \frac12 M_{(A}{}^C \star M_{CB)}+\frac15 \eta_{AB}C_2 = 0~,\qquad V_{ABCD} := M_{(AB}\star M_{CD)} = 0~,
\end{align}
implying the Casimir constraint \cite{fibre}
\begin{align}
    C_2 := \frac12 M^{AB}\star M_{AB} = -\frac54~.
\end{align}
Equivalently, the states forming the (anti-)singleton representation can be obtained from the one-sided star-product action 
\be {\cal D}^\pm(\pm 1/2) := {\rm Env}(\mso(2,3))\star {\cal P}_{\pm 1/2|\pm 1/2} \ee
of the enveloping algebra of $\mso(2,3)$ on the projectors
\begin{align} \label{eq:vac proj}
{\cal P}_{\pm 1/2|\pm 1/2} \ = 4e^{\mp 4E} \ , \qquad {\cal P}_{\pm 1/2|\pm 1/2} \star {\cal P}_{\pm 1/2|\pm 1/2} ={\cal P}_{\pm 1/2|\pm 1/2}~, 
\end{align}
which are the images of the Wigner-Ville map applied to the projectors onto the singleton lowest-weight $(+)$ and anti-singleton highest-weight $(-)$ states $|\pm 1/2,(0)\rangle$, viz.,
\begin{align}
    \cP_{\pm 1/2|\pm 1/2} = | \pm 1/2,(0)\rangle \langle \pm 1/2,(0)| ~.\label{compsing2}
\end{align}
Such projectors carry quantum numbers of the compact subalgebra such that
\begin{align} E\star\cP_{\pm 1/2|\pm 1/2} \ ={}& \ \cP_{\pm 1/2|\pm 1/2} \star E \ = \ \pm\frac12 \cP_{\pm 1/2|\pm 1/2}  \ , \\
 M_{rs}\star\cP_{\pm 1/2|\pm 1/2} \ ={}& \ 0 \ = \ \cP_{\pm 1/2|\pm 1/2}\star M_{rs}  \ , \end{align}
and their lowest- and highest-weight properties are manifested by 
\bea  L^\mp_r\star\cP_{\pm 1/2|\pm 1/2} \ = \ 0 \ = \ \cP_{\pm 1/2|\pm 1/2}\star L^\pm_r \ . \eea
The Lie algebra also admits a \emph{conformal basis} $(D,M_{mn},T_m, K_m)$, viz.,
\begin{align}\label{dt}
[D,T_m] =  iT_m\ ,\quad [D,K_m] = -iK_m \ ,\quad 
[K_m,T_n] = 2i(\eta_{mn}D-M_{mn}) \ , 
\end{align}
\vspace{-1cm}
\begin{align}
[M_{mn},M_{pq}] = 4i\eta_{[p|[n}M_{m]|q]} \ ,\quad   [M_{mn},T_p] = 2i\eta_{p[n}T_{m]}\ , \quad [M_{mn},K_p] = 2i\eta_{p[n}K_{m]}\ .\label{mt}
\end{align}
which is 3-graded with respect to the dilation operator
$D$ of the non-compact subalgebra $\mso(1,1)_D\oplus\mso(1,2)_{M_{mn}}$, and exhibits the translations $T_m$ and special conformal transformations $K_m$ of 3D conformal Minkowski spacetime. Embedding the boundary conformal algebra in such a way that all its generators are hermitian, the dilation generator $D$ can be identified with any spacelike transvection. 
%
For oscillator realizations, and with our conventions on van der Waerden symbols, it is convenient to identify the (boundary) dilation generator as
\begin{align}\label{DasP2}
   D=P_2~, 
\end{align}
and thus the (boundary) Lorentz generators $M_{mn}$, $m,n=0,1,3$, and $D$-raising and $D$-lowering combinations\footnote{Of course, the identification \eq{DasP2} is purely a convenient choice, and we could have rather embedded both compact and conformal slicings by introducing a normalized frame $(L_i^a,L^a)$ obeying 
\begin{align}
L^a L_a=\epsilon~, \qquad L_i^a L_a =0~,\qquad L^a_i L_{ja}=\eta_{ij}=(+,+,-\epsilon)~,\nonumber
\end{align}
and letting 
\begin{align}
K:=L^a P_a~,\qquad K^\pm_i := (\e L^ b M_{ab} \mp \sqrt{\epsilon} P_a)L^a_i~,\qquad M_{ij}:=\e L^a_i L^b_j M_{ab}~, \nonumber
\end{align}
where $K$ is referred to as the principal Cartan generator, and the compact and conformal bases arise for $\epsilon = -1$ and $\epsilon = 1$, respectively; for example see \cite{Sezgin:2005pv,cosmo}. The specific realizations above used thus correspond to the particular choices 
\bea
   & \epsilon = -1~:\qquad  L^a=(1,0,0,0)~, \qquad  K = P_0 = E\ , &\nonumber\\ 
   & \epsilon = 1~:\qquad   L^a=(0,0,1,0)~, \qquad K= P_2 = D & 
    \ . \nonumber
\eea}. 
\be T_m \ = \ M_{m2}-P_m \ ,\qquad K_m \ = \ M_{m2}+P_m \ .\label{TmKm}\ee
The construction of lowest/highest-vector modules induced from $\mso(1,1)\oplus\mso(1,2)$-modules proceeds in a completely parallel fashion to the compact-basis case, with the only difference that, in order to compensate for the non-compact nature of $D$, an extra factor of $i$ enters the metaplectic realization of the lowest/highest weight state projectors. Indeed, the conformal analogue of \eq{compsing2} is the realization of the conformal (anti-)singleton highest-weight (lowest-weight) projector 
\be |\pm i/2,(0)\rangle\langle \pm i/2,(0)| \ \equiv \ \cP_{\pm i/2|\pm i/2 } \ = \ 4e^{\pm 4 iD}  \ , \ee
satisfying
\bea & D\star\cP_{\pm i/2|\pm i/2 } \ = \ \pm \frac{i}{2}\cP_{\pm i/2|\pm i/2 } = \cP_{\pm i/2|\pm i/2 }\star D\ , & \\
 & M_{mn}\star \cP_{\pm i/2|\pm i/2} \ = \ 0  \ = \ \cP_{\pm i/2|\pm i/2} \star M_{mn}  \ ,& \eea
and respectively annihilated by $K_m$ from the left (and $T_m$ from the right) 
\bea  K_m\star \cP_{i/2|i/2} \ = \  0 \ = \ \cP_{i/2|i/2} \star T_m\ , \eea
and $T_m$ from the left (and $K_m$ from the right),
\bea  T_m\star\cP_{-i/2|-i/2} \  = \ 0 \ = \ \cP_{-i/2|-i/2}\star K_m \ , \eea
where we note that $\pi^\ast(K_m)=T_m$.  
In the body of the paper we have frequently used the shorthand notation 
\be \ket{(\pm i/2)} := \ket{\pm i/2; (0)} \ . \ee
All states created via the one-sided action of the enveloping algebra of $\mso(2,3)$ on $ 4e^{-4iD}$ ($ 4e^{4iD}$) are $\mso(1,2)$-tensors of left $D$-eigenvalue $-i(2s+1)/2$ ($i(2s+1)/2$) and rank $s$, $s=0,1,2,...$, corresponding to states $\ket{-i(2s+1)/2;(s)}$ ($\ket{i(2s+1)/2;(s)}$), and give rise to the \emph{conformal (anti-)singleton} representation\footnote{The reason why we refer to ${\cal T}^-(-i/2)$ as conformal singleton, instead of anti-singleton --- reversing the convention used in compact basis --- is because we conventionally choose to realize the 3D Minkowski translation $T_m$ as $D$-raising operator, which singles out $\ket{(-i/2)}$ as 3D Poincar\'e invariant vacuum, ``breaking the symmetry'' in the definition od conformal singleton and anti-singleton.} of $\mso(2,3)$,
\be  {\cal T}^\pm(\pm i/2):={\rm Env}(\mso(2,3))\star \cP_{\pm i/2|\pm i/2 } \ . \ee
States in ${\cal T}^\pm(\pm i/2)$ are bounded from below ($+$) and above ($-$) in the eigenvalue $i\D$ of $D$. Note that the $\pi$-map exchanges highest- and lowest-weight modules, i.e., reverses the sign of $\D$.

\subsection{Spinor conventions and oscillator realizations of $\mathfrak{so}(2,3)$}

In terms of the Majorana oscillators $Y_{\underline\a}$ satisfying the commutation relations \eq{3.2}, the realization of the generators of $\mso(2,3)$ is taken to be
\be M_{AB}~=~ -\ft18  (\C_{AB})_{\underline{\a\b}}\,Y^{\underline\a}\star Y^{\underline\b}\ ,\label{MAB}
\ee
using real Dirac matrices $(\Gamma_{A})^{\underline{\alpha\beta}}$ obeying $(\C_A)_{\underline\a}{}^{\underline\b}(\C_B C)_{\underline{\b\c}}=
\eta_{AB}C_{\underline{\a\c}}+(\C_{AB} C)_{\underline{\a\c}}$.   
Going to a Weyl basis $Y_{(W)}^{\underline\alpha} = (y^{\alpha},\bar{y}^{\dot\alpha})$ that diagonalizes  $\Gamma^5:=i\Gamma^{0123}$, the Dirac matrices decompose as follows 
\begin{align}
C_{\underline{\alpha\beta}}={}&\left(
\begin{array}{cc}
    \epsilon_{\alpha\beta} & 0 \\
     0 & \epsilon_{\dot\alpha\dot\beta}\end{array}\right)~, && \left(\Gamma^5_{(W)}\right)_{\underline\alpha}{}^{\underline{\beta}} = \left( 
\begin{array}{cc}
\delta_{\alpha}^\beta & 0 \\ 
0 & -\delta_{\dot\alpha}^{\dot\beta}%
\end{array}
\right)~, \\ \left(\Gamma_{(W)}^{0'}\right)_{\underline\alpha}{}^{\underline\beta} ={}& \left( 
\begin{array}{cc}
i\delta_{\alpha}^\beta & 0 \\ 
0 & -i\delta_{\dot\alpha}^{\dot\beta}%
\end{array}
\right)~, && \left( \Gamma ^{a}_{(W)}\right)_{\underline{\alpha }}^{\ \ \underline{\beta }
}=\left( 
\begin{array}{cc}
0 & -i\left( \sigma ^{a}\right) _{\alpha }^{\ \ \dot{\beta}} \\ 
i\left( \bar{\sigma}^{a}\right) _{\dot{\alpha}}^{\ \ \beta } & 0%
\end{array}
\right) \ , \\ 
\left( \Gamma^{0'a}_{(W)}\right) _{\underline{\alpha }}^{\ \ \underline{\beta }
}={}&\left( 
\begin{array}{cc}
0 & \left( \sigma ^{a}\right) _{\alpha }^{\ \ \dot{\beta}} \\ 
\left( \bar{\sigma}^{a}\right) _{\dot{\alpha}}^{\ \ \beta } & 0%
\end{array}
\right) \ , && \left( \Gamma^{ab}_{(W)}\right)_{\underline{\alpha }}^{\ \ \underline{\beta}} =\left( 
\begin{array}{cc}
\left( \sigma ^{ab}\right) _{\alpha }^{\ \ \dot{\beta}} & 0 \\ 0 &
\left( \bar{\sigma}^{ab}\right) _{\dot{\alpha}}^{\ \ \dot\beta } %
\end{array}
\right) \ ,
\end{align}
one has
\be
 M_{ab}\ =\ -\frac18 \left[~ (\s_{ab})^{\a\b}y_\a\star y_\b+
 (\sb_{ab})^{\ad\bd}\bar y_{\ad}\star \yb_{\bd}~\right]\ ,\qquad P_{a}\ =\
 \frac{1}4 (\s_a)^{\a\bd}y_\a \star \yb_{\bd}\ ,\label{mab}
 \ee
where the van der Waerden symbols obey
 \bea
 & (\s^{a})_{\a}{}^{\ad}(\sb^{b})_{\ad}{}^{\b} ={} \y^{ab}\d_{\a}^{\b}\
 +\ (\s^{ab})_{\a}{}^{\b} \ , \qquad
 (\sb^{a})_{\ad}{}^{\a}(\s^{b})_{\a}{}^{\bd}~=~\y^{ab}\d^{\bd}_{\ad}\
 +\ (\sb^{ab})_{\ad}{}^{\bd} \ ,& \label{so4a}\\
 &\ft12 \e_{abcd}(\s^{cd})_{\a\b} = {} i (\s_{ab})_{\a\b}\ ,\qquad  \ft12
 \e_{abcd}(\sb^{cd})_{\ad\bd}~=~ -i (\sb_{ab})_{\ad\bd}\ , &\label{so4b}
\\ & \e^{\a\b}\e_{\c\d} \ ={} \ 2 \d^{\a\b}_{\c\d} \ , \qquad 
\e^{\a\b}\e_{\a\c} \ = \ \d^\b_\c \ , & \\ 
 &(\s^a)_{\a\bd})^\dagger={}
(\sb^a)_{\ad\b} ~=~ (\s^a)_{\b\ad} \ ,\qquad  ((\s^{ab})_{\a\b})^\dagger\ =\ (\sb^{ab})_{\ad\bd} \ , \qquad  (\e_{\a\b})^\dagger \ = \ \e_{\ad\bd} \ . &
\eea
and two-component spinor indices are raised and lowered according to the
conventions $A^\a=\epsilon^{\a\b}A_\b$ and $A_\a=A^\b\epsilon_{\b\a}$. 
The van der Waerden symbols are realized as 
\begin{align}\label{A.35}
    \epsilon_{\alpha\beta} = i\left(\sigma^2\right)_{\alpha\beta}~,\qquad 
    \left(\sigma^a\right)_{\alpha}{}^{\dot\alpha} = \left(-i \sigma^2,-i\sigma^r\sigma^2\right)_{\alpha}{}^{\dot\alpha}~,\qquad \left(\bar{\sigma}^a\right)_{\dot\alpha}{}^{\alpha} = \left(-i \sigma^2,i\sigma^2\sigma^r\right)_{\alpha}{}^{\dot\alpha}~,
\end{align}
where $\s^r$, $r=1,2,3$, are the Pauli matrices.

In the text we frequently employ the implicit-index notation for contracted indices, in which case we always juxtapose  spinors and spinor-tensors from left to right according to so-called NorthWest-SouthEast rule, e.g.,
\begin{align}\label{implicit}
VW
  &:=
  V^\a W_\a
  =
  -WV\ , \qquad  VABW:=
  V^\a A_\a{}^{\b} B_\b^{\phantom{\b}\gamma} W_\gamma \ .
\end{align}

Every slicing of the $\mso(2,3)$-algebra, like the compact or the conformal basis, has a corresponding grading generator ($E$ and $D$, in the examples above shown) and adapted choice of oscillator basis. Indeed, the $\Gamma_{AB}$ matrix that selects the grading Cartan generator --- $\Gamma_{0'0}$ in the compact case, $\Gamma_{0'2}$ in the conformal one according to the realization \eq{DasP2} --- can be used to define projectors inducing a split of the symplectic coordinates $Y_{\ua}$ into canonical pairs $Y_{\ua}^\pm$. In the non-compact case the latter can be extracted as
\be \widetilde Y_{\ua}^\pm \ = \ \sqrt{2}\, \Pi^\pm_{\ua}{}^{\underline{\b}}Y_{\underline{\b}} \ = \  \frac1{\sqrt{2}}\left(\d_{\ua}{}^{\underline{\b}}\pm \Gamma_{0'2\ua}{}^{\underline{\b}}\right)Y_{\underline{\b}} \ , \label{Ypmdef}\ee
with commutation relations
\be [\widetilde Y_{\ua}^\e,\widetilde Y_{\underline{\b}}^{\e'}]_\star \ = \ 4i\d^{\e,-\e'}\Pi_{\underline{\a\b}}^{\e} \ , \qquad \e,\e'=\pm \ ,\ee
where $\Pi^\pm$ are projectors (see \cite{2011} for the general construction of adapted oscillator bases) and the factor of $\sqrt{2}$ in the definition \eq{Ypmdef} has been added for convenience. More explicitly, according to the realization \eq{A.35} of the van der Waerden symbols, the independent canonical pairs are 
\bea \widetilde y^{\pm}_\a  =  \frac1{\sqrt{2}}\left(y\pm\s_2\yb\right)_\a =\frac1{\sqrt{2}}(y_\a\mp i\yb_{\dot \a})\ , \label{ypms}\eea
satisfying the commutation relations
\be [\widetilde y^\e_\a,\widetilde y^{\e'}_\b]_\star  =  2i\e_{\a\b}\delta^{\e,-\e'} \ .\label{commypm}\ee
Clearly, these oscillators are not real, $(\widetilde y^\pm_\a)^\dagger=\pm i\widetilde y^\pm_\a$. A real pair can be easily defined as $y^\pm_\a:=\exp(\pm i\pi/4)\widetilde y^\pm_\a$, i.e.,
\bea  y^{\pm}_\a  =  \frac{e^{\pm i\pi/4}}{\sqrt{2}}\left(y\pm\s_2\yb\right)_\a =\frac{e^{\pm i\pi/4}}{\sqrt{2}}(y_\a\mp i\yb_{\dot \a})\ , \qquad (y^{\pm}_\a)^\dagger=y^{\pm}_\a \ ,\label{ypmreal}\eea
which definition leaves the commutation relations \eq{commypm} unmodified\footnote{To our knowledge, this basis for the oscillator realization of the 3D conformal group was first used in \cite{F&L}.},
\be [y^\e_\a, y^{\e'}_\b]_\star  =  2i\e_{\a\b}\delta^{\e,-\e'} \ .\label{commypmreal}\ee
As follows from \eq{piyyb}, 
\be \pi^\ast_y(y^\pm_\a) = \mp i y^{\mp}_\a \ , \qquad \bar\pi^\ast_{\yb}(y^\pm_\a) = \pm i y^{\mp}_\a\ .\ee
It is possible to fix the relation between the generators of the 3D conformal group in vectorial basis, viz., $(D,M_{mn},K_m,T_m)$ and spinorial basis, viz., $(D,M_{\a\b},K_{\a\b},T_{\a\b})$, as 
\bea   T_{\a\b}=(\gamma_m)_{\a\b}T^m \ , \qquad K_{\a\b}=(\gamma_m)_{\a\b}K^m\ , \qquad  M_{\a\b}  =  -\frac12(\gamma_{mn})_{\a\b}M^{mn} \ ,
\eea
where $\gamma_{mn}=\frac12[\gamma_m,\gamma_n]$, and
\be T_m = -\frac12 (\gamma_m)_{\a\b}T^{\a\b}\ , \qquad K_m = -\frac12 (\gamma_m)_{\a\b}K^{\a\b} \ , \qquad M_{mn}  =  -\frac12(\gamma_{mn})_{\a\b}M^{\a\b} \ , \ee
where, having selected $P_2$ as transvection generator along the direction of foliation, it is natural to define $\e_{\a\bd}:=i(\s_2)_{\a\bd}$ as the element that breaks $AdS_4$-covariance, and thus, clearly,
\bea (\gamma_m)_{\a\b} & := & (\s_m)_{\a}{}^{\bd}\e_{\bd \b} = i(\s_{2m})_{\a\b} \nonumber \\ & = &  \left\{\left(\begin{array}{cc}-1 & 0 \\ 0 & -1 \end{array}\right), \left(\begin{array}{cc}0 & 1 \\ 1 & 0 \end{array}\right), \left(\begin{array}{cc} 1 & 0 \\ 0 & -1 \end{array}\right)\right\} = (\s_m)_{\a\bd} ,\label{gammasigma}\eea
all real, as expected from boundary Lorentz algebra $\mso(1,2)\sim \msp(2,\Real)$ generators, and satisfying
\be (\gamma_{mn})_{\a\b} \ = \ \epsilon_{mn}{}^r(\gamma_r)_{\a\b} \ ,\qquad \{\gamma_m,\gamma_n\}=2\eta_{mn} \ . \ee
with $\e_{023}=1$. In these conventions, the spinor realization of the conformal group generators is
\be T_{\alpha\beta}=\frac12\,y^+_\alpha y^+_\beta\ , \qquad K_{\alpha\beta}=-\frac12\,y^-_\alpha y^-_\beta \label{PKapp}\ee
\be M_{\alpha\beta}=\frac12\,y^+_{(\alpha} y^-_{\beta)}\ ,\qquad D= \frac14 \,y^{+\alpha} y^-_{\alpha} \ .\label{MDapp}\ee

Analogously, one can define the combinations
\bea  z^{\pm}_\a  =  \frac{e^{\pm i\pi/4}}{\sqrt{2}}\left(z\pm\s_2\zb\right)_\a =\frac{e^{\pm i\pi/4}}{\sqrt{2}}(z_\a\mp i\zb_{\dot \a})\ , \qquad (z^{\pm}_\a)^\dagger=-z^{\pm}_\a \ ,\label{zpmreal}\eea
satisfying the commutation relations
\be [\widetilde z^\e_\a,\widetilde z^{\e'}_\b]_\star \ = \ -2i\e_{\a\b}\delta^{\e,-\e'} \ .\ee
where we note that the extra sign in the reality conditions, making $z^\pm_\a$ purely imaginary, is a direct consequence of the reality conditions \eq{realyz}. 

Finally, the split $\widetilde Y_{\ua}^\pm =  \frac1{2}\left(\d_{\ua}{}^{\underline{\b}}\pm i\Gamma_{0'0\ua}{}^{\underline{\b}}\right)Y_{\underline{\b}} $ yields canonical coordinates in compact basis, leading to the definition of the $SU(2)$ creation/annihilation doublets
\be
     a^{\dagger i} = \frac12 \delta^{i\a}\left(y - i \sigma_0\bar{y}\right)_\a ,\qquad a_i = (a^{+i})^\dagger~,\qquad [a_i,a^{\dagger j}]_\star = \delta_i^j~,
\ee
where we have defined the mixed, intertwining symbol $\delta^{i\a}=(\sigma^{0})^{i\a}$, i.e.,
\be
     a^{\dagger 1} = \frac12 \left(y - i \sigma_0\bar{y}\right)_1\ , \qquad a^{\dagger 2} = \frac12 \left(y - i \sigma_0\bar{y}\right)_2  \ .
\ee
\end{appendix}

\providecommand{\href}[2]{#2}\begingroup\raggedright\endgroup


\begin{thebibliography}{100}

\bibitem{snowmass}
X.~Bekaert, N.~Boulanger, A.~Campoleoni, M.~Chiodaroli, D.~Francia,
  M.~Grigoriev, E.~Sezgin, and E.~Skvortsov, {\itshape {Snowmass White Paper:
  Higher Spin Gravity and Higher Spin Symmetry}},
  \href{http://arxiv.org/abs/2205.01567}{{\ttfamily arXiv:2205.01567}}.

\bibitem{Misha}
M.~A. Vasiliev, {\itshape {Holography, Unfolding and Higher-Spin Theory}},
  {\em J. Phys. A} {\bfseries 46} (2013) 214013,
  [\href{http://arxiv.org/abs/1203.5554}{{\ttfamily arXiv:1203.5554}}].

\bibitem{Arias:2015wha}
C.~Arias, N.~Boulanger, P.~Sundell, and A.~Torres-Gomez, {\itshape {2D sigma
  models and differential Poisson algebras}},  {\em JHEP} {\bfseries 08} (2015)
  095, [\href{http://arxiv.org/abs/1503.05625}{{\ttfamily arXiv:1503.05625}}].

\bibitem{Bonezzi:2015lfa}
R.~Bonezzi, P.~Sundell, and A.~Torres-Gomez, {\itshape {2D Poisson Sigma Models
  with Gauged Vectorial Supersymmetry}},  {\em JHEP} {\bfseries 08} (2015) 047,
  [\href{http://arxiv.org/abs/1505.04959}{{\ttfamily arXiv:1505.04959}}].

\bibitem{Sharapov:2023erv}
A.~Sharapov, E.~Skvortsov, and R.~Van~Dongen, {\itshape {Strong Homotopy
  Algebras for Higher Spin Gravity via Stokes Theorem}},
  \href{http://arxiv.org/abs/2312.16573}{{\ttfamily arXiv:2312.16573}}.

\bibitem{Bengtsson1}
A.~Bengtsson, {\em {Higher Spin Field Theory. Volume 1: Free Theory}}.
\newblock Texts and Monographs in Theoretical Physics. De Gruyter, 9, 2023.

\bibitem{Bengtsson2}
A.~Bengtsson, {\em {Higher Spin Field Theory. Volume 2: Interactions}}.
\newblock Texts and Monographs in Theoretical Physics. De Gruyter, 9, 2023.

\bibitem{Bryant}
R.~L. Bryant, S.~Chern, R.~B. Gardner, H.~L. Goldschmidt, and P.~A. Griffiths,
  {\em {Exterior differential systems}}.
\newblock Springer New York, NY, 1991.

\bibitem{DAuria:1982uck}
R.~D'Auria and P.~Fre, {\itshape {Geometric Supergravity in d = 11 and Its
  Hidden Supergroup}},  {\em Nucl. Phys. B} {\bfseries 201} (1982) 101--140.
  [Erratum: Nucl.Phys.B 206, 496 (1982)].

\bibitem{vanNieuwenhuizen:1982zf}
P.~van Nieuwenhuizen, {\itshape {FREE GRADED DIFFERENTIAL SUPERALGEBRAS}},  in
  {\em {11th International Colloquium on Group Theoretical Methods in
  Physics}}, pp.~228--247, 12, 1982.

\bibitem{DAuria:1982mkx}
R.~D'Auria, P.~Fre, P.~K. Townsend, and P.~van Nieuwenhuizen, {\itshape
  {Invariance of Actions, Rheonomy and the New Minimal $N=1$ Supergravity in
  the Group Manifold Approach}},  {\em Annals Phys.} {\bfseries 155} (1984)
  423.

\bibitem{Vasiliev:1988xc}
M.~A. Vasiliev, {\itshape {Equations of Motion of Interacting Massless Fields
  of All Spins as a Free Differential Algebra}},  {\em Phys. Lett. B}
  {\bfseries 209} (1988) 491--497.

\bibitem{MV}
M.~A. Vasiliev, {\itshape {Consistent Equations for Interacting Massless Fields
  of All Spins in the First Order in Curvatures}},  {\em Annals Phys.}
  {\bfseries 190} (1989) 59--106.

\bibitem{Vasiliev:1992gr}
M.~A. Vasiliev, {\itshape {Unfolded representation for relativistic equations
  in (2+1) anti-De Sitter space}},  {\em Class. Quant. Grav.} {\bfseries 11}
  (1994) 649--664.

\bibitem{Vasiliev:2005zu}
M.~A. Vasiliev, {\itshape {Actions, charges and off-shell fields in the
  unfolded dynamics approach}},  {\em Int. J. Geom. Meth. Mod. Phys.}
  {\bfseries 3} (2006) 37--80,
  [\href{http://arxiv.org/abs/hep-th/0504090}{{\ttfamily hep-th/0504090}}].

\bibitem{review99}
M.~A. Vasiliev, {\itshape {Higher spin gauge theories: Star product and AdS
  space}},  \href{http://arxiv.org/abs/hep-th/9910096}{{\ttfamily
  hep-th/9910096}}.

\bibitem{Bekaert:2005vh}
X.~Bekaert, S.~Cnockaert, C.~Iazeolla, and M.~A. Vasiliev, {\itshape {Nonlinear
  higher spin theories in various dimensions}},  in {\em {Higher spin gauge
  theories: Proceedings, 1st Solvay Workshop: Brussels, Belgium, 12-14 May,
  2004}}, pp.~132--197, 2004.
\newblock \href{http://arxiv.org/abs/hep-th/0503128}{{\ttfamily
  hep-th/0503128}}.

\bibitem{Fronsdal:1978rb}
C.~Fronsdal, {\itshape {Massless Fields with Integer Spin}},  {\em Phys. Rev.
  D} {\bfseries 18} (1978) 3624.

\bibitem{Fronsdal:1978vb}
C.~Fronsdal, {\itshape {Singletons and Massless, Integral Spin Fields on de
  Sitter Space (Elementary Particles in a Curved Space. 7.}},  {\em Phys. Rev.
  D} {\bfseries 20} (1979) 848--856.

\bibitem{Bekaert:2015tva}
X.~Bekaert, J.~Erdmenger, D.~Ponomarev, and C.~Sleight, {\itshape {Quartic AdS
  Interactions in Higher-Spin Gravity from Conformal Field Theory}},  {\em
  JHEP} {\bfseries 11} (2015) 149,
  [\href{http://arxiv.org/abs/1508.04292}{{\ttfamily arXiv:1508.04292}}].

\bibitem{Sleight:2017pcz}
C.~Sleight and M.~Taronna, {\itshape {Higher-Spin Gauge Theories and Bulk
  Locality}},  {\em Phys. Rev. Lett.} {\bfseries 121} (2018), no.~17 171604,
  [\href{http://arxiv.org/abs/1704.07859}{{\ttfamily arXiv:1704.07859}}].

\bibitem{Vasiliev90}
M.~A. Vasiliev, {\itshape {Consistent equation for interacting gauge fields of
  all spins in (3+1)-dimensions}},  {\em Phys. Lett. B} {\bfseries 243} (1990)
  378--382.

\bibitem{properties}
M.~A. Vasiliev, {\itshape {Properties of equations of motion of interacting
  gauge fields of all spins in (3+1)-dimensions}},  {\em Class. Quant. Grav.}
  {\bfseries 8} (1991) 1387--1417.

\bibitem{more}
M.~A. Vasiliev, {\itshape {More on equations of motion for interacting massless
  fields of all spins in (3+1)-dimensions}},  {\em Phys. Lett. B} {\bfseries
  285} (1992) 225--234.

\bibitem{Vasiliev03}
M.~A. Vasiliev, {\itshape {Nonlinear equations for symmetric massless higher
  spin fields in (A)dS(d)}},  {\em Phys. Lett. B} {\bfseries 567} (2003)
  139--151, [\href{http://arxiv.org/abs/hep-th/0304049}{{\ttfamily
  hep-th/0304049}}].

\bibitem{Didenko:2014dwa}
V.~E. Didenko and E.~D. Skvortsov, {\itshape {Elements of Vasiliev theory}},
  \href{http://arxiv.org/abs/1401.2975}{{\ttfamily arXiv:1401.2975}}.

\bibitem{Vasiliev:2017cae}
M.~A. Vasiliev, {\itshape {On the Local Frame in Nonlinear Higher-Spin
  Equations}},  {\em JHEP} {\bfseries 01} (2018) 062,
  [\href{http://arxiv.org/abs/1707.03735}{{\ttfamily arXiv:1707.03735}}].

\bibitem{Gelfond:2018vmi}
O.~A. Gelfond and M.~A. Vasiliev, {\itshape {Homotopy Operators and Locality
  Theorems in Higher-Spin Equations}},  {\em Phys. Lett. B} {\bfseries 786}
  (2018) 180--188, [\href{http://arxiv.org/abs/1805.11941}{{\ttfamily
  arXiv:1805.11941}}].

\bibitem{Didenko:2018fgx}
V.~E. Didenko, O.~A. Gelfond, A.~V. Korybut, and M.~A. Vasiliev, {\itshape
  {Homotopy Properties and Lower-Order Vertices in Higher-Spin Equations}},
  {\em J. Phys. A} {\bfseries 51} (2018), no.~46 465202,
  [\href{http://arxiv.org/abs/1807.00001}{{\ttfamily arXiv:1807.00001}}].

\bibitem{Didenko:2019xzz}
V.~E. Didenko, O.~A. Gelfond, A.~V. Korybut, and M.~A. Vasiliev, {\itshape
  {Limiting Shifted Homotopy in Higher-Spin Theory and Spin-Locality}},  {\em
  JHEP} {\bfseries 12} (2019) 086,
  [\href{http://arxiv.org/abs/1909.04876}{{\ttfamily arXiv:1909.04876}}].

\bibitem{Gelfond:2019tac}
O.~A. Gelfond and M.~A. Vasiliev, {\itshape {Spin-Locality of Higher-Spin
  Theories and Star-Product Functional Classes}},  {\em JHEP} {\bfseries 03}
  (2020) 002, [\href{http://arxiv.org/abs/1910.00487}{{\ttfamily
  arXiv:1910.00487}}].

\bibitem{Didenko:2020bxd}
V.~E. Didenko, O.~A. Gelfond, A.~V. Korybut, and M.~A. Vasiliev, {\itshape
  {Spin-locality of $\eta^{2}$ and $ {\overline{\eta}}^2 $ quartic higher-spin
  vertices}},  {\em JHEP} {\bfseries 12} (2020) 184,
  [\href{http://arxiv.org/abs/2009.02811}{{\ttfamily arXiv:2009.02811}}].

\bibitem{Vasiliev:2022med}
M.~A. Vasiliev, {\itshape {Projectively-compact spinor vertices and space-time
  spin-locality in higher-spin theory}},  {\em Phys. Lett. B} {\bfseries 834}
  (2022) 137401, [\href{http://arxiv.org/abs/2208.02004}{{\ttfamily
  arXiv:2208.02004}}].

\bibitem{Kontsevich:1997vb}
M.~Kontsevich, {\itshape {Deformation quantization of Poisson manifolds. 1.}},
  {\em Lett. Math. Phys.} {\bfseries 66} (2003) 157--216,
  [\href{http://arxiv.org/abs/q-alg/9709040}{{\ttfamily q-alg/9709040}}].

\bibitem{Cattaneo:1999fm}
A.~S. Cattaneo and G.~Felder, {\itshape {A Path integral approach to the
  Kontsevich quantization formula}},  {\em Commun. Math. Phys.} {\bfseries 212}
  (2000) 591--611, [\href{http://arxiv.org/abs/math/9902090}{{\ttfamily
  math/9902090}}].

\bibitem{Alexandrov:1995kv}
M.~Alexandrov, A.~Schwarz, O.~Zaboronsky, and M.~Kontsevich, {\itshape {The
  Geometry of the master equation and topological quantum field theory}},  {\em
  Int. J. Mod. Phys. A} {\bfseries 12} (1997) 1405--1429,
  [\href{http://arxiv.org/abs/hep-th/9502010}{{\ttfamily hep-th/9502010}}].

\bibitem{Grigoriev}
M.~Grigoriev, {\itshape {Off-shell gauge fields from BRST quantization}},
  \href{http://arxiv.org/abs/hep-th/0605089}{{\ttfamily hep-th/0605089}}.

\bibitem{Barnich:2006pc}
G.~Barnich and M.~Grigoriev, {\itshape {Parent form for higher spin fields on
  anti-de Sitter space}},  {\em JHEP} {\bfseries 08} (2006) 013,
  [\href{http://arxiv.org/abs/hep-th/0602166}{{\ttfamily hep-th/0602166}}].

\bibitem{BarnichGrigoriev}
G.~Barnich and M.~Grigoriev, {\itshape {A Poincare lemma for sigma models of
  AKSZ type}},  {\em J. Geom. Phys.} {\bfseries 61} (2011) 663--674,
  [\href{http://arxiv.org/abs/0905.0547}{{\ttfamily arXiv:0905.0547}}].

\bibitem{Sezgin:2011hq}
E.~Sezgin and P.~Sundell, {\itshape {Geometry and Observables in Vasiliev's
  Higher Spin Gravity}},  {\em JHEP} {\bfseries 07} (2012) 121,
  [\href{http://arxiv.org/abs/1103.2360}{{\ttfamily arXiv:1103.2360}}].

\bibitem{Boulanger:2011dd}
N.~Boulanger and P.~Sundell, {\itshape {An action principle for Vasiliev's
  four-dimensional higher-spin gravity}},  {\em J. Phys. A} {\bfseries 44}
  (2011) 495402, [\href{http://arxiv.org/abs/1102.2219}{{\ttfamily
  arXiv:1102.2219}}].

\bibitem{Boulanger:2015kfa}
N.~Boulanger, E.~Sezgin, and P.~Sundell, {\itshape {4D Higher Spin Gravity with
  Dynamical Two-Form as a Frobenius-Chern-Simons Gauge Theory}},
  \href{http://arxiv.org/abs/1505.04957}{{\ttfamily arXiv:1505.04957}}.

\bibitem{Bonezzi:2016ttk}
R.~Bonezzi, N.~Boulanger, E.~Sezgin, and P.~Sundell, {\itshape
  {Frobenius\textendash{}Chern\textendash{}Simons gauge theory}},  {\em J.
  Phys. A} {\bfseries 50} (2017), no.~5 055401,
  [\href{http://arxiv.org/abs/1607.00726}{{\ttfamily arXiv:1607.00726}}].

\bibitem{Colombo:2010fu}
N.~Colombo and P.~Sundell, {\itshape {Twistor space observables and
  quasi-amplitudes in 4D higher spin gravity}},  {\em JHEP} {\bfseries 11}
  (2011) 042, [\href{http://arxiv.org/abs/1012.0813}{{\ttfamily
  arXiv:1012.0813}}].

\bibitem{Colombo:2012jx}
N.~Colombo and P.~Sundell, {\itshape {Higher Spin Gravity Amplitudes From
  Zero-form Charges}},  \href{http://arxiv.org/abs/1208.3880}{{\ttfamily
  arXiv:1208.3880}}.

\bibitem{Didenko:2012tv}
V.~E. Didenko and E.~D. Skvortsov, {\itshape {Exact higher-spin symmetry in
  CFT: all correlators in unbroken Vasiliev theory}},  {\em JHEP} {\bfseries
  04} (2013) 158, [\href{http://arxiv.org/abs/1210.7963}{{\ttfamily
  arXiv:1210.7963}}].

\bibitem{Bonezzi:2017vha}
R.~Bonezzi, N.~Boulanger, D.~De~Filippi, and P.~Sundell, {\itshape
  {Noncommutative Wilson lines in higher-spin theory and correlation functions
  of conserved currents for free conformal fields}},  {\em J. Phys. A}
  {\bfseries 50} (2017), no.~47 475401,
  [\href{http://arxiv.org/abs/1705.03928}{{\ttfamily arXiv:1705.03928}}].

\bibitem{COMST}
D.~De~Filippi, C.~Iazeolla, and P.~Sundell, {\itshape {Fronsdal fields from
  gauge functions in Vasiliev\textquoteright{}s higher spin gravity}},  {\em
  JHEP} {\bfseries 10} (2019) 215,
  [\href{http://arxiv.org/abs/1905.06325}{{\ttfamily arXiv:1905.06325}}].

\bibitem{Sleight:2016dba}
C.~Sleight and M.~Taronna, {\itshape {Higher Spin Interactions from Conformal
  Field Theory: The Complete Cubic Couplings}},  {\em Phys. Rev. Lett.}
  {\bfseries 116} (2016), no.~18 181602,
  [\href{http://arxiv.org/abs/1603.00022}{{\ttfamily arXiv:1603.00022}}].

\bibitem{Douglas:2010rc}
M.~R. Douglas, L.~Mazzucato, and S.~S. Razamat, {\itshape {Holographic dual of
  free field theory}},  {\em Phys. Rev. D} {\bfseries 83} (2011) 071701,
  [\href{http://arxiv.org/abs/1011.4926}{{\ttfamily arXiv:1011.4926}}].

\bibitem{Jevicki:2011ss}
A.~Jevicki, K.~Jin, and Q.~Ye, {\itshape {Collective Dipole Model of AdS/CFT
  and Higher Spin Gravity}},  {\em J. Phys. A} {\bfseries 44} (2011) 465402,
  [\href{http://arxiv.org/abs/1106.3983}{{\ttfamily arXiv:1106.3983}}].

\bibitem{bilocal2}
R.~de~Mello~Koch, A.~Jevicki, J.~a.~P. Rodrigues, and J.~Yoon, {\itshape
  {Canonical Formulation of $O(N)$ Vector/Higher Spin Correspondence}},  {\em
  J. Phys. A} {\bfseries 48} (2015), no.~10 105403,
  [\href{http://arxiv.org/abs/1408.4800}{{\ttfamily arXiv:1408.4800}}].

\bibitem{bilocalrecent}
R.~de~Mello~Koch, A.~Jevicki, K.~Suzuki, and J.~Yoon, {\itshape {AdS Maps and
  Diagrams of Bi-local Holography}},  {\em JHEP} {\bfseries 03} (2019) 133,
  [\href{http://arxiv.org/abs/1810.02332}{{\ttfamily arXiv:1810.02332}}].

\bibitem{Aharony:2020omh}
O.~Aharony, S.~M. Chester, and E.~Y. Urbach, {\itshape {A Derivation of AdS/CFT
  for Vector Models}},  {\em JHEP} {\bfseries 03} (2021) 208,
  [\href{http://arxiv.org/abs/2011.06328}{{\ttfamily arXiv:2011.06328}}].

\bibitem{Neiman:2023orj}
Y.~Neiman, {\itshape {Quartic locality of higher-spin gravity in de Sitter and
  Euclidean anti-de Sitter space}},  {\em Phys. Lett. B} {\bfseries 843} (2023)
  138048, [\href{http://arxiv.org/abs/2302.00852}{{\ttfamily
  arXiv:2302.00852}}].

\bibitem{formal}
A.~Sharapov and E.~Skvortsov, {\itshape {Formal Higher Spin Gravities}},  {\em
  Nucl. Phys. B} {\bfseries 941} (2019) 838--860,
  [\href{http://arxiv.org/abs/1901.01426}{{\ttfamily arXiv:1901.01426}}].

\bibitem{Metsaev}
R.~R. Metsaev, {\itshape {Poincare invariant dynamics of massless higher spins:
  Fourth order analysis on mass shell}},  {\em Mod. Phys. Lett. A} {\bfseries
  6} (1991) 359--367.

\bibitem{Ponomarev:2016lrm}
D.~Ponomarev and E.~D. Skvortsov, {\itshape {Light-Front Higher-Spin Theories
  in Flat Space}},  {\em J. Phys. A} {\bfseries 50} (2017), no.~9 095401,
  [\href{http://arxiv.org/abs/1609.04655}{{\ttfamily arXiv:1609.04655}}].

\bibitem{Sharapov:2022faa}
A.~Sharapov, E.~Skvortsov, A.~Sukhanov, and R.~Van~Dongen, {\itshape {Minimal
  model of Chiral Higher Spin Gravity}},  {\em JHEP} {\bfseries 09} (2022) 134,
  [\href{http://arxiv.org/abs/2205.07794}{{\ttfamily arXiv:2205.07794}}].
  [Erratum: JHEP 02, 183 (2023)].

\bibitem{Sharapov:2022awp}
A.~Sharapov and E.~Skvortsov, {\itshape {Chiral higher spin gravity in (A)dS4
  and secrets of Chern\textendash{}Simons matter theories}},  {\em Nucl. Phys.
  B} {\bfseries 985} (2022) 115982,
  [\href{http://arxiv.org/abs/2205.15293}{{\ttfamily arXiv:2205.15293}}].

\bibitem{Didenko:2022qga}
V.~E. Didenko, {\itshape {On holomorphic sector of higher-spin theory}},  {\em
  JHEP} {\bfseries 10} (2022) 191,
  [\href{http://arxiv.org/abs/2209.01966}{{\ttfamily arXiv:2209.01966}}].

\bibitem{Bergshoeff:1988jm}
E.~Bergshoeff, A.~Salam, E.~Sezgin, and Y.~Tanii, {\itshape {Singletons, Higher
  Spin Massless States and the Supermembrane}},  {\em Phys. Lett. B} {\bfseries
  205} (1988) 237--244.

\bibitem{Sezgin1998}
E.~Sezgin and P.~Sundell, {\itshape {Higher spin N=8 supergravity}},  {\em
  JHEP} {\bfseries 11} (1998) 016,
  [\href{http://arxiv.org/abs/hep-th/9805125}{{\ttfamily hep-th/9805125}}].

\bibitem{Sundborg:2000wp}
B.~Sundborg, {\itshape {Stringy gravity, interacting tensionless strings and
  massless higher spins}},  {\em Nucl. Phys. B Proc. Suppl.} {\bfseries 102}
  (2001) 113--119, [\href{http://arxiv.org/abs/hep-th/0103247}{{\ttfamily
  hep-th/0103247}}].

\bibitem{WittenJHS60}
E.~Witten, {\itshape {Talk given at J.H. Schwarz’ 60th Birthday Conference}},
   {Nov 2-3, 2001}.
\newblock Caltech, http://theory.caltech.edu/jhs60/witten/1.html.

\bibitem{Sezgin:2002rt}
E.~Sezgin and P.~Sundell, {\itshape {Massless higher spins and holography}},
  {\em Nucl. Phys. B} {\bfseries 644} (2002) 303--370,
  [\href{http://arxiv.org/abs/hep-th/0205131}{{\ttfamily hep-th/0205131}}].
  [Erratum: Nucl.Phys.B 660, 403--403 (2003)].

\bibitem{Henneaux:2010xg}
M.~Henneaux and S.-J. Rey, {\itshape {Nonlinear $W_{infinity}$ as Asymptotic
  Symmetry of Three-Dimensional Higher Spin Anti-de Sitter Gravity}},  {\em
  JHEP} {\bfseries 12} (2010) 007,
  [\href{http://arxiv.org/abs/1008.4579}{{\ttfamily arXiv:1008.4579}}].

\bibitem{Campoleoni:2010zq}
A.~Campoleoni, S.~Fredenhagen, S.~Pfenninger, and S.~Theisen, {\itshape
  {Asymptotic symmetries of three-dimensional gravity coupled to higher-spin
  fields}},  {\em JHEP} {\bfseries 11} (2010) 007,
  [\href{http://arxiv.org/abs/1008.4744}{{\ttfamily arXiv:1008.4744}}].

\bibitem{Gaberdiel:2010pz}
M.~R. Gaberdiel and R.~Gopakumar, {\itshape {An AdS$_{3}$ Dual for Minimal
  Model CFTs}},  {\em Phys. Rev. D} {\bfseries 83} (2011) 066007,
  [\href{http://arxiv.org/abs/1011.2986}{{\ttfamily arXiv:1011.2986}}].

\bibitem{Gaberdiel:2011wb}
M.~R. Gaberdiel and T.~Hartman, {\itshape {Symmetries of Holographic Minimal
  Models}},  {\em JHEP} {\bfseries 05} (2011) 031,
  [\href{http://arxiv.org/abs/1101.2910}{{\ttfamily arXiv:1101.2910}}].

\bibitem{Gaberdiel:2011zw}
M.~R. Gaberdiel, R.~Gopakumar, T.~Hartman, and S.~Raju, {\itshape {Partition
  Functions of Holographic Minimal Models}},  {\em JHEP} {\bfseries 08} (2011)
  077, [\href{http://arxiv.org/abs/1106.1897}{{\ttfamily arXiv:1106.1897}}].

\bibitem{Chang:2011mz}
C.-M. Chang and X.~Yin, {\itshape {Higher Spin Gravity with Matter in $AdS_3$
  and Its CFT Dual}},  {\em JHEP} {\bfseries 10} (2012) 024,
  [\href{http://arxiv.org/abs/1106.2580}{{\ttfamily arXiv:1106.2580}}].

\bibitem{Kraus:2011ds}
P.~Kraus and E.~Perlmutter, {\itshape {Partition functions of higher spin black
  holes and their CFT duals}},  {\em JHEP} {\bfseries 11} (2011) 061,
  [\href{http://arxiv.org/abs/1108.2567}{{\ttfamily arXiv:1108.2567}}].

\bibitem{Ammon:2011ua}
M.~Ammon, P.~Kraus, and E.~Perlmutter, {\itshape {Scalar fields and three-point
  functions in D=3 higher spin gravity}},  {\em JHEP} {\bfseries 07} (2012)
  113, [\href{http://arxiv.org/abs/1111.3926}{{\ttfamily arXiv:1111.3926}}].

\bibitem{Perlmutter:2012ds}
E.~Perlmutter, T.~Prochazka, and J.~Raeymaekers, {\itshape {The semiclassical
  limit of $W_N$ CFTs and Vasiliev theory}},  {\em JHEP} {\bfseries 05} (2013)
  007, [\href{http://arxiv.org/abs/1210.8452}{{\ttfamily arXiv:1210.8452}}].

\bibitem{Campoleoni:2013lma}
A.~Campoleoni, T.~Prochazka, and J.~Raeymaekers, {\itshape {A note on conical
  solutions in 3D Vasiliev theory}},  {\em JHEP} {\bfseries 05} (2013) 052,
  [\href{http://arxiv.org/abs/1303.0880}{{\ttfamily arXiv:1303.0880}}].

\bibitem{Iazeolla:2015tca}
C.~Iazeolla and J.~Raeymaekers, {\itshape {On big crunch solutions in
  Prokushkin-Vasiliev theory}},  {\em JHEP} {\bfseries 01} (2016) 177,
  [\href{http://arxiv.org/abs/1510.08835}{{\ttfamily arXiv:1510.08835}}].

\bibitem{KP}
I.~R. Klebanov and A.~M. Polyakov, {\itshape {AdS dual of the critical O(N)
  vector model}},  {\em Phys. Lett. B} {\bfseries 550} (2002) 213--219,
  [\href{http://arxiv.org/abs/hep-th/0210114}{{\ttfamily hep-th/0210114}}].

\bibitem{LeighPetkou}
R.~G. Leigh and A.~C. Petkou, {\itshape {Holography of the N=1 higher spin
  theory on AdS(4)}},  {\em JHEP} {\bfseries 06} (2003) 011,
  [\href{http://arxiv.org/abs/hep-th/0304217}{{\ttfamily hep-th/0304217}}].

\bibitem{Sezgin:2003pt}
E.~Sezgin and P.~Sundell, {\itshape {Holography in 4D (super) higher spin
  theories and a test via cubic scalar couplings}},  {\em JHEP} {\bfseries 07}
  (2005) 044, [\href{http://arxiv.org/abs/hep-th/0305040}{{\ttfamily
  hep-th/0305040}}].

\bibitem{GiombiYin}
S.~Giombi and X.~Yin, {\itshape {Higher Spin Gauge Theory and Holography: The
  Three-Point Functions}},  {\em JHEP} {\bfseries 09} (2010) 115,
  [\href{http://arxiv.org/abs/0912.3462}{{\ttfamily arXiv:0912.3462}}].

\bibitem{GiombiYin2}
S.~Giombi and X.~Yin, {\itshape {Higher Spins in AdS and Twistorial
  Holography}},  {\em JHEP} {\bfseries 04} (2011) 086,
  [\href{http://arxiv.org/abs/1004.3736}{{\ttfamily arXiv:1004.3736}}].

\bibitem{Boulanger:2015ova}
N.~Boulanger, P.~Kessel, E.~D. Skvortsov, and M.~Taronna, {\itshape {Higher
  spin interactions in four-dimensions: Vasiliev versus Fronsdal}},  {\em J.
  Phys. A} {\bfseries 49} (2016), no.~9 095402,
  [\href{http://arxiv.org/abs/1508.04139}{{\ttfamily arXiv:1508.04139}}].

\bibitem{Didenko:2017lsn}
V.~E. Didenko and M.~A. Vasiliev, {\itshape {Test of the local form of
  higher-spin equations via AdS / CFT}},  {\em Phys. Lett. B} {\bfseries 775}
  (2017) 352--360, [\href{http://arxiv.org/abs/1705.03440}{{\ttfamily
  arXiv:1705.03440}}].

\bibitem{Sezgin:2017jgm}
E.~Sezgin, E.~D. Skvortsov, and Y.~Zhu, {\itshape {Chern-Simons Matter Theories
  and Higher Spin Gravity}},  {\em JHEP} {\bfseries 07} (2017) 133,
  [\href{http://arxiv.org/abs/1705.03197}{{\ttfamily arXiv:1705.03197}}].

\bibitem{Skvortsov:2018uru}
E.~Skvortsov, {\itshape {Light-Front Bootstrap for Chern-Simons Matter
  Theories}},  {\em JHEP} {\bfseries 06} (2019) 058,
  [\href{http://arxiv.org/abs/1811.12333}{{\ttfamily arXiv:1811.12333}}].

\bibitem{MZh}
J.~Maldacena and A.~Zhiboedov, {\itshape {Constraining Conformal Field Theories
  with A Higher Spin Symmetry}},  {\em J. Phys. A} {\bfseries 46} (2013)
  214011, [\href{http://arxiv.org/abs/1112.1016}{{\ttfamily arXiv:1112.1016}}].

\bibitem{Giombi:2011ya}
S.~Giombi and X.~Yin, {\itshape {On Higher Spin Gauge Theory and the Critical
  O(N) Model}},  {\em Phys. Rev. D} {\bfseries 85} (2012) 086005,
  [\href{http://arxiv.org/abs/1105.4011}{{\ttfamily arXiv:1105.4011}}].

\bibitem{Mzh2}
J.~Maldacena and A.~Zhiboedov, {\itshape {Constraining conformal field theories
  with a slightly broken higher spin symmetry}},  {\em Class. Quant. Grav.}
  {\bfseries 30} (2013) 104003,
  [\href{http://arxiv.org/abs/1204.3882}{{\ttfamily arXiv:1204.3882}}].

\bibitem{BTZ}
R.~Aros, C.~Iazeolla, P.~Sundell, and Y.~Yin, {\itshape {Higher spin
  fluctuations on spinless 4D BTZ black hole}},  {\em JHEP} {\bfseries 08}
  (2019) 171, [\href{http://arxiv.org/abs/1903.01399}{{\ttfamily
  arXiv:1903.01399}}].

\bibitem{paperII}
F.~Diaz, C.~Iazeolla, and P.~Sundell, {\itshape {Fractional Spin, Unfolding,
  and Holography: II. 4D Higher Spin Gravity and 3D Conformal Dual}},
  \href{http://arxiv.org/abs/2403.02301}{{\ttfamily arXiv:2403.02301}}.

\bibitem{Didenko:2021vui}
V.~E. Didenko and A.~V. Korybut, {\itshape {Planar solutions of higher-spin
  theory. Part I. Free field level}},  {\em JHEP} {\bfseries 08} (2021) 144,
  [\href{http://arxiv.org/abs/2105.09021}{{\ttfamily arXiv:2105.09021}}].

\bibitem{corfu19}
C.~Iazeolla, {\itshape {On boundary conditions and spacetime/fibre duality in
  Vasiliev's higher-spin gravity}},  {\em PoS} {\bfseries CORFU2019} (2020)
  181, [\href{http://arxiv.org/abs/2004.14903}{{\ttfamily arXiv:2004.14903}}].

\bibitem{bdhm}
T.~Banks, M.~R. Douglas, G.~T. Horowitz, and E.~J. Martinec, {\itshape {AdS
  dynamics from conformal field theory}},
  \href{http://arxiv.org/abs/hep-th/9808016}{{\ttfamily hep-th/9808016}}.

\bibitem{svr}
K.~Skenderis and B.~C. van Rees, {\itshape {Real-time gauge/gravity duality}},
  {\em Phys. Rev. Lett.} {\bfseries 101} (2008) 081601,
  [\href{http://arxiv.org/abs/0805.0150}{{\ttfamily arXiv:0805.0150}}].

\bibitem{svr2}
K.~Skenderis and B.~C. van Rees, {\itshape {Real-time gauge/gravity duality:
  Prescription, Renormalization and Examples}},  {\em JHEP} {\bfseries 05}
  (2009) 085, [\href{http://arxiv.org/abs/0812.2909}{{\ttfamily
  arXiv:0812.2909}}].

\bibitem{silva}
M.~Botta-Cantcheff, P.~Mart\'\i{}nez, and G.~A. Silva, {\itshape {On excited
  states in real-time AdS/CFT}},  {\em JHEP} {\bfseries 02} (2016) 171,
  [\href{http://arxiv.org/abs/1512.07850}{{\ttfamily arXiv:1512.07850}}].

\bibitem{celestial}
A.~Strominger, {\em {Lectures on the Infrared Structure of Gravity and Gauge
  Theory}}.
\newblock 3, 2017.

\bibitem{Neiman:2022enh}
Y.~Neiman, {\itshape {New Diagrammatic Framework for Higher-Spin Gravity}},
  {\em Phys. Rev. Lett.} {\bfseries 130} (2023), no.~17 171601,
  [\href{http://arxiv.org/abs/2209.02185}{{\ttfamily arXiv:2209.02185}}].

\bibitem{Bekaert:2012vt}
X.~Bekaert and M.~Grigoriev, {\itshape {Notes on the ambient approach to
  boundary values of AdS gauge fields}},  {\em J. Phys. A} {\bfseries 46}
  (2013) 214008, [\href{http://arxiv.org/abs/1207.3439}{{\ttfamily
  arXiv:1207.3439}}].

\bibitem{Bekaert:2013zya}
X.~Bekaert and M.~Grigoriev, {\itshape {Higher order singletons, partially
  massless fields and their boundary values in the ambient approach}},  {\em
  Nucl. Phys. B} {\bfseries 876} (2013) 667--714,
  [\href{http://arxiv.org/abs/1305.0162}{{\ttfamily arXiv:1305.0162}}].

\bibitem{Bekaert:2017bpy}
X.~Bekaert, M.~Grigoriev, and E.~D. Skvortsov, {\itshape {Higher Spin Extension
  of Fefferman-Graham Construction}},  {\em Universe} {\bfseries 4} (2018),
  no.~2 17, [\href{http://arxiv.org/abs/1710.11463}{{\ttfamily
  arXiv:1710.11463}}].

\bibitem{LP2}
R.~G. Leigh and A.~C. Petkou, {\itshape {SL(2,Z) action on three-dimensional
  CFTs and holography}},  {\em JHEP} {\bfseries 12} (2003) 020,
  [\href{http://arxiv.org/abs/hep-th/0309177}{{\ttfamily hep-th/0309177}}].

\bibitem{Giombi:2013yva}
S.~Giombi, I.~R. Klebanov, S.~S. Pufu, B.~R. Safdi, and G.~Tarnopolsky,
  {\itshape {AdS Description of Induced Higher-Spin Gauge Theory}},  {\em JHEP}
  {\bfseries 10} (2013) 016, [\href{http://arxiv.org/abs/1306.5242}{{\ttfamily
  arXiv:1306.5242}}].

\bibitem{Pope:1989vj}
C.~N. Pope and P.~K. Townsend, {\itshape {Conformal Higher Spin in
  (2+1)-dimensions}},  {\em Phys. Lett. B} {\bfseries 225} (1989) 245--250.

\bibitem{F&L}
E.~S. Fradkin and V.~Y. Linetsky, {\itshape {A Superconformal Theory of
  Massless Higher Spin Fields in $D$ = (2+1)}},  {\em Mod. Phys. Lett. A}
  {\bfseries 4} (1989) 731.

\bibitem{Grigoriev:2019xmp}
M.~Grigoriev, I.~Lovrekovic, and E.~Skvortsov, {\itshape {New Conformal Higher
  Spin Gravities in $3d$}},  {\em JHEP} {\bfseries 01} (2020) 059,
  [\href{http://arxiv.org/abs/1909.13305}{{\ttfamily arXiv:1909.13305}}].

\bibitem{Segal}
A.~Y. Segal, {\itshape {Conformal higher spin theory}},  {\em Nucl. Phys. B}
  {\bfseries 664} (2003) 59--130,
  [\href{http://arxiv.org/abs/hep-th/0207212}{{\ttfamily hep-th/0207212}}].

\bibitem{Tseytlin:2002gz}
A.~A. Tseytlin, {\itshape {On limits of superstring in AdS(5) x S**5}},  {\em
  Theor. Math. Phys.} {\bfseries 133} (2002) 1376--1389,
  [\href{http://arxiv.org/abs/hep-th/0201112}{{\ttfamily hep-th/0201112}}].

\bibitem{Nilsson:2013tva}
B.~E.~W. Nilsson, {\itshape {Towards an exact frame formulation of conformal
  higher spins in three dimensions}},  {\em JHEP} {\bfseries 09} (2015) 078,
  [\href{http://arxiv.org/abs/1312.5883}{{\ttfamily arXiv:1312.5883}}].

\bibitem{Nilsson}
B.~E.~W. Nilsson, {\itshape {On the conformal higher spin unfolded equation for
  a three-dimensional self-interacting scalar field}},  {\em JHEP} {\bfseries
  08} (2016) 142, [\href{http://arxiv.org/abs/1506.03328}{{\ttfamily
  arXiv:1506.03328}}].

\bibitem{Boulanger:2015uha}
N.~Boulanger, P.~Sundell, and M.~Valenzuela, {\itshape {Gravitational and gauge
  couplings in Chern-Simons fractional spin gravity}},  {\em JHEP} {\bfseries
  01} (2016) 173, [\href{http://arxiv.org/abs/1504.04286}{{\ttfamily
  arXiv:1504.04286}}]. [Erratum: JHEP 03, 075 (2016)].

\bibitem{OLC}
F.~Diaz, C.~Iazeolla, and P.~Sundell. work in progress.

\bibitem{Boulanger:2013naa}
N.~Boulanger, P.~Sundell, and M.~Valenzuela, {\itshape {Three-dimensional
  fractional-spin gravity}},  {\em JHEP} {\bfseries 02} (2014) 052,
  [\href{http://arxiv.org/abs/1312.5700}{{\ttfamily arXiv:1312.5700}}].
  [Erratum: JHEP 03, 076 (2016)].

\bibitem{2017}
C.~Iazeolla and P.~Sundell, {\itshape {4D Higher Spin Black Holes with
  Nonlinear Scalar Fluctuations}},  {\em JHEP} {\bfseries 10} (2017) 130,
  [\href{http://arxiv.org/abs/1705.06713}{{\ttfamily arXiv:1705.06713}}].

\bibitem{Gaberdiel:1997ia}
M.~R. Gaberdiel and B.~Zwiebach, {\itshape {Tensor constructions of open string
  theories. 1: Foundations}},  {\em Nucl. Phys. B} {\bfseries 505} (1997)
  569--624, [\href{http://arxiv.org/abs/hep-th/9705038}{{\ttfamily
  hep-th/9705038}}].

\bibitem{Vasiliev:1990bu}
M.~A. Vasiliev, {\itshape {Algebraic aspects of the higher spin problem}},
  {\em Phys. Lett. B} {\bfseries 257} (1991) 111--118.

\bibitem{Sezgin:2005pv}
E.~Sezgin and P.~Sundell, {\itshape {An Exact solution of 4-D higher-spin gauge
  theory}},  {\em Nucl. Phys.} {\bfseries B762} (2007) 1--37,
  [\href{http://arxiv.org/abs/hep-th/0508158}{{\ttfamily hep-th/0508158}}].

\bibitem{2011}
C.~Iazeolla and P.~Sundell, {\itshape {Families of exact solutions to
  Vasiliev's 4D equations with spherical, cylindrical and biaxial symmetry}},
  {\em JHEP} {\bfseries 12} (2011) 084,
  [\href{http://arxiv.org/abs/1107.1217}{{\ttfamily arXiv:1107.1217}}].

\bibitem{Iazeolla:2022dal}
C.~Iazeolla and P.~Sundell, {\itshape {Unfolding, higher spins, metaplectic
  groups and resolution of classical singularities}},  {\em PoS} {\bfseries
  CORFU2021} (2022) 276, [\href{http://arxiv.org/abs/2205.00296}{{\ttfamily
  arXiv:2205.00296}}].

\bibitem{sullivan}
D.~Sullivan, {\itshape {Infinitesimal computations in topology}},  {\em
  Publications mathématiques de l’I.H.É.S.} {\bfseries 47} (1977) 269--331.

\bibitem{Iazeolla:2007wt}
C.~Iazeolla, E.~Sezgin, and P.~Sundell, {\itshape {Real forms of complex higher
  spin field equations and new exact solutions}},  {\em Nucl. Phys.} {\bfseries
  B791} (2008) 231--264, [\href{http://arxiv.org/abs/0706.2983}{{\ttfamily
  arXiv:0706.2983}}].

\bibitem{review}
C.~Iazeolla, E.~Sezgin, and P.~Sundell, {\itshape {On Exact Solutions and
  Perturbative Schemes in Higher Spin Theory}},  {\em Universe} {\bfseries 4}
  (2018), no.~1 5, [\href{http://arxiv.org/abs/1711.03550}{{\ttfamily
  arXiv:1711.03550}}].

\bibitem{cosmo}
R.~Aros, C.~Iazeolla, J.~Noreña, E.~Sezgin, P.~Sundell, and Y.~Yin, {\itshape
  {FRW and domain walls in higher spin gravity}},  {\em JHEP} {\bfseries 03}
  (2018) 153, [\href{http://arxiv.org/abs/1712.02401}{{\ttfamily
  arXiv:1712.02401}}].

\bibitem{paper0}
F.~Diaz, C.~Iazeolla, and P.~Sundell, {\itshape {Harmonic Expansions of 3D
  Conformal Scalars and the Holomorphic Metaplectic Group}}, . In preparation.

\bibitem{Konstein:1989ij}
S.~E. Konstein and M.~A. Vasiliev, {\itshape {Extended Higher Spin
  Superalgebras and Their Massless Representations}},  {\em Nucl. Phys. B}
  {\bfseries 331} (1990) 475--499.

\bibitem{Vasiliev:2004cm}
M.~A. Vasiliev, {\itshape {Higher spin superalgebras in any dimension and their
  representations}},  {\em JHEP} {\bfseries 12} (2004) 046,
  [\href{http://arxiv.org/abs/hep-th/0404124}{{\ttfamily hep-th/0404124}}].

\bibitem{fibre}
C.~Iazeolla and P.~Sundell, {\itshape {A Fiber Approach to Harmonic Analysis of
  Unfolded Higher-Spin Field Equations}},  {\em JHEP} {\bfseries 10} (2008)
  022, [\href{http://arxiv.org/abs/0806.1942}{{\ttfamily arXiv:0806.1942}}].

\bibitem{Engquist:2005yt}
J.~Engquist and P.~Sundell, {\itshape {Brane partons and singleton strings}},
  {\em Nucl. Phys. B} {\bfseries 752} (2006) 206--279,
  [\href{http://arxiv.org/abs/hep-th/0508124}{{\ttfamily hep-th/0508124}}].

\bibitem{Engquist:2007pr}
J.~Engquist, P.~Sundell, and L.~Tamassia, {\itshape {On Singleton Composites in
  Non-compact WZW Models}},  {\em JHEP} {\bfseries 02} (2007) 097,
  [\href{http://arxiv.org/abs/hep-th/0701051}{{\ttfamily hep-th/0701051}}].

\bibitem{Vasiliev:2018zer}
M.~A. Vasiliev, {\itshape {From Coxeter Higher-Spin Theories to Strings and
  Tensor Models}},  {\em JHEP} {\bfseries 08} (2018) 051,
  [\href{http://arxiv.org/abs/1804.06520}{{\ttfamily arXiv:1804.06520}}].

\bibitem{Leinaas:1977fm}
J.~M. Leinaas and J.~Myrheim, {\itshape {On the theory of identical
  particles}},  {\em Nuovo Cim. B} {\bfseries 37} (1977) 1--23.

\bibitem{Lewkowycz:2013nqa}
A.~Lewkowycz and J.~Maldacena, {\itshape {Generalized gravitational entropy}},
  {\em JHEP} {\bfseries 08} (2013) 090,
  [\href{http://arxiv.org/abs/1304.4926}{{\ttfamily arXiv:1304.4926}}].

\bibitem{Dong:2016fnf}
X.~Dong, {\itshape {The Gravity Dual of Renyi Entropy}},  {\em Nature Commun.}
  {\bfseries 7} (2016) 12472,
  [\href{http://arxiv.org/abs/1601.06788}{{\ttfamily arXiv:1601.06788}}].

\bibitem{Arias:2019pzy}
C.~Arias, F.~Diaz, and P.~Sundell, {\itshape {De Sitter Space and
  Entanglement}},  {\em Class. Quant. Grav.} {\bfseries 37} (2020), no.~1
  015009, [\href{http://arxiv.org/abs/1901.04554}{{\ttfamily
  arXiv:1901.04554}}].

\bibitem{Carlip:2002be}
S.~Carlip, {\itshape {Near horizon conformal symmetry and black hole entropy}},
   {\em Phys. Rev. Lett.} {\bfseries 88} (2002) 241301,
  [\href{http://arxiv.org/abs/gr-qc/0203001}{{\ttfamily gr-qc/0203001}}].

\bibitem{Holst:1997tm}
S.~Holst and P.~Peldan, {\itshape {Black holes and causal structure in anti-de
  Sitter isometric space-times}},  {\em Class. Quant. Grav.} {\bfseries 14}
  (1997) 3433--3452, [\href{http://arxiv.org/abs/gr-qc/9705067}{{\ttfamily
  gr-qc/9705067}}].

\bibitem{meta}
D.~De~Filippi, C.~Iazeolla, and P.~Sundell, {\itshape {Metaplectic
  representation and ordering (in)dependence in Vasiliev\textquoteright{}s
  higher spin gravity}},  {\em JHEP} {\bfseries 07} (2022) 003,
  [\href{http://arxiv.org/abs/2111.09288}{{\ttfamily arXiv:2111.09288}}].

\bibitem{Mishasiegel}
O.~A. Gelfond and M.~A. Vasiliev, {\itshape {Higher Spin Fields in Siegel
  Space, Currents and Theta Functions}},  {\em JHEP} {\bfseries 03} (2009) 125,
  [\href{http://arxiv.org/abs/0801.2191}{{\ttfamily arXiv:0801.2191}}].

\bibitem{Didenko:2009td}
V.~E. Didenko and M.~A. Vasiliev, {\itshape {Static BPS black hole in 4d
  higher-spin gauge theory}},  {\em Phys. Lett.} {\bfseries B682} (2009)
  305--315, [\href{http://arxiv.org/abs/0906.3898}{{\ttfamily
  arXiv:0906.3898}}]. [Erratum: Phys. Lett.B722,389(2013)].

\bibitem{Iazeolla:2012nf}
C.~Iazeolla and P.~Sundell, {\itshape {Biaxially symmetric solutions to 4D
  higher-spin gravity}},  {\em J. Phys. A} {\bfseries 46} (2013) 214004,
  [\href{http://arxiv.org/abs/1208.4077}{{\ttfamily arXiv:1208.4077}}].

\bibitem{Sundell:2016mxc}
P.~Sundell and Y.~Yin, {\itshape {New classes of bi-axially symmetric solutions
  to four-dimensional Vasiliev higher spin gravity}},  {\em JHEP} {\bfseries
  01} (2017) 043, [\href{http://arxiv.org/abs/1610.03449}{{\ttfamily
  arXiv:1610.03449}}].

\bibitem{Didenko:2021vdb}
V.~E. Didenko and A.~V. Korybut, {\itshape {Planar solutions of higher-spin
  theory. Part II. Nonlinear corrections}},
  \href{http://arxiv.org/abs/2110.02256}{{\ttfamily arXiv:2110.02256}}.

\bibitem{Didenko:2023txr}
V.~E. Didenko and A.~V. Korybut, {\itshape {Towards higher-spin symmetry
  breaking in the bulk}},  \href{http://arxiv.org/abs/2312.11096}{{\ttfamily
  arXiv:2312.11096}}.

\bibitem{Folland}
G.~Folland, {\em Harmonic Analysis in Phase Space}.
\newblock Annals of Mathematics Studies. Princeton University Press, 1989.

\bibitem{CSM}
R.~Carter, G.~Segal, and I.~MacDonald, {\em {Lectures on Lie groups and Lie
  algebras}}.
\newblock Cambridge University Press, 1995.

\bibitem{Guillemin:1990ew}
V.~Guillemin and S.~Sternberg, {\em Symplectic Techniques in Physics}.
\newblock Cambridge University Press, 1990.

\bibitem{Woit:2017vqo}
P.~Woit, {\em {Quantum Theory, Groups and Representations}}.
\newblock Springer, 2017.

\bibitem{Arias:2016agc}
C.~Arias, P.~Sundell, and A.~Torres-Gomez, {\itshape {Differential Poisson
  Sigma Models with Extended Supersymmetry}},
  \href{http://arxiv.org/abs/1607.00727}{{\ttfamily arXiv:1607.00727}}.

\bibitem{Aros:2017ror}
R.~Aros, C.~Iazeolla, J.~Nore\~na, E.~Sezgin, P.~Sundell, and Y.~Yin, {\itshape
  {FRW and domain walls in higher spin gravity}},  {\em JHEP} {\bfseries 03}
  (2018) 153, [\href{http://arxiv.org/abs/1712.02401}{{\ttfamily
  arXiv:1712.02401}}].

\bibitem{Prokushkin:1998vn}
S.~Prokushkin and M.~A. Vasiliev, {\itshape {3-d higher spin gauge theories
  with matter}},  in {\em {2nd International Seminar on Supersymmetries and
  Quantum Symmetries}: {Dedicated to the Memory of Victor I. Ogievetsky}}, 12,
  1998.
\newblock \href{http://arxiv.org/abs/hep-th/9812242}{{\ttfamily
  hep-th/9812242}}.

\bibitem{Gunaydin:1999jb}
M.~Gunaydin, {\itshape {AdS / CFT dualities and the unitary representations of
  noncompact groups and supergroups: Wigner versus Dirac}},  in {\em {6th
  International Wigner Symposium (WIGSYM 6)}}, 8, 1999.
\newblock \href{http://arxiv.org/abs/hep-th/0005168}{{\ttfamily
  hep-th/0005168}}.

\bibitem{Bars}
I.~Bars, {\itshape {Conformal symmetry and duality between free particle, H -
  atom and harmonic oscillator}},  {\em Phys. Rev. D} {\bfseries 58} (1998)
  066006, [\href{http://arxiv.org/abs/hep-th/9804028}{{\ttfamily
  hep-th/9804028}}].

\bibitem{FF}
M.~Flato and C.~Fronsdal, {\itshape {One Massless Particle Equals Two Dirac
  Singletons: Elementary Particles in a Curved Space. 6.}},  {\em Lett. Math.
  Phys.} {\bfseries 2} (1978) 421--426.

\bibitem{wip}
F.~Diaz, C.~Iazeolla, and P.~Sundell. work in progress.

\end{thebibliography}
\end{document}